\documentclass[
  aps,
  prx,
  twocolumn,
  superscriptaddress,
  longbibliography,
  nofootinbib
]{revtex4-2}

\usepackage{amsmath}
\usepackage{amssymb}
\usepackage{bm}
\usepackage{graphicx}
\usepackage{braket}
\usepackage[colorlinks=true,
            linkcolor=blue,
            citecolor=blue,
            urlcolor=blue]{hyperref}

\usepackage[dvipsnames]{xcolor}

\usepackage{hyperref}
\usepackage[nameinlink]{cleveref} 

\crefname{section}{Sec.}{Secs.}
\Crefname{section}{Sec.}{Secs.}

\newtheorem{theorem}{Theorem}[section]

\newtheorem{lemma}[theorem]{Lemma}

\usepackage[normalem]{ulem}

\begin{document}

\title{Designing Robust Quantum Gates}

\author{Yuanxing Hao}
\email{yuanxing@pks.mpg.de}
\affiliation{Max Planck Institute for the Physics of Complex Systems, N\"{o}thnitzer Str.~38, 01187 Dresden, Germany}
\author{Marin Bukov}
\email{mgbukov@pks.mpg.de}
\affiliation{Max Planck Institute for the Physics of Complex Systems, N\"{o}thnitzer Str.~38, 01187 Dresden, Germany}
\date{\today}

\newcommand{\staticcoe}{\ensuremath{u_0}}
\newcommand{\staticterm}{\ensuremath{H_0}}
\newcommand{\schedulecoe}{\ensuremath{u_j(t)}}
\newcommand{\scheduleterm}{\ensuremath{H_j}}
\newcommand{\identity}{\mathbb{I}}
\newcommand{\dimension}{d}
\newcommand{\target}{U_\ast}
\newcommand{\maxlind}{r_{\mathrm{Lind}}}

\begin{abstract}
For given control channels, quantum control theory determines whether quantum gates can be realized, but not their resilience to errors. We address this problem by establishing existence criteria for robust gate implementation in $\dimension$-level systems. Our criteria reveal that implementability of all gates is not enough: a universal family of robust gates exists if and only if, in addition, the driven Hamiltonian operators share no common symmetry. When these criteria hold, we can make every gate resilient to any combination of static perturbations, uniform Hamiltonian rescaling, perturbations at a finite set of selected frequencies, and perturbations with arbitrary selected waveforms. We can impose robustness conditions broadly with limited information about errors, e.g., against any static perturbations; we can also tailor these conditions to the platform, e.g., against environment-induced noise with dominant spectral peaks, requiring no spectral details beyond the peak frequencies.  We numerically construct explicit protocols for robust gates, reducing average gate infidelity by approximately two to seven orders of magnitude under Hamiltonian perturbations and environment-induced noise. Together, our existence criteria and numerical constructions provide practical guidance for evaluating the potential of quantum platforms and designing robust quantum gates.
\end{abstract}

\maketitle

\section{Introduction}
\label{sec:introduction}

Quantum computers are predicted to enable the implementation of a wide range of quantum algorithms, which are difficult to execute efficiently on classical computers~\cite{Shor1999, Montanaro_2016}. However, realizing quantum computation requires preserving quantum coherence in the presence of various errors~\cite{Ladd_2010, Preskill_2018}. These errors include, among others, environment-induced noise~\cite{RevModPhys.86.361, Chirolli_2008, Szankowski2017}, control errors~\cite{PhysRevA.90.012316, PhysRevA.70.052318}, and crosstalk~\cite{PhysRevLett.109.240504, Sarovar_2020}. Since such errors can accumulate in quantum circuits, constructing robust quantum gates is of critical importance~\cite{PhysRevLett.132.250604, doi:10.1137/S0097539799359385}.

Quantum control can improve gate robustness without hardware upgrades. Such robustness usually comes from analytical constructions in specific cases or from numerical optimizations~\cite{Glaser2015training, Watanabe2024ZZ, Li2024CrossResonance, SchulteHerbrueggen2011, PhysRevA.85.040302, Dong2016, PhysRevLett.118.150502, PhysRevLett.121.180502, PhysRevA.101.052317, PhysRevA.89.022310, Yang2016, Bando_2013, PhysRevA.100.042305, iSWAPmjzy, PhysRevApplied.21.034042, PhysRevApplied.15.064054}. Analytical constructions typically tailor the controls to suppress the effects of errors; examples include dynamically corrected gates (DCGs) against different perturbations~\cite{DCGViola2009, PhysRevA.80.032314, PhysRevLett.110.070503, PRXQuantum.2.010341, PhysRevA.99.052321} and super-robust geometric gates against the uniform time-stretching error~\cite{EGGoriginal}. However, these constructions are usually restricted to specific control settings or specific errors. Numerical optimization, on the other hand, does not require an analytical solution; it can therefore address more complex settings, as in numerically synthesized DCGs~\cite{PhysRevA.86.042329} and robust gates against environment-induced noise with known spectra~\cite{Ding_2025, PhysRevA.95.062325, Khan2026Thesis}. It even achieves robustness against \textit{arbitrary} static perturbations without knowing the error operators~\cite{DCGpoggi2024}. However, it provides no guarantee that a robust solution exists with the available controls~\cite{PhysRevA.91.043401}.

Without existence criteria, one cannot determine in advance which robust gates the available controls can implement. A failed numerical optimization does not rule out the existence of a robust solution~\cite{PhysRevA.91.043401}. In practice, multiple error sources, possibly of different types, typically act simultaneously~\cite{PhysRevA.108.012407, Floether_2012, Wang2026offresonance}; accordingly, robust gates should satisfy combinations of robustness conditions. These errors are often only partially characterized. For example, spectrum-based methods need spectral details~\cite{Ding_2025, PhysRevA.95.062325, PhysRevApplied.19.014062}, but measurement capabilities can restrict the characterization of noise-power spectra~\cite{PhysRevApplied.11.014064}; in contrast, their peak frequencies are easier to measure~\cite{Vezvaee2024FTNS, PhysRevLett.110.110503}. On the control side, experimental platforms have only limited control channels~\cite{PhysRevResearch.4.023155}, often including static couplings;
limited control further restricts robust gate implementation~\cite{PhysRevApplied.23.054002, PRXQuantum.2.010341}. For gates without the requirement of robustness, controllability theory gives an existence criterion: a gate can be implemented if and only if it is in the group generated by the available control Hamiltonians~\cite{PhysRevA.51.960, Albertini_2003}. Crucially, however, it does not determine whether robust gate implementations exist~\cite{Dey2026robust, Albertini2026robust}. Together, these limitations leave a key open problem: within the available control channels, what determines which robust gates can be implemented?

In this work, we address this question by establishing general existence criteria for robust gates, using the available control Hamiltonians alone. Here, robust gates eliminate the leading-order effects of specific errors. To implement arbitrary gates in $\dimension$-level systems, the available control Hamiltonians need to generate the $\mathfrak{su}(\dimension)$ algebra. One might expect that this capability at least guarantees that every target gate can be implemented robustly against arbitrary static perturbations. However, for implementing every target gate with this robustness, there is an extra necessary and sufficient condition: the driven Hamiltonian operators share no common symmetry. The absence of common symmetry allows us to impose the robustness conditions against static perturbations, and more generally, different forms of robustness and their combinations. These robustness conditions can counter perturbations with selected frequencies~\cite{Hayes2011Walsh} or selected waveforms~\cite{Piltz2014}, and uniform time-stretching errors~\cite{EGGoriginal}. Beyond the $\mathfrak{su}(\dimension)$ case, we use a controllability theorem on semidirect product groups~\cite{Sachkov2009theorems}, giving sufficient existence criteria to robustly implement every gate reachable with the available controls. For Lindblad noise, our analysis separates the suppressible contribution from the fixed depolarizing component, exposing a fundamental limitation on the suppression of Lindblad noise by coherent Hamiltonian control~\cite{Fidlindabad2022}.

Once existence is guaranteed, we develop a numerical algorithm to construct explicit protocols. The possibility of adding different robustness conditions enables different strategies in experiments. With less information about errors, one can impose robustness conditions broadly, such as those against arbitrary static perturbations, as we show in Sec.~\ref{sec:numerical-A}. It is also possible to tailor the robustness conditions according to experimental platforms. Specifically, if heat baths have dominant peaks, one can impose robustness conditions based only on peak frequencies and coupling operators. This robustness can suppress the infidelity by several orders of magnitude, as we show in Sec.~\ref{sec:numerical-B}. In Appendices~\ref{app:additional-gate-examples}--\ref{app:waveform-selective-examples}, we further apply the algorithm to a single-qubit gate, a qutrit gate, the $\sqrt{i\mathrm{SWAP}}$ gate, as well as waveform-selective robust qutrit and CZ gates. These examples illustrate the generality of our framework, which applies to arbitrary quantum gates in general $\dimension$-level systems. Our research links the existence of robust gate implementations to control channels in experiments, guiding the design of robust quantum gates tailored to specific experiments.

\section{Criteria for Realizing Robust Quantum Gates}
\label{sec:theoretical}

In this section, we introduce the criteria for assessing whether a family of robust gates exists with given control channels. For Hamiltonian perturbations, we consider extended geometric gates (EGGs)~\cite{EGGoriginal} and dynamically corrected gates (DCGs)~\cite{DCGViola2009, PhysRevA.80.032314}. In this work, we consider three classes of DCGs: static DCGs, finite-frequency DCGs, and waveform-selective DCGs. We treat Lindblad noise~\cite{Lindblad1976, GKS1976} separately, since its robustness conditions require a superoperator formulation and its suppression by Hamiltonian control has a fundamental limitation. In what follows, we subtract all terms proportional to the identity from all Hamiltonians, restricting the unitary dynamics to the $\mathrm{SU}(\dimension)$ group.\footnote{We use the convention in which the Lie algebra $\mathfrak{su}(\dimension)$ consists of traceless anti-Hermitian matrices.}

To construct these robust gates in a $\dimension$-level system, we choose the instantaneous Hamiltonian as
\begin{equation}
H(t)=\staticcoe\staticterm+\sum_{j=1}^{m}\schedulecoe\scheduleterm,\ \ i\staticterm,\ i\scheduleterm\in\mathfrak{su}(\dimension),
\label{eq:control_resources}
\end{equation}
where the dimensionless Hermitian operators $\staticterm$ and $\scheduleterm$, $j=1,\cdots,m$, are the static and driven Hamiltonian operators, respectively. The static coefficient $\staticcoe$ and the schedules $\schedulecoe$ have units of energy. In the main setting, we prescribe the gate duration $T$, choose a constant $\staticcoe$ separately for each gate protocol, and take the schedules $\schedulecoe$ to be smooth $T$-periodic functions. This setting allows us to synthesize the corresponding control pulses using Fourier components at the integer harmonics $2\pi k/T$. Similar structures have appeared in control protocols for cavity-based systems~\cite{FourierRoque2021}, superconducting circuits~\cite{FourierSong2022,FourierHyyppa2024}, neutral atoms~\cite{FourierWang2025}, and trapped-ion processors~\cite{FourierBlumel2021}. A given target gate may be implementable using different choices of $\staticcoe$ and $\schedulecoe$; we exploit this freedom below to impose robustness conditions.

This setting is not the only choice, though: in Appendix~\ref{app:reachability-results}, we prove that piecewise-constant schedules share the same existence criteria as the smooth $T$-periodic driven schedules $\schedulecoe$. We obtain the same criteria for continuous schedules on $[0, T]$, with or without periodicity. In Sec.~\ref{sec:theoryB}, we further show that, except for general waveform-selective robustness, the same existence criteria hold when we fix the static coefficient $\staticcoe$ at any nonzero value and choose the gate duration $T$ separately for each protocol. Previous work has also implemented piecewise-constant pulse parameterizations experimentally in superconducting-qubit systems~\cite{Werninghaus2021}. We also construct a robust single-qubit gate using piecewise-constant controls in Appendix~\ref{app:single-qubit-example}, with the corresponding results shown in Fig.~\ref{fig-single-qubit-robustness}.

In the main setting, the choice of $T$ does not affect the existence of gates satisfying the robustness conditions against static perturbations, uniform time-stretching errors, and Lindblad noise. Rescaling a protocol in time, together with a reciprocal rescaling of $\staticcoe$ and $\schedulecoe$, can only change the perturbation strengths of these errors. For perturbations at selected frequencies or with selected waveforms, the rescaling changes these frequencies and waveforms, and requiring all robust gate implementations at a prescribed $T$ is a stronger condition than allowing $T$ to be chosen freely. Nevertheless, the existence criteria for static perturbations with the same error operators suffice to construct all robust gate implementations at every prescribed $T$, as we will show later in this section.

The following subsections introduce the robustness conditions, present concise existence criteria under the unitary controllability condition, and establish the existence criteria for general cases. Together, these results give sufficient algebraic conditions for implementing a family of robust gates with the available static and driven operators. The family covers every target gate that these operators can implement in the absence of errors. In the main setting, each robust gate uses the same prescribed duration $T$ and satisfies the same robustness constraints. We also prove that the full family of DCGs against arbitrary static perturbations exists only if the driven operators share no common symmetry. For specific perturbation directions and most target gates, common symmetries impose leading-order error floors, and no choice of schedules or gate duration can lower them. These criteria guide our choice of robustness constraints for the numerical optimization in Sec.~\ref{sec:numerical}.

\subsection{Robust-Gate Conditions}
\label{sec:theoryA}

We first specify the robust-gate conditions. Each Hamiltonian robustness condition eliminates the leading-order effect of a specified class of perturbations: static perturbations for the static DCG conditions, perturbations at selected frequencies for the finite-frequency DCG conditions, perturbations with selected waveforms for the waveform-selective DCG conditions, and uniform Hamiltonian rescaling for the EGG condition. These robustness conditions allow us to push the average gate infidelity to higher orders in the perturbation strength.


\textbf{\textit{Static DCGs}} can suppress static perturbations and low-frequency noise components, including those from a dynamical quantum environment~\cite{DCGViola2009, PhysRevA.80.032314}. For systems subject to the perturbation $H(t)\to H(t)+\epsilon X$ with $|\epsilon T|\ll1$, the average gate infidelity takes the form~\cite{fidelitynielsen2002249, DCGpoggi2024}
\begin{equation}
\begin{aligned}
1-F_{\text{avg}}&=\frac{(\epsilon T)^2}{\dimension+1}\left(\chi^{(\mathrm D)}\right)^2+O(\epsilon^3T^3),\\
\chi^{(\mathrm D)}&=\left\|\frac{1}{T}\int_0^T\mathrm{d}t\,U_0^{\dagger}(t)XU_0(t)\right\|,
\end{aligned}
\label{eq:infidelity-DCG}
\end{equation}
where $U_0(t)=\mathcal{T}\exp[-i\int_0^t\mathrm{d}\tau\,H(\tau)]$ is the ideal unperturbed evolution operator, $\dimension$ is the Hilbert space dimension, $\|\cdot\|$ denotes the Frobenius norm, and $\chi^{(\mathrm D)}$ is the susceptibility to the perturbation $X$. The static DCG condition for first-order error cancellation~\cite{FrequencyPazSilva2014} is $\chi^{(\mathrm D)}=0$, i.e.,
\begin{equation}
\frac{1}{T}\int_0^T\mathrm{d}t\,U_0^{\dagger}(t)XU_0(t)=0,
\label{eq:DCG-condition}
\end{equation}
which directly eliminates the leading-order contribution to the average gate infidelity in Eq.~\eqref{eq:infidelity-DCG}, pushing the dominant infidelity term to higher orders. Spin echo~\cite{spinechoPress2010, Hahn1950}, which uses a special control sequence to cancel the accumulated effects of slowly varying perturbations in a single qubit, is closely related to this mechanism. Furthermore, concatenated DCGs can in principle extend this cancellation to arbitrarily high order~\cite{Khodjasteh2010Concatenated}.

In experiments, the precise operator form of a static perturbation may be unknown, even when it can be restricted to a known operator subspace~\cite{DCGpoggi2024}. If no such restriction is available, we should take the perturbation space to be the space of all traceless Hermitian operators. Let $X_1, \cdots, X_r$ be traceless Hermitian operators specifying $r$ perturbation directions that span the chosen perturbation space. Correspondingly, we introduce $\chi_p^{(\mathrm D)}$, which denotes the susceptibility in Eq.~\eqref{eq:infidelity-DCG} evaluated at $X=X_p$. A static DCG is resilient to all perturbations in $\operatorname{span}_{\mathbb{R}}\{X_1,\cdots,X_r\}$ if $\chi_p^{(\mathrm D)}=0$, i.e.,
\begin{equation}
\frac{1}{T}\int_0^T\mathrm{d}t\,U_0^{\dagger}(t)X_pU_0(t)=0,\quad 
\forall p\in\{1,\cdots,r\}.
\label{eq:DCG-many-condition}
\end{equation}
Due to linearity of the integral operation in $X$, these conditions imply that the susceptibility vanishes for every $X\in\operatorname{span}_{\mathbb{R}}\{X_1, \cdots, X_r\}$. Consequently, for the entire class of static Hamiltonian perturbations in this span, these conditions eliminate all leading-order contributions to the average gate infidelity.
In particular, this universal robustness relinquishes the need to identify and quantify (to leading order) the precise perturbation channel in experiments.


\textbf{\textit{Finite-frequency DCGs}} extend the static DCG condition by adding oscillations to the perturbations. For each perturbation direction $X_p$, we associate a finite set $\Omega_p=\{\omega_{p,k}\}$ of selected nonnegative frequencies. The positive elements $\omega_{p,k}$ of this set represent oscillation modes, while the zero element, if present, represents a static perturbation. We now consider the perturbation
\begin{equation}
H(t)\to H(t)+\sum_{p,k}\epsilon_{p,k}X_p
\cos\!\left(\omega_{p,k}t+\phi_{p,\omega_{p,k}}\right);
\label{eq:finite-frequency-perturbation}
\end{equation}
for each $p\in\{1,\cdots,r\}$, the sum over $k$ runs over all $\omega_{p,k}\in\Omega_p$. We remain in the perturbative regime, where $|\epsilon_{p,k}T|\ll1$ for every $p$ and $k$. Under this perturbation, the leading-order average gate infidelity is
\begin{equation}
\begin{aligned}
1-F_{\mathrm{avg}}&\simeq
\frac{T^2}{4(\dimension+1)}\Bigg\|\sum_{p,k}\epsilon_{p,k}
e^{i\phi_{p,\omega_{p,k}}}
\mathcal A_p^{(\mathrm D)}(\omega_{p,k})
+\mathrm{h.c.}\Bigg\|^2,
\end{aligned}
\label{eq:infidelity-finite-frequency-Hamiltonian}
\end{equation}
where 
\begin{equation}
    \mathcal A_p^{(\mathrm D)}(\omega) =\frac{1}{T}\int_0^T\mathrm{d}t\, U_0^\dagger(t)X_pU_0(t)e^{i\omega t}
\end{equation}
determines the leading-order error generated by $X_p$ at frequency $\omega$, and we define the corresponding susceptibility as $\chi_p^{(\mathrm D)}(\omega)=\|\mathcal A_p^{(\mathrm D)}(\omega)\|$. Since $\mathcal A_p^{(\mathrm D)}(-\omega)=\mathcal A_p^{(\mathrm D)\dagger}(\omega)$, we can take every $\omega_{p,k}$ to be nonnegative. 

For arbitrary values of the phases $\phi_{p,\omega_{p,k}}$, $T^2/(\dimension+1)\big[\sum_{p,k}|\epsilon_{p,k}|\chi_p^{(\mathrm D)}(\omega_{p,k})\big]^2$ forms an upper bound on the leading-order average gate infidelity, as we show in Appendix~\ref{app:infidelity-Hamiltonian}. Consequently, the finite-frequency DCG conditions are $\chi_p^{(\mathrm D)}(\omega_{p,k})=0$, i.e.,
\begin{equation}
\mathcal A_p^{(\mathrm D)}(\omega_{p,k})
=\frac{1}{T}\int_0^T\mathrm{d}t\,
U_0^\dagger(t)X_pU_0(t)e^{i\omega_{p,k}t}=0,
\label{eq:DCG-many-with-frequency}
\end{equation}
for every $p\in\{1,\cdots,r\}$ and every $\omega_{p,k}\in\Omega_p$. These conditions eliminate the leading-order effect of the perturbations, with multiple perturbation terms and multiple frequencies acting simultaneously. 

At $\omega_{p,k}=0$, we have $\chi_p^{(\mathrm D)}(0)=\chi_p^{(\mathrm D)}$, and we recover the static DCG conditions. At nonzero frequencies, dynamical decoupling~\cite{ViolaLloyd1998, ViolaKnillLloyd1999} provides a related frequency-selective mechanism~\cite{FrequencyPazSilva2014, Malinowski2017}, as do DCGs with a filter-function notch~\cite{Hayes2011Walsh}. Such discrete-frequency noise has been observed on a trapped-ion qubit~\cite{Kotler2011}.

For a system weakly coupled to a stationary environment and subject to static perturbations, finite-frequency DCG conditions can also eliminate or suppress the leading-order effects of the gate errors. We use the one-sided spectral matrix $S_{pq}(\nu)$ to represent the symmetrized two-point correlations of the stationary heat bath, where $\nu>0$ is the angular frequency and $p,q\in\{1,\cdots,r\}$ label the system coupling operators $X_p$ and $X_q$. The diagonal entries $S_{pp}(\nu)$ are the noise-power spectra associated with the system coupling operator $X_p$, while the off-diagonal entries $S_{pq}(\nu)$ $(p\neq q)$ describe the correlations between the environmental fluctuations coupled through $X_p$ and $X_q$~\cite{PazSilva2017,vonLupke2020}. The definition of $S_{pq}(\nu)$ includes the system--environment coupling strengths; we also include the static perturbation directions $X_p$ with amplitudes $\epsilon_{p,0}$ for which $0\in\Omega_p$. For the couplings and static perturbations considered here, the leading-order average gate infidelity takes the spectral-overlap form~\cite{FrequencyGreen2013, FrequencyPazSilva2014}
\begin{equation}
\begin{aligned}
1-F_{\mathrm{avg}}
&\simeq\frac{T^2}{\dimension+1}\Bigg\{
\left\|\sum_p\epsilon_{p,0}
\mathcal A_p^{(\mathrm D)}(0)\right\|^2\\
&\quad+\int_0^\infty\mathrm{d}\nu\sum_{p,q}
S_{pq}(\nu)\operatorname{tr}\!\left[
\mathcal A_p^{(\mathrm D)\dagger}(\nu)
\mathcal A_q^{(\mathrm D)}(\nu)\right]\Bigg\}.
\end{aligned}
\label{eq:infidelity-finite-frequency}
\end{equation}
In the static term (top line), the sum runs over the static perturbation directions $X_p$ with $0\in\Omega_p$; in the spectral term, the sum runs over all pairs of system coupling operators $X_p$ and $X_q$. The details of the heat bath, the definition of $S_{pq}(\nu)$, and the derivation of Eq.~\eqref{eq:infidelity-finite-frequency} are given in Appendix~\ref{app:infidelity-environment}.

For discrete spectra, we have $S_{pq}(\nu)=\sum_jG_{pq}^{(j)}\delta[\nu-\omega_j]$, where $\omega_j>0$ is the frequency of spectral line $j$, and each matrix $G^{(j)}=[G_{pq}^{(j)}]$ is Hermitian and positive semidefinite. By associating the frequency set $\Omega_p$ with each system coupling operator $X_p$ such that $\omega_j\in\Omega_p\cup\Omega_q$ whenever $G_{pq}^{(j)}\neq0$, we can cover all spectral-line frequencies by the robustness conditions. At each such frequency, setting the response of at least one of $X_p$ and $X_q$ to zero eliminates the corresponding spectral contribution for any $p$ and $q$. Covering the corresponding static and spectral-line frequencies with the conditions $\chi_p^{(\mathrm D)}(\omega)=0$ fully eliminates the leading-order effects in Eq.~\eqref{eq:infidelity-finite-frequency}, including the static term, self-spectral terms, and cross-spectral terms. The correlations between the environment-induced noise coupled through $X_p$ and $X_q$ ($p\neq q$) do not alter the form of these sufficient robustness conditions for eliminating the leading-order term.

Experimental noise spectra are often continuous but concentrated around distinct peak frequencies~\cite{Akerman2015, Yan2013, Malinowski2017}. Therefore, we can impose the robustness conditions at the peak frequencies, which significantly reduce the dominant leading-order contribution to the average gate infidelity, as we show numerically in Sec.~\ref{sec:numerical}. This can effectively improve gate robustness under realistic experimental conditions.

\textbf{\textit{Waveform-selective DCGs}}. By considering perturbations described by selected time-dependent functions, we further generalize the static DCG and finite-frequency DCG conditions to the waveform-selective DCG conditions. In experiments, time-dependent perturbations, whose specific forms depend on the platform, can be important sources of gate errors~\cite{PhysRevLett.109.240504, Piltz2014}:
\begin{equation}
H(t)\to H(t)+\sum_{p,\ell}\epsilon_{p,\ell}f_{p,\ell}(t)X_p,
\label{eq:waveform-selective-perturbation}
\end{equation}
where $f_{p,\ell}(t)$ are selected smooth real functions, which may come from experimental characterization and need not be periodic (e.g., $e^{-t/\tau}$ with decay time $\tau>0$). These functions do not depend on the control coefficients $\staticcoe$ and $\schedulecoe$. We consider a finite number of functions $f_{p,\ell}(t)$, and $\epsilon_{p,\ell}$ are small unknown real parameters. To leading order, the average gate infidelity is
\begin{equation}
\begin{aligned}
1-F_{\mathrm{avg}}
&\simeq
\frac{T^2}{\dimension+1}
\left\|
\sum_{p,\ell}\epsilon_{p,\ell}
\mathcal A_p^{(\mathrm D)}[f_{p,\ell}]
\right\|^2,\\
\mathcal A_p^{(\mathrm D)}[f_{p, \ell}]
&=
\frac{1}{T}\int_0^T\mathrm{d}t\,
f_{p, \ell}(t)U_0^\dagger(t)X_pU_0(t),
\end{aligned}
\label{eq:infidelity-waveform-selective}
\end{equation}
where $\mathcal A_p^{(\mathrm D)}[f_{p, \ell}]$ determines the leading-order error associated with the perturbation direction $X_p$ and selected function $f_{p, \ell}(t)$, and we define the corresponding susceptibility as $\chi_p^{(\mathrm D)}[f_{p,\ell}]=\|\mathcal A_p^{(\mathrm D)}[f_{p,\ell}]\|$.

Consequently, the waveform-selective DCG conditions are
$\chi_p^{(\mathrm D)}[f_{p,\ell}]=0$, i.e.,
\begin{equation}
\mathcal A_p^{(\mathrm D)}[f_{p,\ell}]
=
\frac{1}{T}\int_0^T\mathrm{d}t\,
f_{p,\ell}(t)U_0^\dagger(t)X_pU_0(t)=0,
\label{eq:waveform-selective-condition}
\end{equation}
for every $p$ and $\ell$. These conditions eliminate the leading-order effect of any perturbation formed from the selected functions, independently of the coefficients $\epsilon_{p,\ell}$. The static DCG condition corresponds to the constant function $f(t)=1$, while the finite-frequency DCG condition at a positive frequency $\omega$ is recovered by imposing Eq.~\eqref{eq:waveform-selective-condition} for both $f(t)=\cos(\omega t)$ and $f(t)=\sin(\omega t)$. This broader condition allows us to construct robust gates for experimentally relevant perturbations with general time dependence, including nonperiodic functions, beyond the static and purely oscillatory cases, as we show in Appendix~\ref{app:waveform-selective-examples}.


\textbf{\textit{Extended\footnote{We extend the term to emphasize that the condition extends the ordinary geometric gate condition by imposing restrictions on off-diagonal terms.} geometric gates}} (EGGs) correspond to the super-robust geometric gates introduced in previous work~\cite{EGGoriginal}. For systems subject to uniform time-stretching errors $\mathrm{d}t\to(1+\delta)\mathrm{d}t\, (|\delta|\ll1)$, or equivalently, to uniform Hamiltonian-rescaling errors $H(t)\to(1+\delta)H(t)$, the average gate infidelity has the expansion 
\begin{equation}
\begin{aligned}
    1-F_{\text{avg}}&=\frac{\delta^2}{\dimension+1}\left(\chi^{(\mathrm E)}\right)^2+O(\delta^3),\\
    \chi^{(\mathrm E)}&=\left\|\int_0^T\mathrm{d}t\, U_0^{\dagger}(t)H(t)U_0(t)\right\|.
\end{aligned}
\label{eq:infidelity_EGG}
\end{equation}
As in the preceding examples, the extended geometric gate condition is $\chi^{(\mathrm E)}=0$, i.e.,
\begin{equation}
\int_0^T\mathrm{d}t\, U_0^{\dagger}(t)H(t)U_0(t)=0.
\label{eq:EGG}
\end{equation}
In contrast, ordinary geometric gates satisfy
\begin{equation}
\int_0^T\mathrm{d}t\bra{\psi_n}U_0^{\dagger}(t)H(t)U_0(t)\ket{\psi_n}=0,\quad n=1,\cdots, \dimension,
\end{equation}
where $\{\ket{\psi_n}\}$ is an eigenbasis of the gate operator $U_0(T)$. The geometric condition means that the accumulated phases are purely geometric, with vanishing dynamical phases. These geometric phases might help protect the gates against certain perturbations~\cite{PhysRevA.72.020301}, but the ordinary geometric condition alone provides no explicit guarantee of robustness~\cite{PhysRevA.93.032313}. In particular, the ordinary geometric condition only sets the diagonal elements of the integral in Eq.~\eqref{eq:EGG} to zero in the eigenbasis $\{\ket{\psi_n}\}$, while the off-diagonal elements still contribute to $\chi^{(\mathrm E)}$ in Eq.~\eqref{eq:infidelity_EGG}. Therefore, ordinary geometric gates generally have a nonzero leading-order contribution to the average gate infidelity under uniform Hamiltonian rescaling. The extended geometric gate condition also sets these off-diagonal elements to zero, so it keeps the gate geometric while eliminating the leading-order contribution to the average gate infidelity~\cite{EGGoriginal}.

\begin{table*}[t!]
\centering
\small
\renewcommand{\arraystretch}{1.16}
\setlength{\tabcolsep}{3pt}
\setlength{\arrayrulewidth}{0.5pt}
\begin{tabular}{|l|l|l|l|}
\hline
\parbox[t]{0.145\textwidth}{\raggedright \textbf{Type}} &
\parbox[t]{0.240\textwidth}{\raggedright \textbf{Robustness constraint}} &
\parbox[t]{0.245\textwidth}{\raggedright \textbf{Existence criterion under full unitary controllability}} &
\parbox[t]{0.315\textwidth}{\raggedright \textbf{Main consequences}} \\
\hline
\parbox[t]{0.145\textwidth}{\raggedright Static DCGs (\textit{arbitrary})} &
\raisebox{-1pt}{\parbox[t]{0.240\textwidth}{\raggedright
$\chi_p^{(\mathrm D)}=0$, i.e.,
$\int_0^T \frac{\mathrm{d}t}{T}\, U_0^\dagger(t)X_pU_0(t)=0$}} &
\parbox[t]{0.245\textwidth}{\raggedright
Exists for every target gate $\target\in\mathrm{SU}(\dimension)$ with a given finite period $T>0$, if and only if there is no common symmetry of the driven Hamiltonian operators $\scheduleterm$, i.e., $\mathcal C_{\mathrm{dyn}}=\{0\}$.} &
\parbox[t]{0.315\textwidth}{\raggedright
Eliminates the leading-order contribution of \textit{arbitrary static Hamiltonian perturbations} acting on the $\dimension$-level system to the average gate infidelity, pushing the dominant term to higher orders.} \\
\hline
\parbox[t]{0.145\textwidth}{\raggedright Finite-frequency DCGs (\textit{arbitrary})} &
\raisebox{-1pt}{\parbox[t]{0.240\textwidth}{\raggedright
$\chi_p^{(\mathrm D)}(\omega_{p,k})=0$, i.e.,
$\int_0^T \frac{\mathrm{d}t}{T}\, U_0^\dagger(t)X_pU_0(t)e^{i\omega_{p,k}t}=0$}} &
\parbox[t]{0.245\textwidth}{\raggedright
Same sufficient condition as that for static DCGs (\textit{arbitrary}), independent of the selected frequency sets $\Omega_p$.\footnotemark[1]} &
\parbox[t]{0.315\textwidth}{\raggedright
Eliminates the leading-order contribution of \textit{selected discrete spectral lines} to the average gate infidelity; can suppress the contribution from continuous spectral peaks near the selected frequencies.} \\
\hline
\parbox[t]{0.145\textwidth}{\raggedright Waveform-selective DCGs (\textit{arbitrary})} &
\raisebox{-1pt}{\parbox[t]{0.240\textwidth}{\raggedright
$\chi_p^{(\mathrm D)}[f_{p,\ell}]=0$, i.e.,
$\int_0^T \frac{\mathrm{d}t}{T}\, f_{p,\ell}(t)U_0^\dagger(t)X_pU_0(t)=0$}} &
\parbox[t]{0.245\textwidth}{\raggedright
Same sufficient condition as that for static DCGs (\textit{arbitrary}), independent of the number and forms of the selected functions $f_{p,\ell}$.} &
\parbox[t]{0.315\textwidth}{\raggedright
Eliminates the leading-order contribution of \textit{perturbations with the selected waveforms}, including nonperiodic ones, to the average gate infidelity.} \\
\hline
\parbox[t]{0.145\textwidth}{\raggedright Extended geometric gates} &
\raisebox{-1pt}{\parbox[t]{0.240\textwidth}{\raggedright
$\chi^{(\mathrm E)}=0$, i.e.,
$\int_0^T \mathrm{d}t\, U_0^\dagger(t)H(t)U_0(t)=0$}} &
\parbox[t]{0.245\textwidth}{\raggedright
Always exists under full unitary controllability.} &
\parbox[t]{0.315\textwidth}{\raggedright
Eliminates the leading-order effect of \textit{uniform Hamiltonian rescaling} on the average gate infidelity, pushing the dominant term to higher orders.} \\
\hline
\parbox[t]{0.145\textwidth}{\raggedright Lindblad-robust gates} &
\raisebox{-1pt}{\parbox[t]{0.240\textwidth}{\raggedright
$\chi_a^{(\mathrm L)}=0$, i.e.,
$\int_0^T \frac{\mathrm{d}t}{T}\, \mathcal U_0^\dagger(t)D_{a,\mathrm{sup}}\mathcal U_0(t)=0$}} &
\parbox[t]{0.245\textwidth}{\raggedright
Fixed depolarization direction $D_{\mathrm{fix}}$ excluded; leading-order effect of the suppressible components $D_{a,\mathrm{sup}}$ can be eliminated if Eq.~\eqref{EQ:full-driving-term} holds; otherwise, Eq.~\eqref{eq:general-criteria} provides a sufficient condition for its elimination.\par\vspace{3pt}} &
\parbox[t]{0.315\textwidth}{\raggedright
Eliminates the leading-order effects of the \textit{suppressible nondepolarizing components}; can improve worst-case gate fidelity, while the fixed depolarizing contribution remains.} \\
\hline
\end{tabular}
\footnotetext[1]{$\mathcal C_{\mathrm{dyn}}=\{0\}$ is also necessary for the full family if all frequency sets share at least one element, i.e., $\bigcap_{p=1}^{\dimension^2-1}\Omega_p\neq\varnothing$.}
\caption{Summary of the robust-gate conditions and their concise existence results under the unitary controllability condition. Here, DCG means dynamically corrected gate. If the driven Hamiltonian operators share no common symmetry, we can implement every gate with any combination of the first four robustness conditions, at any prescribed gate duration $T$. If they share a common symmetry, not all gates can be resilient to arbitrary static perturbations, even with gate-dependent duration $T$.}
\label{tab:robustness-criteria}
\end{table*}

\textbf{\textit{Lindblad-robust gates}} differ from Hamiltonian-noise robust gates. Lindblad noise describes Markovian dissipative processes and does not require information about the noise spectrum~\cite{Lindblad1976, GKS1976}. We describe the weak-noise dynamics by
\begin{equation}
\begin{aligned}
\dot{\rho}(t)&=-i[H(t),\rho(t)]+\sum_{a=1}^{\maxlind}\epsilon_a\mathcal{D}_a(\rho(t)),
\quad \epsilon_aT\ll1,\\
\mathcal{D}_a(\rho)&=
\sum_{x}\left(L_{ax}\rho L_{ax}^{\dagger}
-\frac{1}{2}\{L_{ax}^{\dagger}L_{ax},\rho\}\right).
\end{aligned}
\label{eq:master-equation}
\end{equation}
Here, $a$ labels the dissipative processes, $\epsilon_a\geq0$ is the corresponding Lindblad rate, and $x$ labels the Lindblad operators in $\mathcal{D}_a$. We assume that the dissipators $\mathcal{D}_a$ and rates $\epsilon_a$ do not depend on time or the Hamiltonian controls.

To formulate leading-order robustness conditions similar to the Hamiltonian case, we rewrite the density matrix $\rho$ in the Hilbert--Schmidt space by reshaping it into a vector. This vector satisfies a Schrödinger-type equation:
\begin{equation}
\frac{\partial|\rho(t)\rangle\rangle}{\partial t}=\mathcal{L}(t)|\rho(t)\rangle\rangle,
\end{equation}
where the superoperator $\mathcal{L}(t)$ takes the form
\begin{equation}
\begin{aligned}
\mathcal{L}(t)&=-i\big[H(t)\otimes\identity_{\dimension}-\identity_{\dimension}\otimes H^T(t)\big]+\sum_{a=1}^{\maxlind}\epsilon_aD_a\\
&=-i\big[H(t)\otimes\identity_{\dimension}-\identity_{\dimension}\otimes H^T(t)\big]+\sum_{a=1}^{\maxlind}\epsilon_a\sum_x\\
&\ \bigg(L_{ax}\otimes L_{ax}^{*}-\frac{1}{2}\big[L_{ax}^{\dagger}L_{ax}\otimes\identity_{\dimension}+\identity_{\dimension}\otimes(L_{ax}^{\dagger}L_{ax})^{T}\big]\bigg).
\end{aligned}
\label{eq:vector-form-master}
\end{equation}
Let $\mathcal{L}_0(t)$ denote the Hamiltonian part in Eq.~\eqref{eq:vector-form-master}, which describes the unitary evolution in the absence of noise. The corresponding noise-free propagator is $\mathcal{U}_0(t)=\mathcal{T}\exp{\int_0^t\mathrm{d}\tau\mathcal{L}_0(\tau)}$.

Na\"ively, one might expect a robustness condition analogous to $\int_0^T\frac{\mathrm{d}t}{T}\,\mathcal{U}_0^{\dagger}(t)D_a\mathcal{U}_0(t)=0$. However, $D_a$ could contain a nonzero component along a fixed direction, which takes the form
\begin{equation}
D_{\mathrm{fix}}
=
\frac{1}{\dimension}|\identity_{\dimension}\rangle\rangle\langle\langle \identity_{\dimension}|-\identity_{\dimension^2}.
\end{equation}
Here, $|\identity_{\dimension}\rangle\rangle$ represents the identity matrix in the Hilbert--Schmidt space.
Since $\int_0^T\frac{\mathrm{d}t}{T}\mathcal{U}_0^{\dagger}(t)D_{\mathrm{fix}}\mathcal{U}_0(t)=D_{\mathrm{fix}}$ holds for any $\mathcal{U}_0(t)$, we can never change the leading-order response of this direction by coherent Hamiltonian controls. This direction corresponds to isotropic decoherence, i.e., depolarization.

As we show in Appendix~\ref{app:fixed-Lindblad-direction}, there is no independent fixed direction under all coherent unitary conjugations other than the depolarization direction. Consequently, we define $\lambda_{a,\mathrm{fix}}D_{\mathrm{fix}}$ as the projection of $D_a$ onto the fixed depolarization direction $\operatorname{span}_{\mathbb R}\{D_{\mathrm{fix}}\}$, and $D_a$ has the decomposition
\begin{equation}
D_a=D_{a,\mathrm{sup}}+\lambda_{a,\mathrm{fix}}D_{\mathrm{fix}},
\label{eq:deconposition-lindblad-superoperators}
\end{equation}
where $D_{a,\mathrm{sup}}$ contains no fixed direction under all coherent unitary conjugations. We impose the Lindblad robustness condition only on this suppressible component:
\begin{equation}
\int_0^T\frac{\mathrm{d}t}{T}\, \mathcal{U}_0^{\dagger}(t)D_{a,\mathrm{sup}}\mathcal{U}_0(t)=0,
\quad a=1,\cdots,\maxlind.
\end{equation}

As in previous examples, we can also define the susceptibility in the Hilbert--Schmidt space:
\begin{equation}
\chi_{a}^{(\mathrm L)}=\left\|
\frac{1}{T}\int_0^T\mathrm{d}t\,
\mathcal U_0^{\dagger}(t)D_{a,\mathrm{sup}}
\mathcal U_0(t)\right\|.
\label{eq:Lindblad-susceptibility}
\end{equation}
This susceptibility measures the residual leading-order response of $D_{a,\mathrm{sup}}$.
The decomposition in Eq.~\eqref{eq:deconposition-lindblad-superoperators} has a direct consequence for the average gate infidelity, whose leading-order expansion is~\cite{Fidlindabad2022}
\begin{equation}
1-F_{\text{avg}}=
\frac{\dimension-1}{\dimension}
T\sum_{a=1}^{\maxlind}
\epsilon_a\lambda_{a,\mathrm{fix}}
+O\!\left((T\max_a\epsilon_a)^2\right).
\label{eq:average-infidelity-Lindblad}
\end{equation}

Thus, coherent Hamiltonian controls cannot eliminate the leading-order term of the average gate infidelity $1-F_{\mathrm{avg}}$, which is set by the fixed depolarizing component alone. However, they can suppress the state-dependent effects of the other nondepolarizing components $D_{a,\mathrm{sup}}$, thereby reducing the worst-case gate infidelity at leading order, as we show in Appendix~\ref{app:fixed-Lindblad-direction}.

\subsection{Concise Criteria under Full Unitary Controllability}
\label{sec:theoryB}

In this subsection, we state concise existence criteria for robust gates under the \textit{unitary controllability condition}, where the available Hamiltonian operators generate the full $\mathfrak{su}(\dimension)$ algebra. 

We first clarify the notation for generated Lie algebras. Given a subset $\{\gamma_k\}$ of a Lie algebra $\mathfrak{w}$, $\operatorname{Lie}(\{\gamma_k\})$ denotes the Lie algebra generated by $\{\gamma_k\}$, namely the smallest Lie subalgebra of $\mathfrak{w}$ containing $\{\gamma_k\}$. Equivalently, it consists of all elements obtained from $\{\gamma_k\}$ by finite applications of Lie brackets and real linear combinations. Below, Lie-algebra generation by Hamiltonian operators refers to generation by their anti-Hermitian counterparts $iH$.

For the Hamiltonian introduced above, the unitary controllability condition states that the available Hamiltonian operators generate the full $\mathfrak{su}(\dimension)$ algebra, i.e.,
\begin{equation}
    \operatorname{Lie}(\{i\staticterm\}\cup\{i\scheduleterm\mid j=1,\cdots,m\})
    =
    \mathfrak{su}(\dimension).
    \label{eq:generation-Hamiltonian-su}
\end{equation}
This condition means the control channels can generate any $\mathrm{SU}(\dimension)$ gate~\cite{Albertini_2003}. Importantly, we can achieve full unitary controllability using a limited set of control channels, and the available Hamiltonian operators need not form a complete basis of $\mathfrak{su}(\dimension)$ by themselves~\cite{Kuznetsov2026classification}.

Na\"ively, one might expect full unitary controllability to suffice for implementing robust gates. However, we show that, because the static coefficient $\staticcoe$ is time-independent, robustness against arbitrary static perturbations requires an additional condition.

We now consider whether robust-gate protocols exist for every target gate $\target\in \mathrm{SU}(\dimension)$ for a prescribed finite period $T$. Their noise-free evolutions must satisfy both $U_0(T)=\target$ and the corresponding robustness conditions. The finite static coefficient $\staticcoe$ and the smooth $T$-periodic schedules $\schedulecoe$ are both subject to adjustment for different gates and constraints. We introduce simple and explicit existence criteria for protocols satisfying the Hamiltonian robustness conditions in Sec.~\ref{sec:theoryA}, as summarized in Table~\ref{tab:robustness-criteria}.


\textbf{\textit{Static DCGs} (\textit{arbitrary})}. This is the case of Eq.~\eqref{eq:DCG-many-condition} with $r=\dimension^2-1$, where $\{iX_1,\cdots,iX_r\}$ is a complete basis of $\mathfrak{su}(\dimension)$. We define the common symmetry algebra of the driven Hamiltonian operators by
\begin{equation}
    \mathcal C_{\mathrm{dyn}}=
    \{Y\in\mathfrak{su}(\dimension)\mid [Y, \scheduleterm]=0,\ j=1,\cdots,m\}.
    \label{eq:C_dyn}
\end{equation}
Note that we do not require elements of $\mathcal C_{\mathrm{dyn}}$ to commute with the static Hamiltonian operator $\staticterm$. This algebra contains elements that are difficult to change: the driven Hamiltonian operators leave them invariant; only the static Hamiltonian operator $\staticterm$ can change them.

Under the unitary controllability condition, protocols with all susceptibilities $\chi_{p}^{(\mathrm D)}$ vanishing exist for every target gate $\target\in\mathrm{SU}(\dimension)$ and every prescribed finite period $T>0$, if and only if
\begin{equation}
    \mathcal C_{\mathrm{dyn}}=\{0\}.
    \label{eq:reachability-all-DCG}
\end{equation}
Physically, this condition means that the driven Hamiltonian operators have no common symmetry. With this condition satisfied, for every target gate and every prescribed finite period $T>0$, dynamically corrected gates against arbitrary static Hamiltonian perturbations acting on the $\dimension$-level system exist. Consequently, we can also implement weaker dynamically corrected gates, which have resilience against any smaller error space $\operatorname{span}_{\mathbb R}\{X^{\prime}_p\}\subset\operatorname{span}_{\mathbb R}\{X_p\}$.

A useful special case arises when the driven Hamiltonian operators alone generate the full algebra,
\begin{equation}
    \operatorname{Lie}(\{i\scheduleterm\mid j=1,\cdots,m\})=
    \mathfrak{su}(\dimension).
    \label{EQ:full-driving-term}
\end{equation}
In this special case, Eq.~\eqref{eq:reachability-all-DCG} automatically holds.

If Eq.~\eqref{eq:reachability-all-DCG} fails, robust protocols may still exist for particular target gates or selected subsets of static perturbations, but they require case-specific constructions. In particular, under the unitary controllability condition, a DCG satisfying the single robustness condition in Eq.~\eqref{eq:DCG-condition}, with $X=\staticterm$, still exists for every target gate and every prescribed finite period $T>0$. However, the failure of Eq.~\eqref{eq:reachability-all-DCG} leads to obstructions to robustness against specific perturbation directions, as we demonstrate below.

\textit{Common-symmetry obstruction}. If the driven Hamiltonian operators have common symmetries, i.e., $\mathcal{C}_{\mathrm{dyn}}\neq\{0\}$, we can explicitly construct static perturbation directions, with their leading-order effects impossible to eliminate. Considering the direction $X_{\mathcal Q}=[\Lambda, H_0]$, where $\Lambda\in\mathcal C_{\mathrm{dyn}}$ and $\Lambda\neq0$,\footnote{Under the unitary controllability condition, $X_{\mathcal Q}\neq0$. Otherwise, $\Lambda$ would commute with $i\staticterm$ and every $i\scheduleterm$, and hence with all of $\mathfrak{su}(\dimension)$, which is impossible for $\Lambda\neq0$.} we have $\frac{\mathrm d}{\mathrm dt}(U_0^{\dagger}(t)\Lambda U_0(t))=-i\staticcoe U_0^{\dagger}(t)X_{\mathcal Q}U_0(t)$. After integration, we obtain
\begin{equation}
    \|\target^{\dagger}\Lambda\target-\Lambda\|=\left\|[\Lambda, \target]\right\|=T|\staticcoe| \chi^{(\mathrm D)}_{\mathcal{Q}}, 
    \label{eq:common-symmetry-floor}
\end{equation}
where $\chi_{\mathcal Q}^{(\mathrm D)}$ is the susceptibility in Eq.~\eqref{eq:infidelity-DCG} evaluated at $X=X_{\mathcal{Q}}$. For any target gate $\target$ that does not commute with $\Lambda$, we cannot eliminate the leading-order effect of the perturbation $H(t)\to H(t)+\epsilon X_{\mathcal Q}$. Equation~\eqref{eq:infidelity-DCG} then gives the leading-order average gate infidelity $\epsilon^2\|[\Lambda,\target]\|^2/[(\dimension+1)\staticcoe^2]$, which is independent of the schedules and of $T$, and only a larger static coefficient $\staticcoe$ can reduce it.

Let us take a single-qubit $\mathfrak{su}(2)$ system as an example. If $H(t)=u_0\sigma_x+u(t)\sigma_y$, where $u_0$ is time-independent, the system satisfies the unitary controllability condition, but $\mathcal C_{\mathrm{dyn}}=\operatorname{span}_{\mathbb R}\{i\sigma_y\}\neq\{0\}$. According to the criterion in Eq.~\eqref{eq:reachability-all-DCG}, the DCG conditions against all static $\mathfrak{su}(2)$ perturbations cannot hold simultaneously for every target gate, but we can always fulfill the static DCG condition for the specific perturbation $H(t)\to H(t)+\epsilon\sigma_x$. Here, $\Lambda=i\sigma_y$ gives $X_{\mathcal Q}=2\sigma_z$, so every target gate that does not commute with $\sigma_y$ keeps a nonzero $\sigma_z$ susceptibility. In contrast, for $H(t)=u_x(t)\sigma_x+u_y(t)\sigma_y$, no nonzero element of $\mathfrak{su}(2)$ commutes with both $i\sigma_x$ and $i\sigma_y$, so $\mathcal C_{\mathrm{dyn}}=\{0\}$, and we can always fulfill the DCG conditions against all static $\mathfrak{su}(2)$ perturbations.

Beyond the $\mathfrak{su}(2)$ example, the obstruction also exists in experimental settings. For example, in a coupled electron--nuclear spin system~\cite{Hegde2020Indirect}, the effective Hamiltonian in the rotating frame is
\begin{equation*}
\begin{aligned}
H(t)={}&-\pi\nu_C\,IZ+u_x(t)XI+u_y(t)YI\\
&-\frac{\pi}{2}\left[A_{zz}(IZ-ZZ)+A_{zx}(IX-ZX)\right].
\end{aligned}
\end{equation*}
Here, the first and the second operators in each product respectively act on the electron and nuclear spins, $\nu_C$ is the nuclear Larmor frequency, $A_{zz}$ and $A_{zx}$ are hyperfine couplings, and $u_x(t)$ and $u_y(t)$ are the microwave schedules. For the experimental parameters, this model satisfies the unitary controllability condition, but the driven Hamiltonian operators have the common symmetries $\mathcal C_{\mathrm{dyn}}=\operatorname{span}_{\mathbb R}\{iIX,iIY,iIZ\}$. Consequently, the DCG conditions against all static $\mathfrak{su}(4)$ perturbations cannot hold simultaneously for every target gate.

\textbf{\textit{Finite-frequency and waveform-selective DCGs}}. These gates share the same sufficient existence criterion as their corresponding static DCGs. Under the unitary controllability condition, if $\mathcal C_{\mathrm{dyn}}=\{0\}$, we can implement every target gate $\target\in\mathrm{SU}(\dimension)$ at any prescribed finite period $T>0$ while satisfying Eq.~\eqref{eq:waveform-selective-condition}. Here, we allow any finite collection of selected smooth real functions $f_{p,\ell}(t)$ on $[0, T]$. We choose a finite static coefficient $u_0$ and smooth $T$-periodic schedules $u_j(t)$ separately for each protocol. This result includes the finite-frequency conditions in Eq.~\eqref{eq:DCG-many-with-frequency} for any family of finite selected frequency sets $\{\Omega_p\}_{p=1}^{\dimension^2-1}$. Section~\ref{sec:theoryC} extends this sufficient result beyond full unitary controllability.

For general selected waveforms, we establish existence only in this fixed-$T$ setting. For finite-frequency robustness, we can also fix any nonzero $u_0$ and choose $T$ separately for each protocol without changing the selected physical frequencies, provided that full unitary controllability and $\mathcal C_{\mathrm{dyn}}=\{0\}$ hold. We prove this result in Appendix~\ref{app:fixed-static-frequency}. Moreover, if all frequency sets share at least one element, i.e., $\bigcap_{p=1}^{\dimension^2-1}\Omega_p\neq\varnothing$, the condition $\mathcal C_{\mathrm{dyn}}=\{0\}$ is also necessary for the existence of a family covering every target gate, even when $T$ is chosen separately for each target gate.


\textbf{\textit{Extended geometric gates}} (EGGs). Under the unitary controllability condition, the extended geometric condition in Sec.~\ref{sec:theoryA} will not obstruct the existence of robust gates. Therefore, we can impose this condition in addition to the static DCG, finite-frequency, and waveform-selective DCG conditions above whenever the corresponding existence condition holds. However, imposing more constraints simultaneously may require robust controls with more harmonics and larger control amplitudes.

In summary, under full unitary controllability and $\mathcal C_{\mathrm{dyn}}=\{0\}$, we can simultaneously impose any chosen collection of static DCG conditions, any prescribed finite set of finite-frequency conditions, any finite collection of waveform-selective DCG conditions, and the EGG condition. If the finite-frequency conditions cover a complete basis $\{iX_p\}_{p=1}^{\dimension^2-1}$ of $\mathfrak{su}(\dimension)$ and share a common frequency, i.e., $\bigcap_{p=1}^{\dimension^2-1}\Omega_p\neq\varnothing$, the condition $\mathcal C_{\mathrm{dyn}}=\{0\}$ is also necessary for implementing every target gate, even when $T$ is chosen separately for each target gate.


\textbf{\textit{Lindblad-robust gates}}. These gates have a separate obstruction. Even under the unitary controllability condition, coherent Hamiltonian control cannot remove the fixed depolarizing component described in Sec.~\ref{sec:theoryA}. Therefore, we can only eliminate the leading-order contribution of the suppressible components $D_{a,\mathrm{sup}}$. If the driven Hamiltonian operators alone generate the full algebra, as in Eq.~\eqref{EQ:full-driving-term}, we can eliminate the leading-order effect of the suppressible components $D_{a,\mathrm{sup}}$. If the driven Hamiltonian operators do not generate the full algebra, the general criterion introduced in the next subsection provides a sufficient condition to eliminate the leading-order effect of the suppressible components.

Except for general waveform-selective robustness, the existence criteria in this subsection also allow us to fix any nonzero static coefficient $\staticcoe$ in advance and choose the duration $T$ separately for each protocol. For every prescribed $\staticcoe\neq0$ and every target gate $\target\in\mathrm{SU}(\dimension)$, we can choose a finite $T>0$ and smooth $T$-periodic schedules $\schedulecoe$ that satisfy the corresponding robustness conditions. Under $\mathcal C_{\mathrm{dyn}}=\{0\}$, we can impose the static DCG, finite-frequency, and EGG conditions simultaneously while keeping the selected physical frequencies unchanged. We prove these results in Appendix~\ref{app:reachability-results} and summarize the four control settings in Table~\ref{tab:control-settings}.

\begin{table}[t!]
\centering
\small
\setlength{\tabcolsep}{3pt}
\setlength{\arrayrulewidth}{0.5pt}
\renewcommand{\arraystretch}{1.15}
\begin{tabular}{|p{0.28\columnwidth}|p{0.35\columnwidth}|p{0.25\columnwidth}|}
\hline
\raggedright Static coefficient, $u_0$ &
\raggedright Driven schedules, $u_j(t)$ &
\raggedright Gate duration, $T$ \tabularnewline
\noalign{\hrule height 1pt}
\raggedright Chosen per protocol &
\raggedright Smooth, $T$-periodic &
\raggedright Fixed \tabularnewline
\hline
\raggedright Chosen per protocol &
\raggedright Piecewise constant &
\raggedright Fixed \tabularnewline
\hline
\raggedright Fixed, nonzero &
\raggedright Smooth, $T$-periodic &
\raggedright Chosen per protocol \tabularnewline
\hline
\raggedright Fixed, nonzero &
\raggedright Piecewise constant &
\raggedright Chosen per protocol \tabularnewline
\hline
\end{tabular}
\caption{Control settings with the same sufficient and, if applicable, necessary conditions as in Sec.~\ref{sec:theoryB}. For general waveform-selective robustness, only the two settings with fixed $T$ apply. In all cases, $u_0$ remains constant within each protocol. In both smooth periodic settings, replacing the schedules with continuous functions on $[0, T]$ without a periodicity requirement leaves the existence results unchanged.}
\label{tab:control-settings}
\end{table}

These concise criteria provide practical guidance for experimental gate design. Many experiments already employ control Hamiltonians of the form of Eq.~\eqref{eq:control_resources} whose control channels satisfy the unitary controllability condition~\cite{PhysRevLett.114.240401, Ringbauer2022, Rong2015, Werninghaus2021}, so that the same platform can in principle implement every gate in $\mathrm{SU}(\dimension)$. Under these experimental settings, once the common symmetry algebra of the driven Hamiltonian operators is $\{0\}$, experimenters can impose any chosen combination of the static DCG conditions, the finite-frequency DCG conditions, the waveform-selective DCG conditions, and the extended geometric condition on their gates, all within the same platform.

The existence criteria provide two strategies for experiments. If the platform tolerates large control amplitudes but the perturbations are not well characterized, experimenters can impose more robustness conditions to cover a wider range of possible perturbations. For example, they can impose the static DCG conditions for a complete basis of $\mathfrak{su}(\dimension)$, eliminating the leading-order effects of arbitrary static perturbations without identifying the perturbation directions. If the noise is well characterized, they can tailor the robustness conditions to the relevant perturbation directions, spectral peaks, and waveforms, before imposing them to improve gate fidelity.

\subsection{General Criterion for Robust Gate Implementations}
\label{sec:theoryC}

The previous subsection states the concise criteria under the unitary controllability condition. Readers interested only in these concise criteria may skip the technical details of this subsection. Here, we state the general criteria for constructing a family of robust gates using the control resources in Eq.~\eqref{eq:control_resources}, without assuming unitary controllability.

Before stating the general criteria, we define the mathematical notions needed to formulate them. These notions allow us to unify the gate implementation and robustness constraints within an algebraic structure, paving the way for explicit existence criteria.

Let $K$ be a connected Lie group acting on the vector space $V$ by conjugation, $v\circ k=k^{-1}vk$, where $v\in V$, $k\in K$, and $k^{-1}vk\in V$. We denote the Lie algebra of $K$ by $\mathfrak{k}$ and define the corresponding semidirect product group
\begin{equation}
G=K\rtimes V=\{(k,v)|k\in K,\ v\in V\},
\label{eq:semidirect-product-group-G}
\end{equation}
with the group multiplication\footnote{We use the right-action convention~\cite{RightColombo2015} for this semidirect product to match the evolution of the quantum propagator $U_0(t)$.}
\begin{equation}
(k_1,v_1)\cdot(k_2,v_2)=(k_1k_2, v_2+k_2^{-1}v_1k_2).
\end{equation}
Here, the component $k\in K$ records the implemented gate, whereas the component $v\in V$ records the accumulated leading-order response. The Lie algebra of $G$ is $\mathfrak{g}=\{(a,v)\mid a\in\mathfrak{k},\ v\in V\}$, with the Lie bracket defined as
\begin{equation}
[(a_1,v_1),(a_2,v_2)]=([a_1,a_2], [a_1,v_2]-[a_2,v_1]),
\end{equation}
where $a_1,a_2\in\mathfrak{k}$ and $v_1, v_2\in V$.

On the group $G$, we encode the constrained dynamics into a right-invariant control system. For a static element $A\in\mathfrak{g}$ and driven elements $B_j\in\mathfrak{g}$, $j=1,\cdots,m$, the corresponding control system is~\cite{Sachkov2009theorems}
\begin{equation}
\Gamma=
\left\{
A+\sum_{j=1}^{m}u_jB_j
\ \middle|\
\ u_j\in\mathbb R
\right\}
\subset\mathfrak{g}.
\end{equation}
A gate protocol corresponds to a trajectory $\varphi(t)=(k(t),v(t))\in G$. Writing $R(t)=(a(t),w(t))\in\Gamma$, the right-invariant dynamics are
\begin{equation}
\begin{aligned}
\dot{\varphi}(t)&=R(t)\varphi(t),\\
\dot{k}(t)&=a(t)k(t),\\
\dot{v}(t)&=k^{-1}(t)w(t)k(t).
\end{aligned}
\end{equation}
Here, the mapping $R:\mathbb R\to\Gamma$ is smooth and $T$-periodic. With $\varphi(0)=(I,0)$ and $\varphi(T)=(k_{\ast},0)$, where $k_{\ast}\in K$, the trajectory $\varphi(t)$ describes the robust gate implementation, and $a(t)$ encodes $-iH(t)$. The condition $k(T)=k_{\ast}$ guarantees the implementation of the target gate. By integrating the equation for $v(t)$, we have
\begin{equation*}
v(T)=\int_0^T\mathrm{d}t\,k^{-1}(t)w(t)k(t).
\end{equation*}
Since $k(t)=U_0(t)$, this integral has the form of the first-order corrections in Sec.~\ref{sec:theoryA}, such as those in Eqs.~\eqref{eq:DCG-many-condition} and \eqref{eq:EGG}. Therefore, in each construction below, the specified $w(t)$ makes the components of $v(T)$ equal to the accumulated response operators in the robustness conditions, up to normalization, so $v(T)=0$ sets these response operators to zero and enforces the corresponding robustness conditions. Moreover, to first order in the perturbation strength $\epsilon$, we can combine two consecutive segments with evolution operators $U_1(I-i\epsilon v_1)$ and $U_2(I-i\epsilon v_2)$ into $U_2U_1[I-i\epsilon(v_1+U_1^\dagger v_2U_1)]$, which is exactly the product $(U_2,v_2)\cdot(U_1,v_1)$ in $G$. Hence, the semidirect product group is the natural description of robust gate implementations.

We now apply this construction to the robustness conditions introduced above. For the Hamiltonian conditions, we can define
\begin{equation}
\mathfrak{k}=\operatorname{Lie}\big(\{i\staticterm\}\cup\{i\scheduleterm\mid j=1,\cdots,m\}\big),
\end{equation}
and let $K$ be the connected group generated by $\mathfrak{k}$. Each construction below follows the same pattern: we choose the response space and specify the static and driven elements $A$ and $B_j$ in the corresponding semidirect-product Lie algebra. In every case, a zero vector component at the endpoint imposes the desired robustness condition.

\textbf{\textit{Static DCGs}}. For the static DCG conditions, the vector space $V$ is
\begin{equation}
V_{\mathrm{DCG}}=\operatorname{span}_{\mathbb R}\{(k^{-1}X_1k,\cdots,k^{-1}X_rk)\mid k\in K\}.
\end{equation}
The component $k(t)\in K$ corresponds to $U_0(t)$, so a path $(I,0)\to(\target,0)$ represents the construction of a static DCG. With $\mathfrak{g}_{\mathrm{DCG}}=\mathfrak{k}\rtimes V_{\mathrm{DCG}}$, the corresponding control system in $\mathfrak{g}_{\mathrm{DCG}}$ is
\begin{equation}
\begin{aligned}
\Gamma_{\mathrm{DCG}}
&=\{A_{\mathrm{DCG}}+\sum_{j=1}^{m}c_jB_{\mathrm{DCG},j}\mid c_j\in\mathbb R\}, \\
A_{\mathrm{DCG}}&=(-i\staticcoe\staticterm,(X_1,\cdots,X_r)), \\
B_{\mathrm{DCG},j}&=(-i\scheduleterm,(0,\cdots,0)).
\end{aligned}
\label{eq:smeidirect-DCG}
\end{equation}

\textbf{\textit{Extended geometric gates}}. For the EGG condition, the response vector space is
\begin{equation}
V_{\mathrm{EGG}}=\operatorname{span}_{\mathbb R}\{k^{-1}Hk\mid k\in K,\ iH\in\mathfrak{k}\}.
\end{equation}
With $\mathfrak{g}_{\mathrm{EGG}}=\mathfrak{k}\rtimes V_{\mathrm{EGG}}$, the corresponding control system in the enlarged algebra $\mathfrak{g}_{\mathrm{EGG}}$ takes the form
\begin{equation}
\begin{aligned}
\Gamma_{\mathrm{EGG}}
&=\{A_{\mathrm{EGG}}+\sum_{j=1}^{m}c_jB_{\mathrm{EGG},j}\mid c_j\in\mathbb R\}, \\
A_{\mathrm{EGG}}&=(-i\staticcoe\staticterm,\staticcoe\staticterm), \\
B_{\mathrm{EGG},j}&=(-i\scheduleterm,\scheduleterm).
\end{aligned}
\label{eq:semidirect-EGG}
\end{equation}

\textbf{\textit{Joint static DCG--EGG conditions}}. For the joint static DCG--EGG conditions, the response space is $V_{\mathrm{joint}}=V_{\mathrm{DCG}}\oplus V_{\mathrm{EGG}}$. Writing elements of $\mathfrak{g}_{\mathrm{joint}}=\mathfrak{k}\rtimes V_{\mathrm{joint}}$ as $(a,v_{\mathrm{DCG}},v_{\mathrm{EGG}})$, the corresponding control system in the enlarged algebra $\mathfrak{g}_{\mathrm{joint}}$ is
\begin{equation}
\begin{aligned}
\Gamma_{\mathrm{joint}}
&=\{A_{\mathrm{joint}}+\sum_{j=1}^{m}c_jB_{\mathrm{joint},j}\mid c_j\in\mathbb R\}, \\
A_{\mathrm{joint}}
&=(-i\staticcoe\staticterm,(X_1,\cdots,X_r),\staticcoe\staticterm), \\
B_{\mathrm{joint},j}
&=(-i\scheduleterm,(0,\cdots,0),\scheduleterm).
\end{aligned}
\label{eq:semidirect-DCG+EGG}
\end{equation}

\begin{table*}[t]
\centering
\small
\setlength{\tabcolsep}{3pt}
\setlength{\arrayrulewidth}{0.5pt}
\renewcommand{\arraystretch}{1.25}
\begin{tabular}{|p{0.25\textwidth}|p{0.385\textwidth}|p{0.32\textwidth}|}
\hline
\raggedright\textbf{Construction} &
\raggedright\textbf{Response space} &
\raggedright\textbf{Control elements} \tabularnewline[2pt]
\hline
\raggedright Static DCGs &
\raggedright
$\displaystyle V_{\mathrm{DCG}}=\operatorname{span}_{\mathbb R}
\{k^{-1}(X_1,\ldots,X_r)k\mid k\in K\}$ &
\raggedright
$\begin{aligned}[t]
A_{\mathrm{DCG}}&=(-i\staticcoe\staticterm,(X_1,\ldots,X_r)),\\
B_{\mathrm{DCG},j}&=(-i\scheduleterm,(0,\ldots,0))
\end{aligned}$ \tabularnewline[2pt]
\hline
\raggedright Extended geometric gates &
\raggedright
$\displaystyle V_{\mathrm{EGG}}=\operatorname{span}_{\mathbb R}
\{k^{-1}Hk\mid k\in K,\ iH\in\mathfrak{k}\}$ &
\raggedright
$\begin{aligned}[t]
A_{\mathrm{EGG}}&=(-i\staticcoe\staticterm,\staticcoe\staticterm),\\
B_{\mathrm{EGG},j}&=(-i\scheduleterm,\scheduleterm)
\end{aligned}$ \tabularnewline[2pt]
\hline
\raggedright Joint static DCG--extended geometric gate conditions &
\raggedright
$\displaystyle V_{\mathrm{joint}}=V_{\mathrm{DCG}}\oplus V_{\mathrm{EGG}}$ &
\raggedright
$\begin{aligned}[t]
A_{\mathrm{joint}}&=(-i\staticcoe\staticterm,(X_1,\ldots,X_r),
\staticcoe\staticterm),\\
B_{\mathrm{joint},j}&=(-i\scheduleterm,(0,\ldots,0),\scheduleterm)
\end{aligned}$ \tabularnewline[2pt]
\hline
\raggedright Lindblad-robust gates &
\raggedright
$\displaystyle V_{\mathrm{Lind}}=\operatorname{span}_{\mathbb R}
\{(k^{-1}D_{a,\mathrm{sup}}k)_{a=1}^{r_{\mathrm{Lind}}}
\mid k\in K_{\mathrm{Lind}}\}$ &
\raggedright
$\begin{aligned}[t]
A_{\mathrm{Lind}}&=(\staticcoe C(\staticterm),d_{\mathrm{sup}}),\\
B_{\mathrm{Lind},j}&=(C(\scheduleterm),0)
\end{aligned}$ \tabularnewline[2pt]
\hline
\end{tabular}
\caption{Data for applying the general criterion in
Eq.~\eqref{eq:general-criteria}. Each case uses
$\mathfrak g=\mathfrak{k}\rtimes V$ and
$\Gamma=\{A+\sum_{j=1}^{m}c_jB_j\mid c_j\in\mathbb R\}$,
with $V$, $A$, and $B_j$ specified above. Here, DCG means dynamically corrected gate.
For the first three cases, $K\subseteq\mathrm{SU}(\dimension)$ is the
connected Lie subgroup; its Lie algebra is
$\mathfrak{k}=\operatorname{Lie}\big(\{i\staticterm\}\cup
\{i\scheduleterm\mid j=1,\ldots,m\}\big)$, and the prescribed target
element in $K$ is $\target$.
For the Lindblad case, $K_{\mathrm{Lind}}$ is the connected Lie group;
its Lie algebra is
$\mathfrak{k}_{\mathrm{Lind}}=\operatorname{Lie}\big(
\{C(\staticterm)\}\cup
\{C(\scheduleterm)\mid j=1,\ldots,m\}\big)$, and the prescribed target
element in $K_{\mathrm{Lind}}$ is
$\target\otimes\target^*$.
Conjugation by $k$ acts componentwise on operator tuples.}
\label{tab:semidirect-control-data}
\end{table*}

\textbf{\textit{Lindblad-robust gates}}. For Lindblad robustness, we apply the same construction in Hilbert--Schmidt space. We exclude the fixed depolarizing component, and only the suppressible components enter the response vector space. Define
\begin{equation}
C(H)=-i(H\otimes\identity_{\dimension}-\identity_{\dimension}\otimes H^T).
\end{equation}
We can define
\begin{equation}
\mathfrak{k}_{\mathrm{Lind}}=\operatorname{Lie}\big(\{C(\staticterm)\}\cup\{C(\scheduleterm)\mid j=1,\cdots,m\}\big),
\end{equation}
and let $K_{\mathrm{Lind}}$ be the connected group generated by $\mathfrak{k}_{\mathrm{Lind}}$. For this construction, the group component is $\mathcal U_0(t)=U_0(t)\otimes U_0^*(t)$, and the target element in $K_{\mathrm{Lind}}$ is $\target\otimes\target^*$.\footnote{Here, the subscript $\ast$ labels the target gate, while the superscript $*$ denotes complex conjugation.} After setting
\begin{equation}
d_{\mathrm{sup}}=(D_{1,\mathrm{sup}},\cdots,D_{r_{\mathrm{Lind}},\mathrm{sup}}),
\end{equation}
the Lindblad response vector space is
\begin{equation}
V_{\mathrm{Lind}}
=
\operatorname{span}_{\mathbb R}
\{(k^{-1}D_{a,\mathrm{sup}}k)_{a=1}^{r_{\mathrm{Lind}}}\mid k\in K_{\mathrm{Lind}}\}.
\end{equation}
With $\mathfrak{g}_{\mathrm{Lind}}=\mathfrak{k}_{\mathrm{Lind}}\rtimes V_{\mathrm{Lind}}$, the corresponding control system in the enlarged algebra $\mathfrak{g}_{\mathrm{Lind}}$ is
\begin{equation}
\begin{aligned}
\Gamma_{\mathrm{Lind}}
&=\{A_{\mathrm{Lind}}+\sum_{j=1}^{m}c_jB_{\mathrm{Lind},j}\mid c_j\in\mathbb R\}, \\
A_{\mathrm{Lind}}
&=(\staticcoe C(\staticterm),d_{\mathrm{sup}}), \\
B_{\mathrm{Lind},j}
&=(C(\scheduleterm),0).
\end{aligned}
\end{equation}

For each construction above, as we summarize in Table~\ref{tab:semidirect-control-data}, let $K$, $V$, $\mathfrak{g}$, and $\Gamma$ denote the corresponding objects. If the conditions
\begin{equation}
K\ \text{is compact},\qquad
V^K=\{0\},\qquad
\operatorname{Lie}(\Gamma)=\mathfrak{g}
\label{eq:general-criteria}
\end{equation}
hold, we can implement all the corresponding robust gates in group $K$, using the control channels in Eq.~\eqref{eq:control_resources}. Here, $V^K=\{v\in V\mid k^{-1}vk=v\ \text{for every }k\in K\}$ denotes the set of fixed points in $V$ under the action of $K$. Compactness is a prerequisite for the controllability theorem applied to semidirect-product groups~\cite{Sachkov2009theorems}. The condition $V^K=\{0\}$ excludes a nonzero response fixed by every element of $K$, while $\operatorname{Lie}(\Gamma)=\mathfrak{g}$ requires the available constrained generators to generate the full enlarged algebra. Under these conditions, for every target element $k_{\ast}\in K$ and every prescribed finite period $T>0$, there exist a finite static coefficient $\staticcoe$ and smooth $T$-periodic (or piecewise-constant) schedules $\schedulecoe$ such that the trajectory reaches $(k_{\ast},0)$.

To provide a concrete example, we apply the general criterion to the single-qubit Hamiltonian $H(t)=u_0\sigma_x+u(t)\sigma_y$, with $u_0\neq0$. We consider the static perturbation $H(t)\to H(t)+\epsilon_x\sigma_x+\epsilon_y\sigma_y$, with independent $\epsilon_x$ and $\epsilon_y$. Here, $K=\mathrm{SU}(2)$ and $V_{\mathrm{DCG}}$ consists of all pairs of traceless Hermitian $2\times2$ matrices. Thus, $K$ is compact, and no nonzero element of $V_{\mathrm{DCG}}$ is fixed under the conjugation action of $K$, i.e., $V_{\mathrm{DCG}}^K=\{0\}$. On the semidirect product group $G$ in Eq.~\eqref{eq:semidirect-product-group-G}, we find
\begin{equation*}
\begin{aligned}
    A&=(-iu_0\sigma_x,(\sigma_x,\sigma_y)),\\
    B_1&=(-i\sigma_y,(0,0)),\\
    \Gamma&=\{A+cB_1\mid c\in\mathbb R\}.
\end{aligned}
\end{equation*}
Consequently, the following relations hold
\begin{equation*}
\begin{aligned}
    A-\frac14[B_1,[A,B_1]]&=(0,(0,\sigma_y)),\\
    [A,[A,[A,B_1]]]+4u_0^2[A,B_1]&=(0,(-16u_0^2\sigma_z,0)).
\end{aligned}
\end{equation*}

Further brackets with $A$ and $B_1$ generate $(0,(\sigma_j,0))$ and $(0,(0,\sigma_j))$ for $j=x,y,z$. Subtracting $(0,(\sigma_x,\sigma_y))$ from $A$ gives $(-iu_0\sigma_x,(0,0))$. Together with $B_1$, this element generates $(-i\sigma_j,(0,0))$ for $j=x,y,z$. Therefore, we have a complete basis for the algebra $\mathfrak{g}_{\mathrm{DCG}}$, giving the relation $\operatorname{Lie}(\Gamma)=\mathfrak{g}_{\mathrm{DCG}}$. For every target gate $\target\in\mathrm{SU}(2)$ and every prescribed finite period $T>0$, we can choose a finite $u_0\neq0$ and a smooth $T$-periodic (or piecewise-constant) schedule $u(t)$ to implement the gate $U_0(T)=\target$ under the static DCG conditions for both $\sigma_x$ and $\sigma_y$. 

For the $\sigma_z$ static DCG condition, we instead use $A=(-iu_0\sigma_x,\sigma_z)$ and $B_1=(-i\sigma_y,0)$ in the static DCG construction, obtaining
\begin{equation*}
\begin{aligned}
    \relax[A,[A,B_1]]&=-4u_0^2B_1,\\
    [B_1,[A,B_1]]&=4A.
\end{aligned}
\end{equation*}
These relations give $\operatorname{Lie}(\Gamma)=\operatorname{span}_{\mathbb R}\{A,B_1,[A,B_1]\}$, with $\dim\operatorname{Lie}(\Gamma)=3$. The response space now consists of all traceless Hermitian $2\times2$ matrices, so $\dim\mathfrak{g}_{\mathrm{DCG}}=6$. Therefore, $\operatorname{Lie}(\Gamma)\neq\mathfrak{g}_{\mathrm{DCG}}$. The failure of the algebraic condition fits well with our conclusions in Sec.~\ref{sec:theoryB}, which clarify the obstruction due to the common symmetry of the driven Hamiltonian operators $\scheduleterm$.

The waveform-selective DCG conditions for the perturbation directions $X_1, \cdots, X_r$ need no algebraic existence criteria beyond those for the static DCG control system in Eq.~\eqref{eq:smeidirect-DCG} built from the same directions $X_1, \cdots, X_r$. For each $X_p$, we may select any finite collection of smooth real functions $f_{p,\ell}(t)$ on $[0, T]$. If the criteria in Eq.~\eqref{eq:general-criteria} hold for this static DCG control system, then, for every target element $k_{\ast}\in K$ and every prescribed finite period $T>0$, there exists at least one robust control satisfying
\begin{equation}
\mathcal A_p^{(\mathrm D)}[f_{p,\ell}]
=\frac{1}{T}\int_0^T\mathrm{d}t\,
f_{p,\ell}(t)U_0^\dagger(t)X_pU_0(t)=0,
\end{equation}
simultaneously for every $p\in\{1,\cdots,r\}$ and every selected function $f_{p,\ell}(t)$. 

If the joint static DCG--EGG control system in Eq.~\eqref{eq:semidirect-DCG+EGG} satisfies the criteria in Eq.~\eqref{eq:general-criteria}, the same waveform-selective conclusion still holds with the EGG condition imposed simultaneously, as we prove in Appendix~\ref{app:waveform-finite-frequency-extension}. The finite-frequency conditions in Eq.~\eqref{eq:DCG-many-with-frequency} follow by selecting $f(t)=1$ for each selected zero frequency and the pair $f(t)=\cos(\omega t)$, $f(t)=\sin(\omega t)$ for each selected positive frequency $\omega$. Thus, the existence criteria depend on the perturbation directions but not on the number or forms of the selected functions, and we do not require a separate semidirect-product construction involving the selected functions.

Note that the criteria in this section do not by themselves determine a pulse shape or the corresponding implementation requirements, such as the required amplitudes and harmonic content. Thus, the theory establishes the algebraic feasibility of robust-gate realization when these conditions are satisfied, and we construct explicit gate protocols in the next numerical section.

\section{Construction of Robust CZ Gates}
\label{sec:numerical}

In this section, we focus on the controlled-Z (CZ) gate, with different robustness conditions enforced individually or simultaneously. The CZ gate is a fundamental building block of quantum computation, and maintaining high fidelity in this two-qubit gate is experimentally more difficult than in single-qubit gates~\cite{Barends2014}.

We choose the following $\mathrm{SU}(4)$ version of the CZ gate as the target gate:
\begin{equation}
\target=e^{-i\pi/4}\operatorname{diag}(1,1,1,-1).
\end{equation}
The extra global phase has no physical effect. To implement the gate, we take the instantaneous Hamiltonian to be
\begin{equation}
\begin{aligned}
    H(t)&=u_0 ZZ+u_{1x}(t)IX+u_{x1}(t)XI\\
    &\quad +u_{1z}(t)IZ+u_{z1}(t)ZI,
\end{aligned}
\end{equation}
where $X,Y,Z$ are the Pauli operators and $I$ is the identity. Here $u_0$ is time-independent, while $u_{1x}(t)$, $u_{x1}(t)$, $u_{1z}(t)$, and $u_{z1}(t)$ are smooth $T$-periodic functions, where $T$ sets the gate duration. 

The available Hamiltonian operators satisfy the unitary controllability condition, since
\begin{equation}
    \operatorname{Lie}\{iZZ,iIX,iXI,iIZ,iZI\}=\mathfrak{su}(4).
\end{equation}
Moreover, the driven Hamiltonian operators $IX$, $XI$, $IZ$, and $ZI$ have no nonzero common symmetry in $\mathfrak{su}(4)$. According to the criteria in Sec.~\ref{sec:theoryB}, for every prescribed gate duration $T>0$, we can implement the CZ gate using Hamiltonians of this form, while satisfying any chosen combination of the static DCG conditions, the finite-frequency DCG conditions, the waveform-selective DCG conditions, and the extended geometric condition. Here, the perturbation directions $\{X_p\}$, the selected frequencies, and the selected functions may be any prescribed finite collections.

For the Lindblad case, we consider the two local dephasing channels labeled by $\mathcal Q_{\mathrm L}=\{IZ, ZI\}$. Their suppressible components satisfy the general criterion in Sec.~\ref{sec:theoryC}, so their leading-order effects can be eliminated.\footnote{Direct computation gives $\dim\operatorname{Lie}(\Gamma_{\mathrm{Lind}})=\dim\mathfrak{g}_{\mathrm{Lind}}=223$.} The fixed depolarizing component remains at leading order, limiting the improvement as stated in Sec.~\ref{sec:theoryA}.

In the numerical implementation, we parametrize each of the four time-dependent schedules by a truncated Fourier series,
\begin{equation}
u_j(t)=a_{j0}+\sum_{\ell=1}^{\ell_{\max}}
\biggl[a_{j\ell}\cos\left(\frac{2\pi\ell t}{T}\right)
+b_{j\ell}\sin\left(\frac{2\pi\ell t}{T}\right)\biggr],
\label{eq:Fourier-control}
\end{equation}
where all Fourier coefficients are real. We keep $u_0$ as a time-independent optimization variable. All controls presented below use the harmonic cutoff $\ell_{\max}=64$. Together with $u_0$, these schedules determine the Hamiltonian $H(t)$ and hence the gate operator $U_0(T)=\mathcal{T}e^{-i\int_0^T\mathrm{d}t\, H(t)}$.

For the Hamiltonian robustness conditions, we define the cost functions to ensure the implementation of CZ gates under different robustness constraints. We set $W_{\mathrm G}$ and $W_{\chi}$ as the weights of the gate and susceptibility terms. These cost functions take the same form
\begin{equation}
\begin{aligned}
\mathcal C_{\mathrm H}={}&\frac{W_{\mathrm G}}{2}\left\|e^{-i\phi}\target^\dagger U_0(T)-I_d\right\|^2+\frac{W_{\chi}}{2}\sum_{q}\chi_q^2,\\
\phi={}&{\arg\left\{\operatorname{Tr}\left[\target^\dagger U_0(T)\right]\right\}},
\end{aligned}
\label{eq:cost-Hamiltonian}
\end{equation}
where $\{\chi_q\}$ is the set of susceptibilities for the chosen robustness conditions. We use the squared-norm form to facilitate numerical optimization. For all numerical results in this paper, we set $W_{\mathrm {G}}=W_{\chi}=1$. We obtain the pulses for the robust gate implementation by numerically minimizing the cost functions over these Fourier coefficients and $u_0$. This cost function works not only for different robust CZ gates, but also for other gates.

For the Lindblad-robust gates, the cost function has the same gate term, evaluated with the unitary $U_0(T)$, but we calculate the susceptibilities with the superoperator $\mathcal U_0(t)$ in the Hilbert--Schmidt space:
\begin{equation}
\begin{aligned}
\mathcal C_{\mathrm{L}}
=&\frac{W_{\mathrm G}}{2}
\left\|
e^{-i\phi}\target^\dagger U_0(T)-I_d
\right\|^2+\frac{W_{\chi}}{2}
\sum_{a}
\big(\chi_a^{(\mathrm L)}\big)^2.
\end{aligned}
\label{eq:cost-Lindblad}
\end{equation}
By minimizing $\mathcal C_{\mathrm H}$ or $\mathcal C_{\mathrm{L}}$, we can determine the coefficient $u_0$ and periodic schedules $u_j(t)$, obtaining the instantaneous Hamiltonians for gate implementations. After omitting the susceptibility terms, we can also obtain the unprotected gates used for comparison from the same parameterization. Details of the numerical implementation are given in Appendix~\ref{app:numerical-algorithm}.

In Sec.~\ref{sec:numerical-A}, we present an extended geometric CZ gate, a dynamically corrected CZ gate against a static $ZZ$ perturbation, and a dynamically corrected CZ gate against arbitrary static perturbations. In Sec.~\ref{sec:numerical-B}, we construct CZ gates robust against bosonic thermal baths and static $ZZ$ drift, using only the peak frequency of the bath spectrum and the system coupling operators. Together, Secs.~\ref{sec:numerical-A} and~\ref{sec:numerical-B} illustrate the two strategies: robustness conditions covering a wider range of possible perturbations with more demanding controls, or robustness conditions tailored to well-characterized noise with less demanding controls. In Sec.~\ref{sec:numerical-C}, we show the limited improvement under Markovian Lindblad noise.

We present further numerical examples in the Appendices. In Appendix~\ref{app:additional-gate-examples}, we construct robust single-qubit, qutrit, and $\sqrt{i\mathrm{SWAP}}$ gates, extending the CZ results to piecewise-constant controls, different systems, and a different static Hamiltonian operator. In Appendix~\ref{app:additional-CZ-results}, we test the bath-robust CZ control for different spectral shapes and combine the EGG condition with further static DCG or finite-frequency conditions. In Appendix~\ref{app:waveform-selective-examples}, we construct waveform-selective DCGs: a qutrit gate against oscillatory noise with a Gaussian envelope, and a CZ gate against exponentially decaying $ZZ$ perturbations. These examples show the generality of our framework.

\begin{figure*}[t!]
    \centering
    \includegraphics[width=1\linewidth]{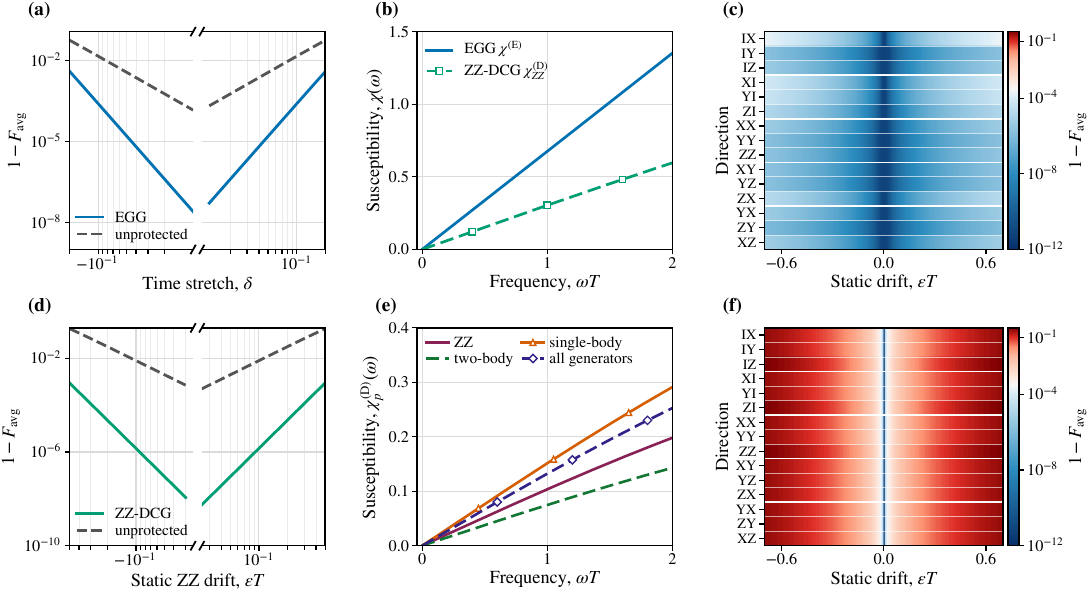}
    \caption{\textbf{Robustness of extended geometric and static dynamically corrected CZ gates.}
\textbf{(a)} Average gate infidelity $1-F_{\mathrm{avg}}$ of the extended geometric gate and the unprotected gate under a uniform time stretch $\delta$. At $|\delta|=0.1$, the extended geometric gate suppresses the infidelity by nearly two orders of magnitude compared to the unprotected gate.
\textbf{(b)} Frequency-dependent susceptibilities $\chi^{(\mathrm {E})}(\omega)$ of the extended geometric gate in (a) and $\chi_{ZZ}^{(\mathrm D)}(\omega)$ of the $ZZ$-static dynamically corrected gate ($ZZ$-DCG) in (d). Both vanish at $\omega=0$, and the corresponding protection remains effective at low frequencies.
\textbf{(c)}, \textbf{(f)} Average gate infidelities $1-F_{\mathrm{avg}}$ of the dynamically corrected gate for all 15 generators $\{X_p\}$ and the unprotected gate, respectively, with different static perturbations applied separately. At $|\epsilon T|=0.1$, the infidelities in (c) are approximately five to seven orders of magnitude lower than those in (f), depending on the perturbation direction.
\textbf{(d)} Average gate infidelity $1-F_{\mathrm{avg}}$ of a $ZZ$-static dynamically corrected gate and the unprotected gate under a static $ZZ$ drift. At $|\epsilon T|=0.1$, the $ZZ$-static dynamically corrected gate suppresses the infidelity by nearly four orders of magnitude relative to the unprotected gate.
\textbf{(e)} Frequency-dependent susceptibilities of the dynamically corrected gate in (c) for $ZZ$ and normalized equal-weight linear combinations of the nine two-body, six single-body, and all 15 generators. All vanish at $\omega=0$, and the suppression persists at low frequencies.
All controls use the harmonic cutoff $\ell_{\max}=64$.
The maximum Hamiltonian norms $\max_t\|H(t)\|$ are $17.19/T$ for the extended geometric gate control in (a,b), $19.70/T$ for the $ZZ$-static dynamically corrected gate in (b,d), $299.35/T$ for the control in (c,e), and $16.67/T$ for the unprotected control, where $T$ is the gate implementation duration.
}
\label{fig-all-static}
\end{figure*}

\subsection{Extended Geometric and Static Dynamically Corrected Gates}
\label{sec:numerical-A}

In this subsection, we consider three examples of robust CZ gates: an extended geometric gate (EGG), a dynamically corrected gate against a static $ZZ$ perturbation ($ZZ$-DCG), and a dynamically corrected gate against arbitrary traceless static Hamiltonian perturbations (all-DCG). We choose the $ZZ$ direction because static and quasi-static $ZZ$ perturbations are ubiquitous in two-qubit platforms~\cite{ZZni2022, ZZkanaar2024}. Since $\staticterm=ZZ$ here, this perturbation also describes an error in the static coefficient $\staticcoe$.

To characterize Hamiltonian rescaling errors at finite frequencies, we consider the control error $H(t)\to[1+\delta\cos(\omega t+\phi)]H(t)$, where $|\delta|\ll1$. Here, $\omega$ denotes the angular frequency of the control error, while $\phi$ is an unknown phase. Correspondingly, we define the susceptibility
\begin{equation}
\chi^{(\mathrm E)}(\omega)=\left\|\int_0^T\mathrm{d}t\,U_0^\dagger(t)H(t)U_0(t)e^{i\omega t}\right\|.
\end{equation}
At zero frequency, $\chi^{(\mathrm E)}(0)=\chi^{(\mathrm E)}$. As we show in Appendix~\ref{app:infidelity-Hamiltonian}, $\delta^2[\chi^{(\mathrm E)}(\omega)]^2/(\dimension+1)$ is an upper bound on the leading-order average gate infidelity for every phase $\phi$. Consequently, this susceptibility can reflect the robustness provided by the extended geometric condition at frequency $\omega$.

Figure~\ref{fig-all-static}(a) shows the average gate infidelity $1-F_{\mathrm{avg}}$ of an EGG under the uniform time-stretching error $\mathrm d t\to\mathrm d t(1+\delta)$. The EGG suppresses the infidelity by nearly two orders of magnitude relative to the unprotected gate at $|\delta|=0.1$, even in a regime where the stretching parameter $\delta$ is no longer perturbatively small, thus rendering the unprotected gate practically unusable. Panel (b) further shows that the EGG remains robust when the time-stretching error varies slowly during the gate, rather than only in the perfectly uniform limit.

Figure~\ref{fig-all-static}(d) shows the average gate infidelity $1-F_{\mathrm{avg}}$ of the $ZZ$-DCG protocol and the unprotected protocol under a static $ZZ$ perturbation. The $ZZ$-DCG protocol suppresses the infidelity by nearly four orders of magnitude relative to the unprotected protocol at $|\epsilon T|=0.1$, while remaining effective over the displayed perturbation range. Its frequency-dependent susceptibility in panel (b) further shows that the protection remains effective for low-frequency $ZZ$ perturbations, rather than only for the static case.

Panels~(c) and (f) compare the all-DCG protocol with the unprotected protocol, respectively, under all 15 independent perturbation directions. The all-DCG protocol is robust against all traceless static perturbations simultaneously, and the infidelities are all strongly suppressed, even when the perturbation strength $\epsilon T$ goes beyond the perturbative regime and drives the unprotected gate out of the high-fidelity regime. Since the static DCG conditions in Eq.~\eqref{eq:DCG-many-condition} are linear in the perturbation, the all-DCG protocol suppresses the effect of static perturbations along any direction, without the need to identify the perturbation directions. This universal construction is convenient for experiments that lack a characterization of the static perturbations but tolerate large control amplitudes.

Figure~\ref{fig-all-static}(e) examines susceptibilities of normalized linear combinations of the two-body generators, the single-body generators, and all 15 generators. Accordingly, the all-DCG protocol also suppresses the susceptibilities of these operators. Their low-frequency behavior further shows that the protection extends from static perturbations to low-frequency perturbations, not limited by the selection of perturbation directions.

Together, these three examples show that we can substantially improve gate fidelity over the displayed error ranges by imposing the corresponding robustness conditions, with protection extending to low-frequency errors. The EGG and $ZZ$-DCG protocols have maximum Hamiltonian norms comparable to that of the unprotected protocol. The all-DCG protocol has a maximum Hamiltonian norm $\max_t\|H(t)\|$ that is approximately one order of magnitude larger than that of the unprotected protocol. However, one can still implement this Hamiltonian either on platforms supporting larger control amplitudes or with a longer gate duration $T$.

\begin{figure*}[t!]
    \centering
    \includegraphics[width=1\linewidth]{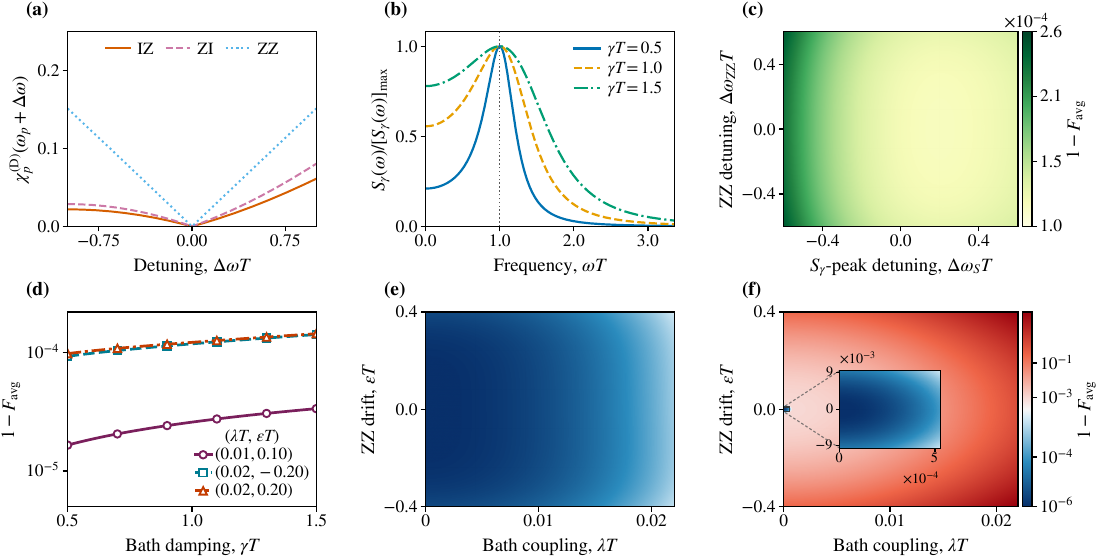}
    \caption{\textbf{Robust CZ gates in the presence of structured thermal baths and a static $ZZ$ drift}. The system is coupled to two identical and independent baths through the local operators $IZ$ and $ZI$.
    \textbf{(a)} Frequency-dependent susceptibilities $\chi_{IZ}^{(\mathrm D)}(\omega)$, $\chi_{ZI}^{(\mathrm D)}(\omega)$, and $\chi_{ZZ}^{(\mathrm D)}(\omega)$ of the robust control, shown as functions of detuning from their suppression frequencies: $\omega_pT=1$ for $IZ$ and $ZI$, and $\omega T=0$ for $ZZ$. All three remain partially suppressed over the detuning range.
    \textbf{(b)} Normalized symmetrized bath spectra $S_{\gamma}(\omega)$. The bath damping $\gamma$ sets the spectral width, with the spectral maximum fixed at $\omega_pT=1$.
    \textbf{(c)} Average gate infidelity $1-F_{\mathrm{avg}}$ under simultaneous detunings of the bath-spectrum maximum and the $ZZ$-perturbation frequency, for $\gamma T=1$, $\lambda T=0.02$, and $\epsilon T=0.2$. The frequency-dependent $ZZ$ perturbation is taken as $\epsilon\cos(\omega t)ZZ$. The average infidelity remains $\sim10^{-4}$ throughout the detuning range.
    \textbf{(d)} Average gate infidelity $1-F_{\mathrm{avg}}$ as the bath damping $\gamma$ varies, under different sets of bath coupling $\lambda T$ and static $ZZ$ drift $\epsilon T$. The spectral maximum is fixed at $\omega_pT=1$. Each average infidelity remains of the same order of magnitude.
    \textbf{(e)}, \textbf{(f)} Average gate infidelities $1-F_{\mathrm{avg}}$ under simultaneous variations of the bath coupling $\lambda T$ and the static $ZZ$ drift $\epsilon T$ for the robust and unprotected controls, at $\gamma T=1$.
    (e) and (f), including the inset, use the same piecewise colorbar, which maps $1-F_{\mathrm{avg}}$ linearly from $1.0\times10^{-6}$ to $1.79\times10^{-4}$ (blue) and from $2.5\times10^{-4}$ to $0.246$ (red), with a smooth transition between the two intervals. The unprotected-control average infidelities lie mainly in the red interval, while the robust-control average infidelities remain in the blue interval; their maxima differ by over three orders of magnitude. The inset in (f) shows a structure similar to that in panel (e), but with much smaller static drift and bath coupling strengths.
    Panels (a), (c), (d), and (e) use the same robust control.
    Both controls use the harmonic cutoff $\ell_{\max}=64$.
    The maximum Hamiltonian norms $\max_t\|H(t)\|$ are $86.30/T$ for the robust control and $16.67/T$ for the unprotected control.}
    \label{fig-DCG(ZZ)+bath(IZ+ZI)}
\end{figure*}

\subsection{Robust CZ Gates in the Presence of a Thermal Bath and Static Drift}
\label{sec:numerical-B}

In this subsection, we consider the implementation of the robust CZ gate in the presence of static $ZZ$ drift and structured bosonic thermal baths. We choose bosonic baths here for computational convenience, although the environment model in Sec.~\ref{sec:theoryA} is not restricted to them. In many experimental two-qubit platforms, static and quasi-static $ZZ$ perturbations are ubiquitous~\cite{ZZni2022, ZZkanaar2024}, while coupling to thermal baths is a common source of decoherence.

As an example of the environment model introduced in Sec.~\ref{sec:theoryA}, we consider two identical and independent structured thermal baths coupled to the two-qubit system through the local operators $IZ$ and $ZI$, respectively. The two baths represent independent local environment couplings to the two qubits. We also include the static perturbation $H(t)\to H(t)+\epsilon ZZ$. Here, $\lambda$ sets the common system--bath coupling strength, while $\epsilon$ sets the static-drift strength. We use units with $\hbar=1$ and $k_B=1$.

We assume that the two baths are initially in independent thermal states at the same temperature $\beta_{\mathrm B}^{-1}$. We choose the following underdamped Brownian noise spectrum~\cite{Spectrumlambert2019}:
\begin{equation}
\begin{aligned}
J_{\gamma}(\omega)&\equiv\frac{\lambda^2\alpha\gamma\omega}{\left[\omega_0^2(\gamma)-\omega^2\right]^2+\gamma^2\omega^2},\\
S_{\gamma}(\omega)&=\lambda^2s_{\gamma}(\omega)
=J_{\gamma}(\omega)\coth\!\left(\frac{\beta_{\mathrm B}\omega}{2}\right),
\end{aligned}
\label{eq:bath-spectrum-model}
\end{equation}
where $J_{\gamma}(\omega)$ is the underdamped Brownian spectral density, and $S_{\gamma}(\omega)$ is the corresponding symmetrized bath spectrum. The parameter $\gamma$ is the bath damping rate, while $\omega_0(\gamma)$ is the characteristic bath frequency. The function $s_{\gamma}(\omega)$ gives the spectrum for $\lambda=1$, while $\alpha$ has units of frequency and sets the overall spectral scale.
In this subsection, we fix $\beta_{\mathrm B}^{-1}T=50\pi$ and $\alpha T=\pi/4$, and choose $\omega_0(\gamma)$ such that the maximum of $S_\gamma(\omega)$ remains at $\omega_pT=1$, where $\omega_p$ denotes the spectral-peak frequency. Varying $\gamma$ therefore changes the linewidth without shifting the spectral maximum. For the gate duration $T=40\,\mathrm{ns}$, this choice corresponds to a physical bath temperature of approximately $30\,\mathrm{mK}$.

Since the two baths are independent, the cross-spectral terms $S_{pq}(\omega)$ with $p\neq q$ vanish, and the leading-order expansion in Eq.~\eqref{eq:infidelity-finite-frequency} reduces to
\begin{equation}
\begin{aligned}
1-F_{\mathrm{avg}}
\simeq{}&
\frac{(\epsilon T)^2}{5}
\left[\chi_{ZZ}^{(\mathrm D)}(0)\right]^2
+\frac{\lambda^2T^2}{5}\int_0^\infty\mathrm d\omega\,s_{\gamma}(\omega)\\
&\times
\left\{
\left[\chi_{IZ}^{(\mathrm D)}(\omega)\right]^2
+\left[\chi_{ZI}^{(\mathrm D)}(\omega)\right]^2
\right\}.
\end{aligned}
\label{eq:bath-leading-infidelity}
\end{equation}
More generally, the two local bath couplings may be correlated, while each coupling keeps the noise-power spectrum $S_\gamma(\omega)$. Since the spectral matrix $S_{pq}(\omega)$ is Hermitian and positive semidefinite, Eq.~\eqref{eq:infidelity-finite-frequency} gives the following bound on the leading-order contributions in $\epsilon$ and $\lambda$:
\begin{equation}
\begin{aligned}
1-F_{\mathrm{avg}}
\leq{}&
\frac{(\epsilon T)^2}{5}
\left[\chi_{ZZ}^{(\mathrm D)}(0)\right]^2
+\frac{2\lambda^2T^2}{5}
\int_0^\infty\mathrm d\omega\,s_{\gamma}(\omega)\\
&\times
\left\{
\left[\chi_{IZ}^{(\mathrm D)}(\omega)\right]^2
+\left[\chi_{ZI}^{(\mathrm D)}(\omega)\right]^2
\right\}.
\end{aligned}
\label{eq:bath-leading-infidelity-correlated}
\end{equation}
The static condition $\chi_{ZZ}^{(\mathrm D)}(0)=0$ eliminates the leading-order effect of the static drift. If the spectrum $S_{\gamma}(\omega)$ is dominated by a peak centered at the frequency $\omega_p$, imposing $\chi_{IZ}^{(\mathrm D)}(\omega_p)=\chi_{ZI}^{(\mathrm D)}(\omega_p)=0$ suppresses the dominant bath contribution and strongly reduces the infidelity. Consequently, we only require knowledge of the position $\omega_p$ of the dominant peak to construct the control, without information about its detailed shape. This is important for experiments, since locating the dominant peak is generally much easier than measuring the full spectrum accurately. 

For the numerical results below, we include the $ZZ$ perturbation directly in the system propagation and evaluate the bath-induced gate infidelity through fourth order in the system--bath coupling $\lambda$. Details of the calculation are given in Appendix~\ref{app:numerical-algorithm}.

We show the spectral origin of this suppression in panels~(a) and (b) of Fig.~\ref{fig-DCG(ZZ)+bath(IZ+ZI)}. In panel~(a), all three susceptibilities remain suppressed over a range of frequency detunings, which is not perturbatively small compared to $1/T$. The bath contribution is therefore strongly suppressed when its dominant spectral weight is concentrated within the corresponding interval, rather than only for an infinitesimally narrow peak.

In panel~(e), the robust control maintains the infidelity at the $10^{-4}$ level throughout the displayed parameter range. In contrast, the infidelity of the unprotected control reaches the $10^{-1}$ level in panel~(f), rendering the unprotected gate practically unusable. Moreover, panel~(c) examines the effect of simultaneously detuning the bath-spectrum maximum from $\omega_pT=1$ and the $ZZ$ perturbation from zero frequency. Throughout the displayed non-perturbative detuning range, the infidelity is still strongly suppressed. Therefore, by designing finite-frequency DCGs at the peak frequency $\omega_p$, we can strongly suppress the effect of a heat bath with a dominant peak at this frequency. Furthermore, the suppression is not confined to the exact design frequencies, remaining effective when the position of the dominant peak is known only within a finite accuracy.

In panel~(d), across $0.5\leq\gamma T\leq1.5$, the infidelities for all three parameter pairs increase by less than a factor of three and remain below approximately $1.5\times10^{-4}$. Furthermore, the robustness of this finite-frequency DCG is not limited to linewidth variations of the underdamped Brownian spectrum in Eq.~\eqref{eq:bath-spectrum-model}. As demonstrated in Appendix~\ref{app:bath-spectral-shapes}, the same fixed gate implementation, without new optimization, remains effective for the tested Brownian, Gaussian, squared-Lorentzian, and split-Gaussian spectra when their dominant peaks remain at the design frequency $\omega_pT=1$. Therefore, we can maintain robustness when the details of the spectra are unknown, but the dominant peak frequencies are known.

For correlated bath couplings with the noise-power spectra $S_\gamma(\omega)=\lambda^2s_\gamma(\omega)$ and the same control, Eq.~\eqref{eq:bath-leading-infidelity-correlated} shows that the leading-order result in Eq.~\eqref{eq:bath-leading-infidelity}, evaluated at the effective coupling $\sqrt{2}\lambda$, bounds the leading-order infidelity for all compatible bath correlations at $\lambda$, with $\epsilon$ unchanged. Thus, bath correlations do not alter the susceptibility conditions used here, and we expect the same robustness under the correlated heat bath.

These results establish that knowledge of the dominant spectral peak position alone, without a detailed reconstruction of the spectrum, is sufficient to construct a single control that can substantially improve gate fidelity while simultaneously suppressing the static $ZZ$ drift. This tailored construction is convenient for experiments in which the perturbation directions and dominant spectral peaks are known, and it keeps the control less demanding than the all-DCG protocol in Sec.~\ref{sec:numerical-A}.

We further extend this construction in three settings: conditions at different peak frequencies for different system coupling operators in a qutrit gate (Appendix~\ref{app:qutrit-example}), conditions at several frequencies for the same perturbation direction (Appendix~\ref{app:joint-CZ-robustness}), and waveform-selective conditions for a $ZZ$ drift with exponentially decaying components (Appendix~\ref{app:waveform-selective-CZ}). In all three settings, the robust controls strongly suppress the targeted errors, showing the generality of our framework.

\subsection{Limited Improvements under Markovian Lindblad Noise}
\label{sec:numerical-C}

\begin{figure}[t!]
    \centering
    \includegraphics[width=1\linewidth]{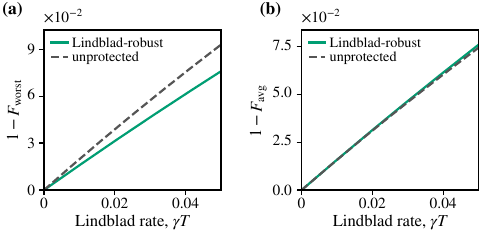}
    \caption{\textbf{Limited improvement of CZ-gate performance under local Lindblad dephasing.} \textbf{(a)} Worst-case gate infidelity $1-F_{\mathrm{worst}}$ and \textbf{(b)} average gate infidelity $1-F_{\mathrm{avg}}$ as functions of the dimensionless Lindblad rate $\gamma T$ for the Lindblad-robust and unprotected gates. The Lindblad-robust gate partially suppresses the worst-case infidelity, while the average infidelity is not suppressed, reflecting the limitation imposed by the fixed depolarizing component, as we show in Sec.~\ref{sec:theoryA}. The maximum Hamiltonian norm $\max_t\|H(t)\|$ is $287.88/T$ for the robust control and $16.67/T$ for the unprotected control.}
    \label{fig-Lindblad}
\end{figure}

Last, we consider the Lindblad-optimized CZ gate in the presence of local dephasing. The two qubits are subject to independent pure-dephasing channels with Lindblad operators $IZ$ and $ZI$ and the same Lindblad rate $\gamma$. We obtain the Lindblad-robust gate by minimizing $\mathcal C_{\mathrm{L}}$ in Eq.~\eqref{eq:cost-Lindblad}. For the unprotected gate, we use the same optimization without the Lindblad susceptibilities $\chi_a^{(\mathrm L)}$ defined in Eq.~\eqref{eq:Lindblad-susceptibility}. We evaluate both gates under the Lindblad dynamics in Eq.~\eqref{eq:master-equation}, using the average gate fidelity $F_{\mathrm{avg}}$ and the worst-case gate fidelity defined as
\begin{equation}
F_{\mathrm{worst}}=\min_{\ket{\psi}}F_\psi=\min_{\ket{\psi}}\bra{\psi}\target^\dagger\rho_\psi(T)\target\ket{\psi},
\end{equation}
where $\rho_\psi(T)$ is the final density matrix for the initial density matrix $\ket{\psi}\bra{\psi}$, and the minimum runs over normalized pure states $\ket{\psi}$.

Figure~\ref{fig-Lindblad} shows a clear difference between the two fidelity measures. The Lindblad-robust gate partially reduces the worst-case infidelity relative to the unprotected gate, whereas the average infidelity is not suppressed. This difference directly reflects the decomposition of dissipation into the fixed depolarizing component and the suppressible nondepolarizing components. The Lindblad-susceptibility constraint suppresses the state-dependent effects of the nondepolarizing components, partially improving the worst-case performance. By contrast, coherent control leaves the fixed depolarizing component unchanged, making the average gate infidelity approximately independent of the control protocol.

The robust control also has a substantially larger maximum Hamiltonian norm compared to the unprotected control. Even with a considerable increase in the available control strength, Lindblad optimization provides only a partial improvement in the worst-case performance, but there is no corresponding suppression of the average infidelity. This result illustrates the intrinsic limitation of coherent control against Markovian dissipation containing a fixed depolarizing component.

\section{Discussion and Outlook}
\label{sec:outlook}

There exist many different protocols to implement the same gate. Using Lie-algebraic control theory, we establish conditions under which this freedom can induce robustness by canceling the leading-order effects of errors on the average gate fidelity. The ability to implement all gates in a $\dimension$-level system is not sufficient: the entire family of robust gates, consisting of every gate under any combination of static DCG conditions, the EGG condition, finite-frequency DCG conditions, and waveform-selective DCG conditions, exists if and only if the driven Hamiltonian operators share no common symmetry. Beyond full unitary controllability, we also develop sufficient algebraic criteria for robust gate implementations. Once existence is guaranteed, we use numerical optimization to construct the robust control protocols, reducing gate infidelities by orders of magnitude under Hamiltonian perturbations and environment-induced noise. In contrast, because of the fixed depolarizing component, coherent control under generic Lindblad noise can only partially suppress the worst-case gate infidelity at a fixed gate duration $T$.

From a broader experimental perspective, the existence criteria allow us to evaluate the potential of a quantum computing platform for implementing robust gates, based on its available control channels. If the original control set does not meet the relevant sufficient conditions, one could also use these criteria to select additional controls to satisfy them. If the criteria are satisfied in experiments, there are different strategies for robust gate implementations: we can impose robustness conditions broadly to guarantee robustness with less information about errors, or tailor robustness conditions according to the information we have about the platform. In particular, to counter environment-induced noise with dominant frequency peaks, we can impose the finite-frequency DCG conditions using only the peak frequencies and system coupling operators. The numerical examples in Sec.~\ref{sec:numerical} and Appendices~\ref{app:additional-gate-examples}--\ref{app:waveform-selective-examples} illustrate this flexibility across different target gates, system dimensions, control parameterizations, and noise environments.

Our examples show that the robust protocols found here can have substantially larger maximum Hamiltonian norms than the corresponding unprotected protocols; hence, their implementation may present additional experimental challenges. These results motivate a systematic search for minimum-cost controls that keep the required robustness. Furthermore, the explicit numerical demonstration is limited to few-body systems. We have not optimized the numerical algorithm for efficiency; more efficient and scalable methods may facilitate the construction of robust gates in larger systems with a growing number of simultaneous robustness constraints.

Our framework is formulated for $\mathfrak{su}(\dimension)$ dynamics. It therefore applies only to finite Hilbert space dimensions, and generalizing it to infinite dimensions, as in bosonic systems, remains to be explored. Further work may also explore other robustness conditions, such as robustness against complicated crosstalk in longer gate sequences. With the existence criteria for a large set of noise-robust gates identified, the stage is now set for experiments to test their capabilities in high-fidelity gate implementations.

\begin{acknowledgments}
We thank Steve Campbell for helpful discussions on robust quantum gates against arbitrary static perturbations. We also thank Hongzheng Zhao for useful suggestions and Lorenza Viola for helpful comments on dynamically corrected gates.
The authors used GPT-5.6 and Claude Opus 5.5 during the preparation of this manuscript to improve readability and assist with code optimization.
The authors assume full responsibility for the scientific content, accuracy, and integrity of this paper.

Funded by the Deutsche Forschungsgemeinschaft (DFG, German Research
Foundation) -- 563114538, and the European Union (ERC, QuSimCtrl, 101113633). Views and opinions expressed are, however, those of the authors only and do not necessarily reflect those of the European Union or the European Research Council Executive Agency. Neither the European Union nor the granting authority can be held responsible for them. 
\end{acknowledgments}

\section*{Data Availability}
The data and code associated with this manuscript are deposited in Zenodo under the DOI {\urlstyle{same}\href{https://doi.org/10.5281/zenodo.23339159}{\nolinkurl{10.5281/zenodo.23339159}}}~\cite{ZenodoData}. Automatic access is restricted; access will be granted to human researchers upon reasonable request via Zenodo for academic and verification purposes.


\appendix

\begingroup
\renewcommand{\tocname}{Contents of the Appendices}
\let\appendixcontentsline\contentsline
\renewcommand{\contentsline}[4]{}
\let\skipcontentsline\contentsline
\addtocontents{toc}{\protect\let\protect\contentsline\protect\appendixcontentsline}
\tableofcontents
\endgroup

\section{Additional Examples of Robust Gates}
\label{app:additional-gate-examples}

In this section, we present three additional numerical examples of robust gates beyond the robust CZ gate considered in Sec.~\ref{sec:numerical}. We consider a non-Clifford single-qubit gate, a spin-1 qutrit Fourier gate, and the $\sqrt{i\mathrm{SWAP}}$ gate. For the single-qubit example, we use both Fourier and piecewise-constant controls.

\subsection{Robust Single-Qubit Gate}
\label{app:single-qubit-example}

We now construct a non-Clifford single-qubit gate using both Fourier and piecewise-constant (PWC) controls. We choose the target gate as
\begin{equation}
    \target=\exp\!\left[-\frac{i\pi}{8}\frac{X+Y+Z}{\sqrt{3}}\right],
\end{equation}
where $X$, $Y$, and $Z$ are the Pauli operators. We use the Hamiltonian
\begin{equation}
    H(t)=u_x(t)X+u_z(t)Z.
\end{equation}
The driven operators satisfy $\operatorname{Lie}\{iX,iZ\}=\mathfrak{su}(2)$ and $\mathcal C_{\mathrm{dyn}}=\{0\}$. Consequently, the criteria in Sec.~\ref{sec:theoryB} allow us to impose the static DCG, finite-frequency, and EGG conditions simultaneously.

We consider two identical and independent bosonic baths coupled to the qubit through $X$ and $Z$, together with a static perturbation $\epsilon Y$. We use the Brownian spectrum in Eq.~\eqref{eq:bath-spectrum-model}, with $\gamma T=1$, $\beta_{\mathrm B}^{-1}T=50\pi$, and $\alpha T=\pi/4$. Both spectra have their maximum at $\omega_pT=1$. Therefore, the joint robustness conditions are
\begin{equation}
\begin{aligned}
    \chi_X^{(\mathrm D)}(\omega_p)&=\chi_Z^{(\mathrm D)}(\omega_p)=0,\\
    \chi_Y^{(\mathrm D)}(0)&=0,\quad \chi^{(\mathrm E)}=0.
\end{aligned}
\end{equation}

For the Fourier controls, we use Eq.~\eqref{eq:Fourier-control} with harmonic cutoff $\ell_{\max}=64$. For the PWC controls, we optimize independent values of $u_x$ and $u_z$ on 128 equal-duration segments. For each parameterization, we minimize $\mathcal C_{\mathrm H}$ with the susceptibilities above to obtain a robust control, and omit the susceptibility terms to obtain an unprotected control. 

\begin{figure*}[t!]
    \centering
    \includegraphics[width=1\linewidth]{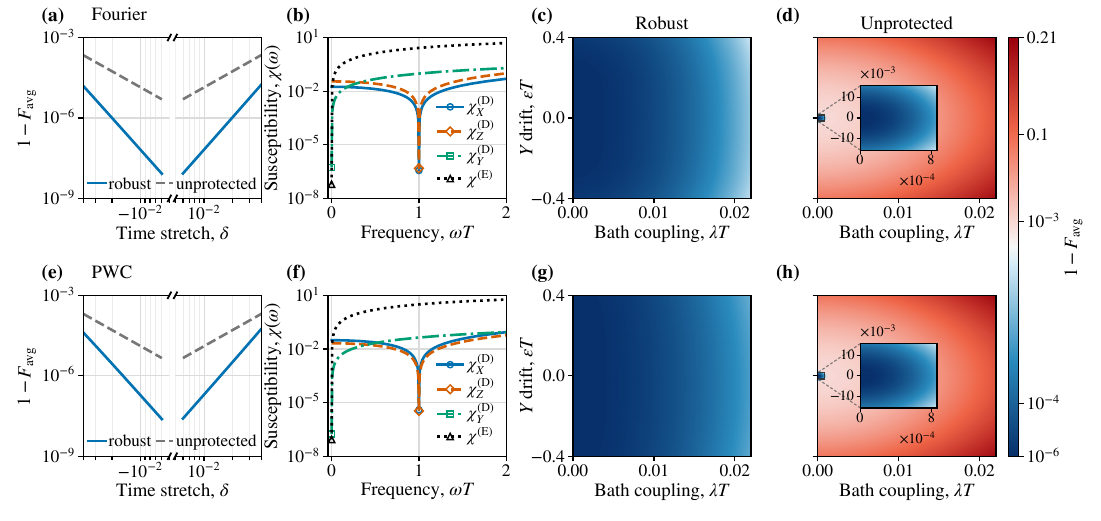}
    \caption{\textbf{Robust single-qubit gate with Fourier and piecewise-constant controls.}
    The upper row shows results for Fourier controls with the harmonic cutoff $\ell_{\max}=64$, and the lower row shows results for piecewise-constant (PWC) controls with 128 equal-duration segments.
    \textbf{(a)}, \textbf{(e)} Average gate infidelities $1-F_{\mathrm{avg}}$ of the robust and unprotected controls under a uniform time stretch $\delta$. At $|\delta|=0.01$, the robust Fourier and PWC controls suppress the infidelity by factors of approximately $210$ and $66$, respectively.
    \textbf{(b)}, \textbf{(f)} Frequency-dependent susceptibilities $\chi_X^{(\mathrm D)}(\omega)$, $\chi_Z^{(\mathrm D)}(\omega)$, $\chi_Y^{(\mathrm D)}(\omega)$, and $\chi^{(\mathrm E)}(\omega)$ of the robust controls. The $X$ and $Z$ susceptibilities vanish at $\omega T=1$, while the $Y$ and extended geometric gate susceptibilities vanish at $\omega=0$.
    \textbf{(c)}, \textbf{(g)} Average gate infidelities of the robust controls under simultaneous static $Y$ drift and coupling to two identical and independent thermal baths through $X$ and $Z$.
    \textbf{(d)}, \textbf{(h)} Corresponding results for the unprotected controls. We use the Brownian spectrum in Eq.~\eqref{eq:bath-spectrum-model}, with $\gamma T=1$, $\omega_pT=1$, $\beta_{\mathrm B}^{-1}T=50\pi$, and $\alpha T=\pi/4$.
    Panels (c), (d), (g), and (h) use the same piecewise colorbar, which maps $1-F_{\mathrm{avg}}$ linearly from $1.0\times10^{-6}$ to $3.6\times10^{-4}$ (blue) and from $3.7\times10^{-4}$ to $0.21$ (red), with a smooth transition between the two intervals.
    The maximum Hamiltonian norms $\max_t\|H(t)\|$ are $39.61/T$ for the robust Fourier control, $9.52/T$ for the unprotected Fourier control, $41.00/T$ for the robust PWC control, and $3.49/T$ for the unprotected PWC control.}
    \label{fig-single-qubit-robustness}
\end{figure*}

Figure~\ref{fig-single-qubit-robustness}(a,e) shows the average gate infidelities under a uniform time stretching $\delta$. At $|\delta|=0.01$, the robust Fourier and PWC controls reduce the infidelity by factors of approximately $210$ and $66$, respectively, compared with their unprotected controls. Panels (b) and (f) show the frequency-dependent susceptibilities of the same robust controls. The $X$ and $Z$ susceptibilities vanish at $\omega T=1$, while the $Y$ and EGG susceptibilities vanish at $\omega=0$. Both controls satisfy the imposed robustness conditions and maintain partial protection under frequency detunings.

Panels (c,d) and (g,h) compare the robust and unprotected controls over $|\epsilon T|\leq0.4$ and $0\leq\lambda T\leq0.022$. The maximum average gate infidelities of the robust Fourier and PWC controls are approximately $3.56\times10^{-4}$ and $2.50\times10^{-4}$, respectively. The corresponding unprotected controls reach $0.200$ and $0.201$. 

As we show using these numerical results, both parameterizations suppress the effects of simultaneous static drift and bath coupling while keeping the EGG condition for the same non-Clifford target gate. Moreover, the suppression remains effective when the width of the dominant spectral peak is far from infinitesimally small, in agreement with the example in Sec.~\ref{sec:numerical-B}.

\subsection{Robust Qutrit Gate}
\label{app:qutrit-example}

We now construct a robust $\mathrm{SU}(3)$ Fourier gate on a spin-1 qutrit system:
\begin{equation}
    \target=\frac{e^{i\pi/6}}{\sqrt{3}}
    \begin{pmatrix}
        1&1&1\\
        1&e^{2\pi i/3}&e^{4\pi i/3}\\
        1&e^{4\pi i/3}&e^{2\pi i/3}
    \end{pmatrix},
\end{equation}
which mixes all three levels. We choose this gate because it is the qutrit analogue of the Hadamard gate. With the spin-1 operators
\begin{equation}
\begin{aligned}
    S_x&=\frac{1}{\sqrt{2}}
    \begin{pmatrix}
        0&1&0\\
        1&0&1\\
        0&1&0
    \end{pmatrix},\quad
    S_y=\frac{1}{\sqrt{2}}
    \begin{pmatrix}
        0&-i&0\\
        i&0&-i\\
        0&i&0
    \end{pmatrix},\\
    S_z&=\operatorname{diag}(1,0,-1),
\end{aligned}
\end{equation}
we define
\begin{equation}
    Q_Z=\sqrt{3}\left(S_z^2-\frac{2}{3}\identity_3\right),
    \quad Q_\perp=S_x^2-S_y^2,
\end{equation}
where $\identity_3$ is the three-dimensional identity. We take the Hamiltonian to be
\begin{equation}
    H(t)=u_0Q_Z+u_x(t)S_x+u_y(t)S_y,
\end{equation}
with $u_0$ time-independent and $u_x(t)$ and $u_y(t)$ smooth and $T$-periodic. The available operators then satisfy
\begin{equation}
    \operatorname{Lie}\{iQ_Z,iS_x,iS_y\}=\mathfrak{su}(3),
\end{equation}
and the driven operators have no nonzero common symmetry, i.e., $\mathcal C_{\mathrm{dyn}}=\{0\}$. Therefore, the criteria in Sec.~\ref{sec:theoryB} allow us to impose the static DCG and finite-frequency DCG conditions below.

We couple the qutrit to two independent bosonic baths through $S_z$ and $Q_\perp$, and include a static perturbation $\epsilon Q_Z$. For each bath, we use the Brownian spectrum in Eq.~\eqref{eq:bath-spectrum-model}, with $\gamma T=1$, $\beta_{\mathrm B}^{-1}T=50\pi$, and $\alpha T=\pi/4$. We choose the characteristic bath frequencies separately to place the spectral maxima at $\omega T=1$ for $S_z$ and $\omega T=2$ for $Q_\perp$, but the two baths share a common coupling strength $\lambda$. With these settings, we impose the following robustness conditions:
\begin{equation}
\begin{aligned}
    \chi_{S_z}^{(\mathrm D)}(1/T)&=0,\qquad
    \chi_{Q_\perp}^{(\mathrm D)}(2/T)=0,\\
    \chi_{Q_Z}^{(\mathrm D)}(0)&=0.
\end{aligned}
\end{equation}
As in Sec.~\ref{sec:numerical}, we parametrize the Hamiltonian with harmonic cutoff $\ell_{\max}=64$, and minimize $\mathcal C_{\mathrm H}$ over $u_0$ and the Fourier coefficients. We obtain the unprotected control by omitting the susceptibility terms.

Figure~\ref{fig-qutrit-sqrtiswap-robustness}(a) shows the two normalized bath spectra. Panel (b) shows the susceptibilities of the robust control, which vanish at their respective design frequencies. We keep each control fixed and evaluate its performance under simultaneous static drift and bath coupling. Over $|\epsilon T|\leq0.4$ and $0\leq\lambda T\leq0.022$, the maximum average gate infidelity is approximately $6.58\times10^{-5}$ for the robust control in panel (c), compared with $5.75\times10^{-2}$ for the unprotected control in panel (d). These maxima differ by nearly three orders of magnitude. We thus suppress static drift and two bath couplings with different peak frequencies in a three-level system.

This example further illustrates the flexibility of the finite-frequency DCG conditions. We can impose these conditions at different frequencies for different system coupling operators, and suppress the effects of thermal baths with different peak frequencies.

\subsection{Robust \texorpdfstring{$\sqrt{i\mathrm{SWAP}}$}{sqrt(iSWAP)} Gate}
\label{app:sqrtiswap-example}

We next construct a robust $\sqrt{i\mathrm{SWAP}}$ gate using an exchange interaction. In the computational basis, the target gate is
\begin{equation}
    \target=
    \begin{pmatrix}
        1&0&0&0\\
        0&1/\sqrt{2}&i/\sqrt{2}&0\\
        0&i/\sqrt{2}&1/\sqrt{2}&0\\
        0&0&0&1
    \end{pmatrix}\in\mathrm{SU}(4).
\end{equation}
To show the generality of our framework, we replace the $ZZ$ interaction by
\begin{equation}
    H_{\mathrm{ex}}=\frac{XX+YY}{\sqrt{2}},
\end{equation}
and use the Hamiltonian
\begin{equation}
\begin{aligned}
    H(t)&=u_0H_{\mathrm{ex}}+u_{x1}(t)XI+u_{y1}(t)YI\\
    &\quad+u_{1x}(t)IX+u_{1y}(t)IY.
\end{aligned}
\end{equation}
Here, $u_0$ is time-independent, and the four driven schedules are smooth and $T$-periodic. The available operators satisfy
\begin{equation}
    \operatorname{Lie}\{iH_{\mathrm{ex}},iXI,iYI,iIX,iIY\}
    =\mathfrak{su}(4),
\end{equation}
and the driven operators satisfy $\mathcal C_{\mathrm{dyn}}=\{0\}$. Therefore, the criteria in Sec.~\ref{sec:theoryB} also apply to this model.

We consider two identical and independent thermal baths coupled through $ZI$ and $IZ$, with a common coupling strength $\lambda$, together with a static perturbation $\epsilon H_{\mathrm{ex}}$. We use the same bath damping, temperature, and spectral scale as in the qutrit example, but place both spectral maxima at $\omega_pT=1$. Then we impose
\begin{equation}
\begin{aligned}
    \chi_{ZI}^{(\mathrm D)}(\omega_p)&=\chi_{IZ}^{(\mathrm D)}(\omega_p)=0,\\
    \chi_{H_{\mathrm{ex}}}^{(\mathrm D)}(0)&=0.
\end{aligned}
\end{equation}
We use the same Fourier parameterization and optimization procedure as above, with harmonic cutoff $\ell_{\max}=64$.

\begin{figure*}[t!]
    \centering
    \includegraphics[width=1\linewidth]{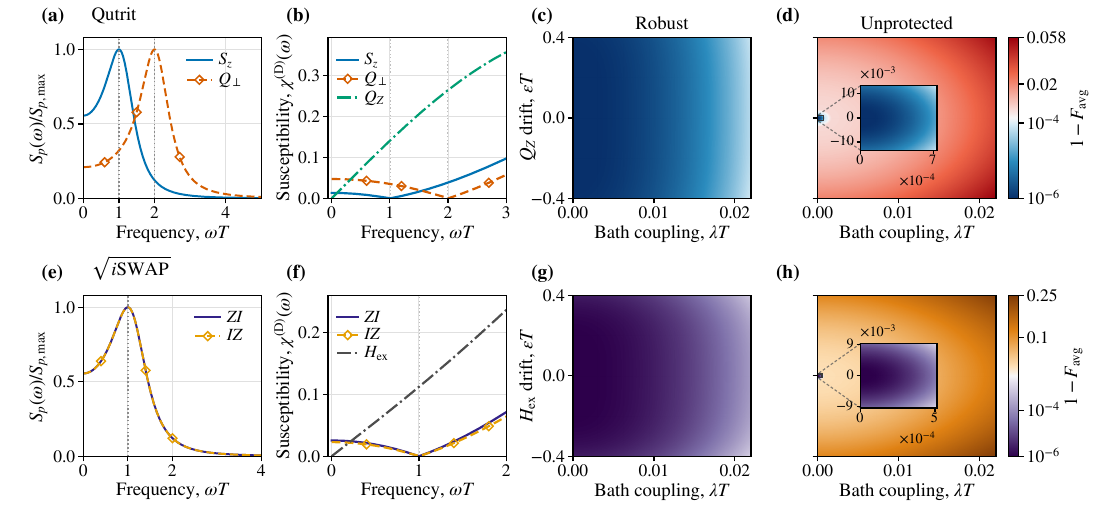}
    \caption{\textbf{Robust qutrit and $\sqrt{i\mathrm{SWAP}}$ gates in the presence of thermal baths and static drift.}
    The upper row shows the qutrit gate, and the lower row shows the $\sqrt{i\mathrm{SWAP}}$ gate.
    \textbf{(a)}, \textbf{(e)} Normalized symmetrized bath spectra. The qutrit spectra peak at $\omega T=1$ and $2$, while both $\sqrt{i\mathrm{SWAP}}$ spectra peak at $\omega T=1$.
    \textbf{(b)}, \textbf{(f)} Frequency-dependent susceptibilities of the robust controls. The susceptibilities for the system coupling operators vanish at the corresponding spectral peak frequencies, while the susceptibilities for the static perturbations $Q_Z$ and $H_{\mathrm{ex}}$ vanish at $\omega=0$.
    \textbf{(c)}, \textbf{(d)} Average gate infidelities $1-F_{\mathrm{avg}}$ of the robust and unprotected qutrit controls, under the combination of static $Q_Z$ drift and bath coupling.
    \textbf{(g)}, \textbf{(h)} Average gate infidelities $1-F_{\mathrm{avg}}$ for the robust and unprotected $\sqrt{i\mathrm{SWAP}}$ controls, under the combination of static $H_{\mathrm{ex}}$ drift and bath coupling.
    Both examples use two independent thermal baths with $\gamma T=1$, $\beta_{\mathrm B}^{-1}T=50\pi$, and $\alpha T=\pi/4$.
    Panels (c) and (d) use the same piecewise colorbar, which maps $1-F_{\mathrm{avg}}$ linearly from $1.0\times10^{-6}$ to $6.6\times10^{-5}$ (blue) and from $2.5\times10^{-4}$ to $0.058$ (red), with a smooth transition between the two intervals.
    Panels (g) and (h) use a separate piecewise colorbar, which maps $1-F_{\mathrm{avg}}$ linearly from $1.0\times10^{-6}$ to $1.6\times10^{-4}$ (purple) and from $2.5\times10^{-4}$ to $0.25$ (orange), with a smooth transition between the two intervals.
    All controls use the harmonic cutoff $\ell_{\max}=64$. The maximum Hamiltonian norms $\max_t\|H(t)\|$ are $123.71/T$ for the robust qutrit control, $22.32/T$ for the unprotected qutrit control, $95.99/T$ for the robust $\sqrt{i\mathrm{SWAP}}$ control, and $15.03/T$ for the unprotected $\sqrt{i\mathrm{SWAP}}$ control.}
    \label{fig-qutrit-sqrtiswap-robustness}
\end{figure*}

Figure~\ref{fig-qutrit-sqrtiswap-robustness}(e) shows the normalized bath spectra, which coincide because the baths are identical. Panel (f) shows that the $ZI$ and $IZ$ susceptibilities vanish at $\omega T=1$, while the susceptibility for $H_{\mathrm{ex}}$ vanishes at zero frequency. As we expect, the robust gate exhibits resilience against the combination of static perturbation and heat bath coupling. Over the same range $|\epsilon T|\leq0.4$ and $0\leq\lambda T\leq0.022$, the maximum average gate infidelity is approximately $1.51\times10^{-4}$ for the robust control in panel (g), compared with $0.244$ for the unprotected control in panel (h). These maxima differ by more than three orders of magnitude. This example shows that the construction also works with an exchange interaction and a target gate different from CZ.

\section{Additional Results for the CZ Gate}
\label{app:additional-CZ-results}

In this section, we present additional numerical results for the CZ gate considered in Sec.~\ref{sec:numerical}. We first test the same robust CZ control under different bath spectral shapes without new optimization. We then construct new robust CZ gates, which provide further examples with multiple robustness conditions satisfied simultaneously.

\subsection{Robust CZ Gates for Different Bath Spectral Shapes}
\label{app:bath-spectral-shapes}

In this subsection, we test whether the robust CZ control in Fig.~\ref{fig-DCG(ZZ)+bath(IZ+ZI)} remains effective for different bath spectral shapes. With both the robust and unprotected controls fixed, we replace the Brownian spectrum with Gaussian, squared-Lorentzian, and split-Gaussian spectra. We keep the two identical and independent thermal baths coupled through $IZ$ and $ZI$, the static perturbation $\epsilon ZZ$, and the temperature $\beta_{\mathrm B}^{-1}T=50\pi$. For calculation convenience, the heat bath is still bosonic.

\begin{figure*}[t!]
    \centering
    \includegraphics[width=1\linewidth]{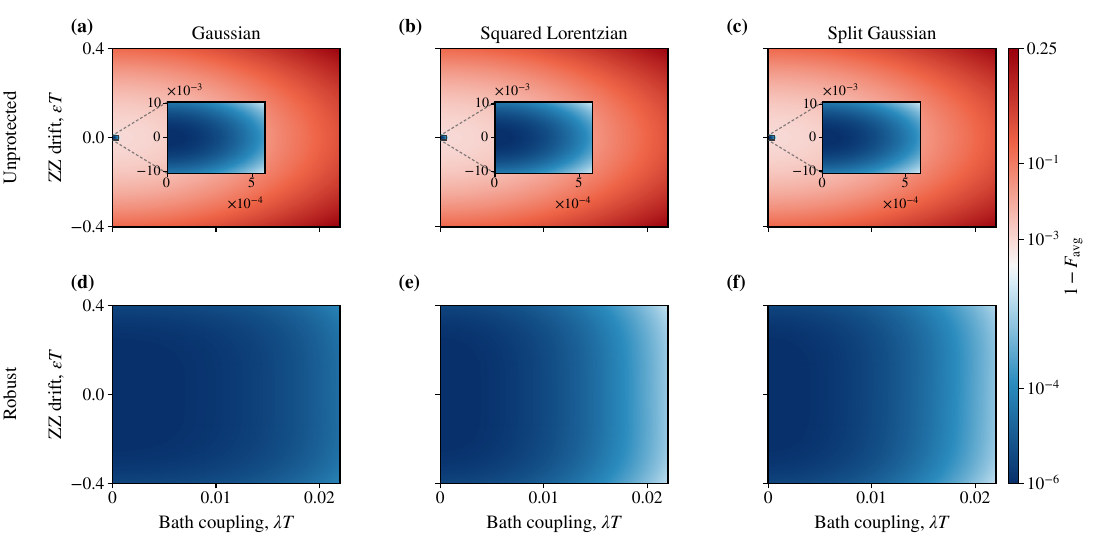}
    \caption{\textbf{Robust CZ gates for different bath spectral shapes.}
Average gate infidelities $1-F_{\mathrm{avg}}$ under simultaneous static $ZZ$ drift and coupling to two identical and independent thermal baths through $IZ$ and $ZI$.
From left to right, the columns correspond to the Gaussian, squared-Lorentzian, and split-Gaussian spectra.
The upper row shows the results for the unprotected control, and the lower row shows the results for the robust control.
We use the same robust and unprotected controls as in Fig.~\ref{fig-DCG(ZZ)+bath(IZ+ZI)}(e) and (f), respectively, without new optimization.
The harmonic cutoff therefore remains $\ell_{\max}=64$ for both controls, while the maximum Hamiltonian norms remain $86.30/T$ for the robust control and $16.67/T$ for the unprotected control.
We choose all three spectra to have the same peak position $\omega_pT=1$, peak height, and positive-frequency spectral weight $\int_0^\infty s(\omega)\,\mathrm d\omega$ as the Brownian spectrum at $\gamma T=1$.
All six panels, including the insets, use the same piecewise colorbar, which maps $1-F_{\mathrm{avg}}$ linearly from $1.0\times10^{-6}$ to $2.1\times10^{-4}$ (blue) and from $2.5\times10^{-4}$ to $0.25$ (red), with a smooth transition between the two intervals. The insets in (a)--(c) show that the unprotected control reaches average infidelities comparable to those in (d)--(f), but with much smaller static drift and bath coupling strengths.}
    \label{fig-bath-spectral-shapes}
\end{figure*}

For each alternative symmetrized spectrum, we assume that $S(\omega)=\lambda^2s(\omega)$ for $\omega\geq0$, where $\lambda$ is the common system--bath coupling strength defined in Sec.~\ref{sec:numerical-B}. We also match its peak position $\omega_pT=1$, peak height, and positive-frequency spectral weight to those of the Brownian spectrum $S_\gamma(\omega)=\lambda^2s_\gamma(\omega)$ in Eq.~\eqref{eq:bath-spectrum-model} at $\gamma T=1$. With $s_\gamma$ evaluated at this value of $\gamma$, the Gaussian and squared-Lorentzian spectra are
\begin{equation}
\begin{aligned}
    s_{\mathrm G}(\omega)
    &=s_\gamma(\omega_p)
    \exp\!\left[-\frac{(\omega-\omega_p)^2}{2\sigma_{\mathrm G}^2}\right],\\
    s_{\mathrm{SL}}(\omega)
    &=s_\gamma(\omega_p)
    \left[1+\frac{(\omega-\omega_p)^2}{\sigma_{\mathrm{SL}}^2}\right]^{-2}.
\end{aligned}
\end{equation}
For the split-Gaussian spectrum, we have
\begin{equation}
\frac{s_{\mathrm{SG}}(\omega)}{s_\gamma(\omega_p)}
=\begin{cases}
\displaystyle \exp\!\left[-\frac{(\omega-\omega_p)^2}{2\sigma_-^2}\right],
&\omega\leq\omega_p,\\[6pt]
\displaystyle \exp\!\left[-\frac{(\omega-\omega_p)^2}{2\sigma_+^2}\right],
&\omega>\omega_p.
\end{cases}
\end{equation}
Here, $\sigma_{\mathrm G}$, $\sigma_{\mathrm{SL}}$, $\sigma_-$, and $\sigma_+$ are positive width parameters. We set $\sigma_+=3\sigma_-$ and determine the remaining widths by requiring
\begin{equation}
    \int_0^\infty s(\omega)\,\mathrm d\omega
    =\int_0^\infty s_\gamma(\omega)\,\mathrm d\omega
\end{equation}
for each alternative spectrum. This gives $\sigma_{\mathrm G}T\simeq0.5394$, $\sigma_{\mathrm{SL}}T\simeq0.9011$, $\sigma_-T\simeq0.2611$, and $\sigma_+T\simeq0.7833$. For each spectrum, we obtain the corresponding spectral density from $J(\omega)=S(\omega)\tanh(\beta_{\mathrm B}\omega/2)$, and evaluate the gate infidelity as in Sec.~\ref{sec:numerical-B}.

In Fig.~\ref{fig-bath-spectral-shapes}, we compare the two controls over $|\epsilon T|\leq0.4$ and $0\leq\lambda T\leq0.022$. For the Gaussian, squared-Lorentzian, and split-Gaussian spectra, the maximum average gate infidelities of the robust control are approximately $1.01\times10^{-4}$, $2.02\times10^{-4}$, and $1.98\times10^{-4}$, respectively. The corresponding maxima for the unprotected control are $0.246$, $0.243$, and $0.238$. For each tested spectrum, the maximum infidelity of the robust control is therefore more than three orders of magnitude lower than that of the unprotected control. Consequently, these results indicate that, when the spectrum remains dominated by a peak at the same frequency, we can maintain robustness across different spectral shapes.

\subsection{Joint Robustness Conditions}
\label{app:joint-CZ-robustness}

\begin{figure*}[t!]
    \centering
    \includegraphics[width=1\linewidth]{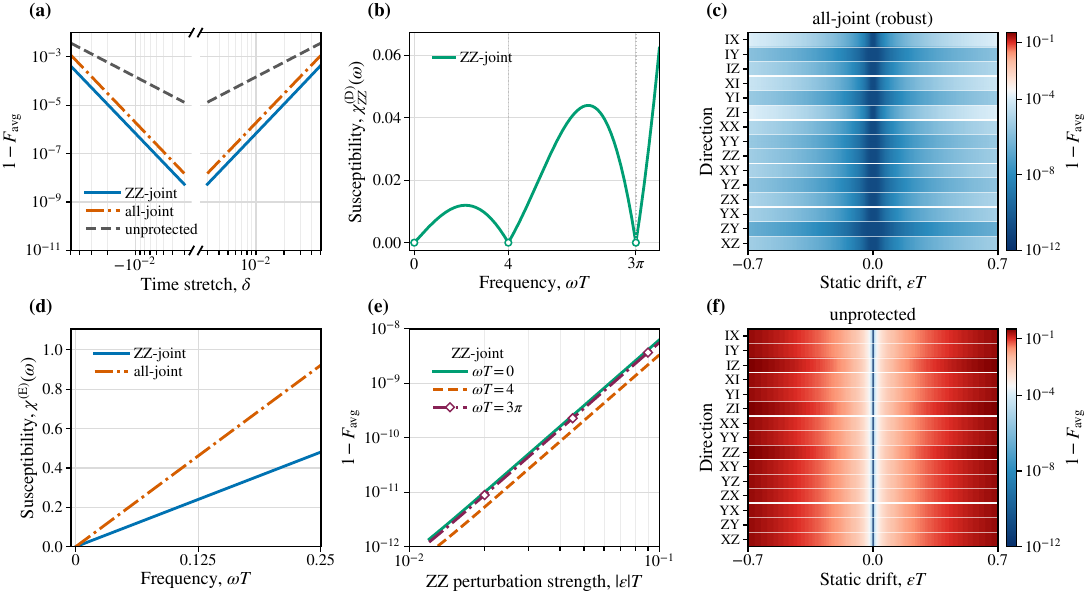}
    \caption{\textbf{CZ gates satisfying joint robustness conditions.}
    The control labeled ``$ZZ$-joint'' satisfies the extended geometric gate condition and the $ZZ$ robustness conditions at $\omega T=0,4,3\pi$. The control labeled ``all-joint'' satisfies the extended geometric gate condition and the static dynamically corrected gate conditions for all 15 generators $\{X_p\}$.
    \textbf{(a)} Average gate infidelities $1-F_{\mathrm{avg}}$ of the two robust controls and the unprotected control under a uniform time stretch $\delta$. At $|\delta|=0.01$, the $ZZ$-joint and all-joint controls suppress the infidelity by factors of approximately $220$ and $78$, respectively.
    \textbf{(b)} Frequency-dependent susceptibility $\chi_{ZZ}^{(\mathrm D)}(\omega)$ of the $ZZ$-joint control. The circles mark the three design frequencies $\omega T=0,4,3\pi$, where the susceptibility vanishes.
    \textbf{(c)}, \textbf{(f)} Average gate infidelities of the all-joint and unprotected controls, respectively, with the 15 static perturbations $\epsilon X_p$ applied separately. Both panels use the same logarithmic color scale.
    \textbf{(d)} Frequency-dependent extended geometric gate susceptibilities $\chi^{(\mathrm E)}(\omega)$ of the two robust controls in (a). Both vanish at $\omega=0$.
    \textbf{(e)} Average gate infidelity of the $ZZ$-joint control under a static perturbation $\epsilon ZZ$ or a finite-frequency perturbation $\epsilon\cos(\omega t+\phi)ZZ$, applied separately at the three design frequencies. For nonzero $\omega$, we maximize the infidelity over $\phi$; for $\omega=0$, we take the larger infidelity between perturbations $|\epsilon|ZZ$ and $-|\epsilon|ZZ$.
    All controls use the harmonic cutoff $\ell_{\max}=64$. The maximum Hamiltonian norms $\max_t\|H(t)\|$ are $144.24/T$ for the $ZZ$-joint control, $297.45/T$ for the all-joint control, and $16.67/T$ for the unprotected control.}
    \label{fig-CZ-joint-robustness}
\end{figure*}

We now construct more examples of robust CZ gates, which satisfy the extended geometric gate condition and other robustness conditions simultaneously. Specifically, we consider two robust controls, namely ``$ZZ$-joint'' and ``all-joint'' controls. The $ZZ$-joint control satisfies
\begin{equation}
\begin{aligned}
    \chi^{(\mathrm E)}&=0,\\
    \chi_{ZZ}^{(\mathrm D)}(\omega)&=0,
    \qquad \omega T\in\{0,4,3\pi\},
\end{aligned}
\end{equation}
and the all-joint control satisfies
\begin{equation}
    \chi^{(\mathrm E)}=0,\qquad
    \chi_p^{(\mathrm D)}=0,\quad p=1,\ldots,15,
\end{equation}
where $\{X_p\}$ are the same 15 generators in Fig.~\ref{fig-all-static}(c,f). 

For numerical constructions of the robust gates, we use the same Hamiltonian control channels and Fourier parameterization as in Sec.~\ref{sec:numerical}, and include all selected susceptibilities in the cost function $\mathcal C_{\mathrm H}$. As a result, both robust controls implement the target CZ gate with noise-free average gate infidelities below $2\times10^{-11}$. For comparison, we use the same unprotected control, as in Fig.~\ref{fig-all-static}(a).

Figure~\ref{fig-CZ-joint-robustness}(a) compares the two robust controls with the unprotected control under a uniform time stretch $\delta$. At $|\delta|=0.01$, the $ZZ$-joint and all-joint controls reduce the infidelity by factors of approximately $220$ and $78$, respectively. Panel (d) shows their frequency-dependent EGG susceptibilities $\chi^{(\mathrm E)}(\omega)$. The two robust gates now share the same property: although we impose the EGG condition with other robustness conditions simultaneously, the protection still extends to slowly varying Hamiltonian-rescaling errors.

Panel (b) shows that the $ZZ$-joint control also suppresses $\chi_{ZZ}^{(\mathrm D)}(\omega)$ at $\omega T=0,4,3\pi$. In panel (e), we obtain average gate infidelities below $7\times10^{-9}$ at $|\epsilon|T=0.1$ for all three frequencies. Since we numerically maximize the infidelity over $\phi$ for each nonzero frequency, this suppression does not rely on a favorable perturbation phase.

Panels (c) and (f) compare the all-joint and unprotected controls under the 15 static perturbations $\epsilon X_p$, applied separately. Over the displayed range $|\epsilon T|\leq0.7$, the largest average gate infidelity of the all-joint control is approximately $1.34\times10^{-4}$, while the unprotected control reaches $0.332$. Thus, the all-joint control suppresses the static perturbations in all directions while satisfying the EGG condition.

These examples show that we can impose the EGG condition and the chosen static DCG or finite-frequency conditions on the same CZ control. The two robust controls use maximum Hamiltonian norms approximately 9 and 18 times that of the unprotected control, respectively, with the same harmonic cutoff $\ell_{\max}=64$.

\section{Waveform-Selective Dynamically Corrected Gates}
\label{app:waveform-selective-examples}

In this section, we present numerical results for qutrit and CZ gates that satisfy the waveform-selective DCG conditions in Eq.~\eqref{eq:waveform-selective-condition}. The qutrit example considers oscillatory noise shaped by a Gaussian envelope, with unknown strength and phase. The CZ example combines static and exponentially decaying $ZZ$ perturbations with environment-induced noise from two thermal baths.

\subsection{Waveform-Selective Robust Qutrit Gate}
\label{app:waveform-selective-qutrit}

We now construct a waveform-selective robust $\mathrm{SU}(3)$ Fourier gate:
\begin{equation}
    \target=\frac{e^{i\pi/6}}{\sqrt{3}}
    \begin{pmatrix}
        1&1&1\\
        1&e^{2\pi i/3}&e^{4\pi i/3}\\
        1&e^{4\pi i/3}&e^{2\pi i/3}
    \end{pmatrix}.
    \label{eqapp:C1-target}
\end{equation}

With the four qutrit operators
\begin{equation}
\begin{aligned}
    H_1&=\begin{pmatrix}
        0&1&0\\
        1&0&0\\
        0&0&0
    \end{pmatrix},&
    H_2&=\begin{pmatrix}
        0&-i&0\\
        i&0&0\\
        0&0&0
    \end{pmatrix},\\
    H_3&=\begin{pmatrix}
        0&0&0\\
        0&0&1\\
        0&1&0
    \end{pmatrix},&
    H_4&=\begin{pmatrix}
        0&0&0\\
        0&0&-i\\
        0&i&0
    \end{pmatrix},
\end{aligned}
    \label{eqapp:C1-operators}
\end{equation}
we use the Hamiltonian
\begin{equation}
    H(t)=\frac{1}{2}\left[u_1(t)H_1+u_2(t)H_2+u_3(t)H_3+u_4(t)H_4\right],
\label{eqapp:C1-Hamiltonian}
\end{equation}
where the four schedules $u_j(t)$ are real, smooth, and $T$-periodic functions. We include no static term, and the driven operators alone generate the full algebra:
\begin{equation}
    \operatorname{Lie}\{iH_1,iH_2,iH_3,iH_4\}=\mathfrak{su}(3).
\end{equation}
This is the special case in Eq.~\eqref{EQ:full-driving-term}, so the driven operators have no common symmetry, i.e., $\mathcal C_{\mathrm{dyn}}=\{0\}$. These control channels therefore allow us to impose the waveform-selective DCG conditions below at any prescribed gate duration $T>0$.

To illustrate a nonperiodic perturbation, we choose a Gaussian envelope, following the Gaussian microwave pulse shapes used in a three-level superconducting circuit~\cite{Vepsalainen2019}:
\begin{equation}
\begin{aligned}
    H(t)\to & H(t)+\frac{\epsilon}{2}e^{-\frac{(t-T/2)^2}{2\sigma^2}}\\
    &\times\big[\cos(\omega t+\phi)H_3+\sin(\omega t+\phi)H_4\big],
\end{aligned}
    \label{eqapp:C1-noise}
\end{equation}
where we fix $\sigma=T/8$ and $\omega T=3.2$, while the noise strength $\epsilon$ and phase $\phi$ are unknown. The two selected smooth real perturbation functions are
\begin{equation}
\begin{aligned}
    f_1(t)&=e^{-\frac{(t-T/2)^2}{2\sigma^2}}\cos(\omega t),\\
    f_2(t)&=e^{-\frac{(t-T/2)^2}{2\sigma^2}}\sin(\omega t).
\end{aligned}
    \label{eqapp:C1-function}
\end{equation}
To construct the robust control, we impose the four waveform-selective DCG conditions:
\begin{equation}
\begin{aligned}
    \chi_{H_3}^{(\mathrm D)}[f_1]&=\chi_{H_3}^{(\mathrm D)}[f_2]=0,\\
    \chi_{H_4}^{(\mathrm D)}[f_1]&=\chi_{H_4}^{(\mathrm D)}[f_2]=0.
\end{aligned}
    \label{eqapp:C1-conditions}
\end{equation}
These four conditions eliminate the leading-order effect of the noise for every phase $\phi$. Since the Gaussian function enters the perturbation functions, the finite-frequency conditions at $\omega$ alone cannot provide the same cancellation.

\begin{figure}[t!]
    \centering
    \vspace{6pt}
    \includegraphics[width=1\linewidth,clip]{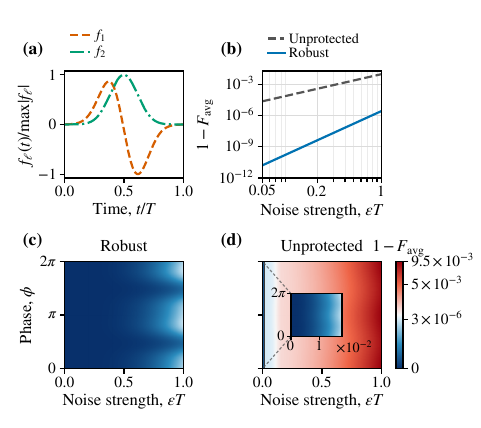}
    \caption{\textbf{Waveform-selective robust qutrit gate.}
    \textbf{(a)} Normalized selected functions $f_1(t)$ and $f_2(t)$ in Eq.~\eqref{eqapp:C1-function}, respectively corresponding to $H_3$ and $H_4$ at $\phi=0$.
    \textbf{(b)} Average gate infidelities $1-F_{\mathrm{avg}}$ of the robust and unprotected controls as the noise strength $\epsilon T$ varies at $\phi=0$.
    \textbf{(c)}, \textbf{(d)} Average gate infidelities of the robust and unprotected controls, respectively, under simultaneous variations of noise strength $\epsilon T$ and phase $\phi$.
    Panels (c) and (d), including the inset, use the same piecewise colorbar, which maps $1-F_{\mathrm{avg}}$ linearly from $0$ to $3.0\times10^{-6}$ (blue) and from $2.5\times10^{-4}$ to $9.5\times10^{-3}$ (red), with a smooth transition between the two intervals.
    The inset in (d) shows that the unprotected control reaches average infidelities comparable to those in (c) at much smaller noise strengths.
    Both controls use the harmonic cutoff $\ell_{\max}=64$.
    The maximum Hamiltonian norms $\max_t\|H(t)\|$ are $47.68/T$ for the robust control and $9.61/T$ for the unprotected control.}
    \label{fig-waveform-selective-qutrit}
\end{figure}

Both controls implement the target gate in the absence of noise. Figure~\ref{fig-waveform-selective-qutrit}(b) shows that, at $\phi=0$, the robust infidelity varies approximately as $(\epsilon T)^4$ over $0.05\leq\epsilon T\leq1$, compared with approximately $(\epsilon T)^2$ for the unprotected control. This agrees with the elimination of the leading-order contribution to the average gate infidelity.

Figure~\ref{fig-waveform-selective-qutrit}(c) shows that the suppression holds for all phases $0\leq\phi<2\pi$. Over $0\leq\epsilon T\leq1$, the maximum average gate infidelity is approximately $2.96\times10^{-6}$ for the robust control in panel (c), compared with $9.49\times10^{-3}$ for the unprotected control in panel (d). These maxima differ by more than three orders of magnitude.

\subsection{Waveform-Selective Robust CZ Gate}
\label{app:waveform-selective-CZ}

We now consider the combination of static DCG, waveform-selective, and finite-frequency DCG conditions for the CZ gate, using the same target CZ gate and control Hamiltonian as in Sec.~\ref{sec:numerical}.

Motivated by the exponential functions used to describe control errors in superconducting circuits~\cite{Rol2020Cryoscope}, we consider the perturbation
\begin{equation}
    H(t)\to H(t)+\left[a_0+a_1e^{-t/\tau_1}+a_2e^{-t/\tau_2}\right]ZZ,
\label{eqapp:C2-perturbation}
\end{equation}
where $\tau_1=T/2$ and $\tau_2=\pi T/2$, while $a_0$, $a_1$, and $a_2$ are independent unknown real coefficients. Moreover, we couple the system to two identical and independent bosonic thermal baths through $IZ$ and $ZI$, with common coupling strength $\lambda$. We use the same Brownian spectrum as in Sec.~\ref{sec:numerical-B}, with $\gamma T=1$, $\beta_{\mathrm B}^{-1}T=50\pi$, and $\alpha T=\pi/4$. Both spectra peak at $\omega_pT=1$. 

To construct the robust control, we impose the robustness conditions:
\begin{equation}
\begin{aligned}
    \chi_{ZZ}^{(\mathrm D)}[1]&=\chi_{ZZ}^{(\mathrm D)}[e^{-t/\tau_1}]=\chi_{ZZ}^{(\mathrm D)}[e^{-t/\tau_2}]=0,\\
    \chi_{IZ}^{(\mathrm D)}(\omega_p)&=\chi_{ZI}^{(\mathrm D)}(\omega_p)=0.
\end{aligned}
    \label{eqapp:C2-waveform-conditions}
\end{equation}
These robustness conditions eliminate the leading-order effect of the perturbation, independently of the ratios between the coefficients. As in Sec.~\ref{sec:numerical-B}, these finite-frequency conditions significantly improve the gate fidelity in the presence of environment-induced noise with dominant spectral peaks at $\omega_p$.

\begin{figure*}[t!]
    \centering
    \includegraphics[width=1\linewidth]{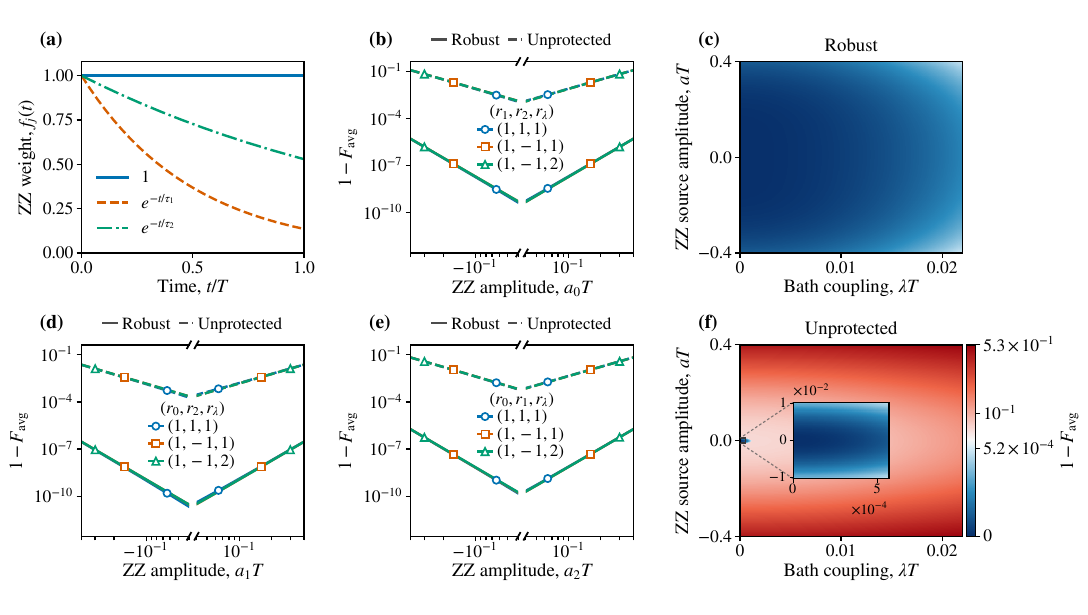}
    \caption{\textbf{Waveform-selective robust CZ gate in the presence of thermal baths.}
    \textbf{(a)} Selected functions $f_0(t)=1$, $f_1(t)=e^{-t/\tau_1}$, and $f_2(t)=e^{-t/\tau_2}$ in the $ZZ$ direction, with $\tau_1=T/2$ and $\tau_2=\pi T/2$.
    \textbf{(b)}, \textbf{(d)}, \textbf{(e)} Average gate infidelities $1-F_{\mathrm{avg}}$ of the robust and unprotected controls as $a_0T$, $a_1T$, and $a_2T$ vary separately. The other two coefficients and the bath coupling remain fixed. We define $r_j=a_jT/10^{-3}$ for $j=0,1,2$ and $r_\lambda=\lambda T/10^{-6}$; the tuples in each legend give the two other fixed coefficients and the bath coupling in that order.
    \textbf{(c)}, \textbf{(f)} Average gate infidelities of the robust and unprotected controls, respectively, with $a_0=a_1=a_2=a$, under simultaneous variations of $aT$ and $\lambda T$.
    We use two identical and independent thermal baths coupled through $IZ$ and $ZI$, with the Brownian spectrum, $\gamma T=1$, $\omega_pT=1$, $\beta_{\mathrm B}^{-1}T=50\pi$, and $\alpha T=\pi/4$.
    Panels (c) and (f), including the inset, use the same piecewise colorbar, which maps $1-F_{\mathrm{avg}}$ linearly from $0$ to $5.2\times10^{-4}$ (blue) and from $5.4\times10^{-4}$ to $0.53$ (red), with a smooth transition between the two intervals.
    The inset in (f) shows average infidelities comparable to those in (c), but at much smaller perturbation amplitudes and bath couplings.
    Both controls use the harmonic cutoff $\ell_{\max}=64$.
    The maximum Hamiltonian norms $\max_t\|H(t)\|$ are $156.48/T$ for the robust control and $16.67/T$ for the unprotected control.}
    \label{fig-waveform-selective-CZ}
\end{figure*}

Both controls implement the target gate in the absence of noise. In Fig.~\ref{fig-waveform-selective-CZ}(b), (d), and (e), we vary $a_0$, $a_1$, and $a_2$ separately, with the other two coefficients fixed at magnitudes $10^{-3}/T$ and the bath coupling fixed at $\lambda T=10^{-6}$ or $2\times10^{-6}$. In the presence of the other two perturbation terms and the thermal baths, the robust control still eliminates the leading-order effect of the perturbation in all three cases.

In Fig.~\ref{fig-waveform-selective-CZ}(c) and (f), we vary the perturbation coefficient $a$ and the coupling strength $\lambda$ simultaneously, with $a_0=a_1=a_2=a$. Over $|aT|\leq0.4$ and $0\leq\lambda T\leq0.022$, the maximum average gate infidelity is approximately $5.12\times10^{-4}$ for the robust control in panel (c), compared with $0.522$ for the unprotected control in panel (f). These maxima differ by more than three orders of magnitude. Therefore, we can obtain robust gates against a combination of different errors, significantly improving gate fidelities in less pristine systems.

\section{Perturbative Gate-Infidelity Expansions}
\label{app:infidelity-expansions}

In this Appendix, we derive the perturbative gate-infidelity expressions used in the main text. We first establish results for Hamiltonian perturbations, and then consider stationary environmental perturbations as well as Lindblad perturbations separately.

\subsection{Hamiltonian Perturbations}
\label{app:infidelity-Hamiltonian}

We keep the system dimension $\dimension$, the gate duration $T$, the ideal evolution $U_0(t)$, and the Frobenius norm as in the main text. The perturbations share the general form
\begin{equation}
H_\eta(t)=H(t)+\eta V(t),
\end{equation}
where $\eta$ is real and $V(t)$ is a traceless Hermitian operator. We denote the corresponding evolution by $U_\eta(t)$, with $U_\eta(0)=\identity_{\dimension}$. The average gate fidelity with respect to $U_0(T)$ is~\cite{fidelitynielsen2002249}
\begin{equation}
F_{\mathrm{avg}}
=\frac{\left|\operatorname{tr}\!\left[U_0^\dagger(T)U_\eta(T)\right]\right|^2+\dimension}
{\dimension(\dimension+1)}.
\label{eqapp:unitary-fidelity}
\end{equation}

We expand the perturbed evolution in the interaction picture, where $V_{\mathrm I}(t)=U_0^\dagger(t)V(t)U_0(t)$. Since $\operatorname{tr}V_{\mathrm I}(t)=0$, the trace has the expansion with respect to $\eta$:
\begin{equation}
\begin{aligned}
\operatorname{tr}\!\left[U_0^\dagger(T)U_\eta(T)\right]
={}&\dimension-\eta^2\int_0^T\mathrm dt\int_0^t\mathrm ds\,
\operatorname{tr}\!\left[V_{\mathrm I}(t)V_{\mathrm I}(s)\right]\\
&+O(\eta^3).
\end{aligned}
\end{equation}
Evidently, we have
\begin{equation}
\begin{aligned}
&\int_0^T\mathrm dt\int_0^t\mathrm ds\,
\operatorname{tr}\!\left[V_{\mathrm I}(t)V_{\mathrm I}(s)\right]
=\frac12\left\|\int_0^T\mathrm dt\,V_{\mathrm I}(t)\right\|^2.
\end{aligned}
\end{equation}
After substituting this expression into Eq.~\eqref{eqapp:unitary-fidelity}, we obtain the general expression for the average gate infidelity:
\begin{equation}
1-F_{\mathrm{avg}}
=\frac{\eta^2}{\dimension+1}
\left\|\int_0^T\mathrm dt\,U_0^\dagger(t)V(t)U_0(t)\right\|^2
+O(\eta^3).
\label{eqapp:general-Hamiltonian-infidelity}
\end{equation}

For perturbations with general time dependence, we need to set $\eta V(t)=\sum_{p,\ell}\epsilon_{p,\ell}f_{p,\ell}(t)X_p$, where $f_{p,\ell}(t)$ are the selected smooth real functions in the main text. The general expression above then gives the average gate infidelity:
\begin{equation}
\begin{aligned}
1-F_{\mathrm{avg}}
&\simeq\frac{T^2}{\dimension+1}
\left\|\sum_{p,\ell}\epsilon_{p,\ell}
\mathcal A_p^{(\mathrm D)}[f_{p,\ell}]\right\|^2,\\
\mathcal A_p^{(\mathrm D)}[f_{p,\ell}]
&=\frac{1}{T}\int_0^T\mathrm dt\,f_{p,\ell}(t)U_0^\dagger(t)X_pU_0(t).
\end{aligned}
\label{eqapp:waveform-selective-Hamiltonian-infidelity}
\end{equation}
Consequently, the waveform-selective DCG conditions $\chi_p^{(\mathrm D)}[f_{p,\ell}]=\|\mathcal{A}_p^{(\mathrm D)}[f_{p,\ell}]\|=0$ eliminate the leading-order effect of the perturbation for every choice of the coefficients $\epsilon_{p,\ell}$. The static and finite-frequency perturbations considered below are special cases of this expression.

For the static perturbation $\epsilon X$, we need to set $\eta=\epsilon$ and $V(t)=X$. Equation~\eqref{eqapp:general-Hamiltonian-infidelity} then gives
\begin{equation}
1-F_{\mathrm{avg}}
=\frac{(\epsilon T)^2}{\dimension+1}
\left(\chi^{(\mathrm D)}\right)^2+O(\epsilon^3T^3),
\end{equation}
with the susceptibility $\chi^{(\mathrm D)}$ defined in the main text. For several static perturbations $\sum_p\epsilon_pX_p$, the same calculation gives the quadratic contribution
\begin{equation}
\begin{aligned}
1-F_{\mathrm{avg}}
&\simeq\frac{T^2}{\dimension+1}
\left\|\sum_p\epsilon_p\mathcal A_p^{(\mathrm D)}(0)\right\|^2,\\
\mathcal A_p^{(\mathrm D)}(0)
&=\frac{1}{T}\int_0^T\mathrm dt\,
U_0^\dagger(t)X_pU_0(t),
\end{aligned}
\label{eqapp:static-Hamiltonian-infidelity}
\end{equation}
where $\simeq$ means we only keep the leading-order effect. Consequently, the static DCG conditions $\chi_p^{(\mathrm D)}=0$, as shown in Eq.~\eqref{eq:DCG-many-condition}, eliminate this leading-order effect for every perturbation in $\operatorname{span}_{\mathbb R}\{X_1,\ldots,X_r\}$.

We next consider the finite-frequency perturbations from the main text:
\begin{equation}
H(t)\to H(t)+\sum_{p,k}\epsilon_{p,k}X_p
\cos\!\left(\omega_{p,k}t+\phi_{p,\omega_{p,k}}\right),
\end{equation}
where $\omega_{p,k}\in\Omega_p$ and every $\Omega_p$ is finite. As in the main text, we use the response operator
\begin{equation}
    \mathcal A_p^{(\mathrm D)}(\omega)=\frac{1}{T}\int_0^T\mathrm dt\,U_0^\dagger(t)X_pU_0(t)e^{i\omega t}.
\end{equation}
Hermiticity of $X_p$ implies $\mathcal A_p^{(\mathrm D)}(-\omega)
=\mathcal A_p^{(\mathrm D)}(\omega)^\dagger$.

After applying Eq.~\eqref{eqapp:general-Hamiltonian-infidelity}, we have the leading-order effect of the finite-frequency perturbations:
\begin{equation}
\begin{aligned}
1-F_{\mathrm{avg}}\simeq{}&\frac{T^2}{4(\dimension+1)}\\
&\times\left\|\sum_{p,k}\epsilon_{p,k}
e^{i\phi_{p,\omega_{p,k}}}\mathcal A_p^{(\mathrm D)}(\omega_{p,k})
+\mathrm{h.c.}\right\|^2.
\end{aligned}
\label{eqapp:finite-frequency-Hamiltonian-infidelity}
\end{equation}
For every phase $\phi$, the triangle inequality gives
\begin{equation}
\left\|\frac{e^{i\phi}\mathcal A_p^{(\mathrm D)}(\omega)
+e^{-i\phi}\mathcal A_p^{(\mathrm D)}(\omega)^\dagger}{2}\right\|
\leq\chi_p^{(\mathrm D)}(\omega).
\end{equation}
Consequently, the quadratic contribution in Eq.~\eqref{eqapp:finite-frequency-Hamiltonian-infidelity} is bounded by
\begin{equation}
\frac{T^2}{\dimension+1}
\left[\sum_{p,k}|\epsilon_{p,k}|\,
\chi_p^{(\mathrm D)}(\omega_{p,k})\right]^2.
\label{eqapp:finite-frequency-Hamiltonian-bound}
\end{equation}
This bound holds for arbitrary phases without averaging over them. Thus, the conditions $\chi_p^{(\mathrm D)}(\omega_{p,k})=0$ eliminate the quadratic contribution for all amplitudes and phases at the selected frequencies. At $\omega=0$, they reduce to the static DCG conditions.

We now consider a uniform time stretching $\mathrm dt\to(1+\delta)\mathrm dt$, which is equivalent to the Hamiltonian rescaling $H(t)\to(1+\delta)H(t)$. After setting $\eta=\delta$ and $V(t)=H(t)$ in Eq.~\eqref{eqapp:general-Hamiltonian-infidelity}, we get the expression
\begin{equation}
\begin{aligned}
1-F_{\mathrm{avg}}
&\simeq{}\frac{\delta^2}{\dimension+1}\left(\chi^{(\mathrm E)}\right)^2,\\
\chi^{(\mathrm E)}
&=\left\|\int_0^T\mathrm dt\,U_0^\dagger(t)H(t)U_0(t)\right\|.
\end{aligned}
\label{eqapp:EGG-infidelity}
\end{equation}
The extended geometric gate condition therefore eliminates the quadratic contribution from uniform Hamiltonian rescaling. The ordinary geometric condition sets only the diagonal elements of this integral to zero in an eigenbasis of $U_0(T)$, and does not in general eliminate the quadratic contribution.

The frequency-dependent susceptibility $\chi^{(\mathrm E)}(\omega)$ used in the numerical results follows from the perturbation $H(t)\to H(t)\big(1+\delta \cos(\omega t+\phi)\big)$, which gives
\begin{equation}
\begin{aligned}
1-F_{\mathrm{avg}}
={}&\frac{\delta^2}{\dimension+1}
\left\|\int_0^T\mathrm dt\,U_0^\dagger(t)H(t)U_0(t)
\cos(\omega t+\phi)\right\|^2\\
&+O(\delta^3).
\end{aligned}
\end{equation}
The same triangle inequality bounds its quadratic contribution by $\delta^2[\chi^{(\mathrm E)}(\omega)]^2/(\dimension+1)$ for every $\phi$, where
\begin{equation}
\chi^{(\mathrm E)}(\omega)
=\left\|\int_0^T\mathrm dt\,U_0^\dagger(t)H(t)U_0(t)e^{i\omega t}\right\|.
\end{equation}
In particular, $\chi^{(\mathrm E)}(0)=\chi^{(\mathrm E)}$. This quantity is useful to evaluate the robustness of gates when the time stretching or Hamiltonian rescaling is not fully uniform.

\subsection{Stationary Environmental Perturbations}
\label{app:infidelity-environment}

We now derive the spectral expression for a system coupled to a stationary environment. If the system is also subject to static perturbations, the total Hamiltonian takes the form
\begin{equation}
\begin{aligned}
H_{\mathrm{tot}}(t)={}&H(t)\otimes\identity_{\mathrm B}
+\identity_{\dimension}\otimes H_{\mathrm B}\\
&+\sum_{p:\,0\in\Omega_p}\epsilon_{p,0}X_p\otimes\identity_{\mathrm B}
+\sum_p\lambda_pX_p\otimes B_p.
\end{aligned}
\label{eqapp:system-bath-Hamiltonian}
\end{equation}
Here, $H_{\mathrm B}$ is the environment Hamiltonian, $B_p$ are Hermitian environment operators, $\lambda_p$ are real coupling strengths, and $\identity_{\dimension}$, $\identity_{\mathrm B}$ are identity operators. The environment is initially in the state $\rho_{\mathrm B}$, with $[\rho_{\mathrm B},H_{\mathrm B}]=0$. We use $\langle O\rangle_{\mathrm B}=\operatorname{Tr}_{\mathrm B}(\rho_{\mathrm B}O)$ for the environment expectation and assume $\langle B_p\rangle_{\mathrm B}=0$. A nonzero mean contributes an additional static system perturbation $\sum_p\lambda_p\langle B_p\rangle_{\mathrm B}X_p$, which must be included in the static term. The operators $B_p$ may be correlated, and the index $p$ labels system coupling directions and need not label independent environments.

After defining the interaction-picture operators,
\begin{equation}
Y_p(t)=U_0^\dagger(t)X_pU_0(t),\qquad
B_p(t)=e^{iH_{\mathrm B}t}B_pe^{-iH_{\mathrm B}t},
\end{equation}
we have the interaction-picture Hamiltonian:
\begin{equation}
H_{\mathrm I}(t)
=\sum_{p:\,0\in\Omega_p}\epsilon_{p,0}Y_p(t)\otimes\identity_{\mathrm B}
+\sum_p\lambda_pY_p(t)\otimes B_p(t),
\end{equation}
which generates the interaction-picture evolution operator $U_{\mathrm I}(T)$.
Now we use a reference system with dimension $\dimension$ that does not evolve. In the corresponding orthonormal bases of the system and reference, we define the maximally entangled state
\begin{equation}
|\phi\rangle=\frac{1}{\sqrt{\dimension}}
\sum_{j=1}^{\dimension}|j\rangle\otimes|j\rangle.
\label{eqapp:max-intangled-state}
\end{equation}
With the initial state $|\phi\rangle\langle\phi|\otimes\rho_{\mathrm B}$, we can define the entanglement fidelity $F_{\mathrm e}$, the probability of recovering $|\phi\rangle$ after the interaction-picture evolution. The average gate fidelity with respect to $U_0(T)$ has the expression~\cite[Eq.~(3)]{fidelitynielsen2002249}
\begin{equation}
F_{\mathrm{avg}}=\frac{\dimension F_{\mathrm e}+1}{\dimension+1}.
\label{eqapp:fidelity-average-entanglement}
\end{equation}
As a result, the calculation of the entanglement fidelity $F_{\mathrm e}$ directly gives us the result of the average gate fidelity. 

We denote the partial trace over the system by $\operatorname{Tr}_{\mathrm S}$. Since $U_{\mathrm I}(T)$ does not act on the reference, we have
\begin{equation}
\langle\phi|U_{\mathrm I}(T)|\phi\rangle
=\frac{1}{\dimension}\sum_{j=1}^{\dimension}
\langle j|U_{\mathrm I}(T)|j\rangle
=\frac{\operatorname{Tr}_{\mathrm S}U_{\mathrm I}(T)}{\dimension}.
\end{equation}
After tracing out the environment, we obtain
\begin{equation}
\begin{aligned}
F_{\mathrm e}
&=\operatorname{Tr}_{\mathrm B}\!\left[
\langle\phi|U_{\mathrm I}(T)|\phi\rangle\rho_{\mathrm B}
\langle\phi|U_{\mathrm I}^\dagger(T)|\phi\rangle\right]\\
&=\frac{1}{\dimension^2}
\left\langle
[\operatorname{Tr}_{\mathrm S}U_{\mathrm I}(T)]^\dagger
[\operatorname{Tr}_{\mathrm S}U_{\mathrm I}(T)]
\right\rangle_{\mathrm B}.
\end{aligned}
\end{equation}
Since $Y_p(t)$ is traceless, $\operatorname{Tr}_{\mathrm S}H_{\mathrm I}(t)=0$. The Dyson expansion therefore gives, to second order,
\begin{equation}
\operatorname{Tr}_{\mathrm S}U_{\mathrm I}(T)
\simeq\dimension\identity_{\mathrm B}
-\int_0^T\mathrm dt\int_0^t\mathrm ds\,
\operatorname{Tr}_{\mathrm S}[H_{\mathrm I}(t)H_{\mathrm I}(s)].
\end{equation}
We substitute this expansion into the expression for $F_{\mathrm e}$ and obtain
\begin{equation}
\begin{aligned}
1-F_{\mathrm e}\simeq{}&\frac{1}{\dimension}
\int_0^T\mathrm dt\int_0^t\mathrm ds\,
\left\langle\operatorname{Tr}_{\mathrm S}
[H_{\mathrm I}(t)H_{\mathrm I}(s)+H_{\mathrm I}(s)H_{\mathrm I}(t)]
\right\rangle_{\mathrm B}\\
={}&\frac{1}{\dimension}
\int_0^T\mathrm dt\int_0^T\mathrm ds\,
\left\langle\operatorname{Tr}_{\mathrm S}
[H_{\mathrm I}(t)H_{\mathrm I}(s)]\right\rangle_{\mathrm B}.
\end{aligned}
\end{equation}
After substituting $H_{\mathrm I}(t)$, the mixed static--environment terms vanish because $\langle B_p(t)\rangle_{\mathrm B}=0$, and we can describe the leading-order effect of static perturbation separately, which is the same as Eq.~\eqref{eqapp:static-Hamiltonian-infidelity}. For the environment contribution, we can always replace $\langle B_p(t)B_q(s)\rangle_{\mathrm B}$ by $\langle\{B_p(t),B_q(s)\}\rangle_{\mathrm B}/2$. Then we have the expression
\begin{equation}
\begin{aligned}
1-F_{\mathrm{avg}}\simeq{}&
\frac{T^2}{\dimension+1}
\left\|\sum_{p}\epsilon_{p,0}
\mathcal A_p^{(\mathrm D)}(0)\right\|^2\\
&+\frac{1}{2(\dimension+1)}\sum_{p,q}\lambda_p\lambda_q
\int_0^T\mathrm dt\int_0^T\mathrm ds\\
&\quad\times\langle\{B_p(t),B_q(s)\}\rangle_{\mathrm B}
\operatorname{tr}[Y_p(t)Y_q(s)],
\end{aligned}
\label{eqapp:environment-time-domain}
\end{equation}
where the first sum runs over the static perturbation directions $X_p$ with $0\in\Omega_p$.

As the bath is stationary, there is $\langle\{B_p(t),B_q(s)\}\rangle_{\mathrm B}=\langle\{B_p(t-s),B_q(0)\}\rangle_{\mathrm B}$. We define the one-sided spectrum by
\begin{equation}
S_{pq}(\omega)=\frac{\lambda_p\lambda_q}{2\pi}
\int_{-\infty}^{\infty}\mathrm d\tau\,e^{i\omega\tau}
\langle\{B_p(\tau),B_q(0)\}\rangle_{\mathrm B},
\ \omega>0.
\label{eqapp:one-sided-spectrum}
\end{equation}
Since the bath is stationary, the symmetry of the anticommutator also gives the relation 
\begin{equation*}
    \langle\{B_p(-\tau), B_q(0)\}\rangle_{\mathrm B}=\langle\{B_q(\tau), B_p(0)\}\rangle_{\mathrm B}.
\end{equation*}
If there is no discrete spectral line at zero frequency, its inverse is
\begin{equation}
\begin{aligned}
&\lambda_p\lambda_q\langle\{B_p(\tau),B_q(0)\}\rangle_{\mathrm B}\\
&\quad=\int_0^\infty\mathrm d\omega\,
\left[S_{pq}(\omega)e^{-i\omega\tau}
+S_{qp}(\omega)e^{i\omega\tau}\right].
\end{aligned}
\end{equation}
As in the main text, $\mathcal A_p^{(\mathrm D)}(\omega)=T^{-1}\int_0^T\mathrm dt\,Y_p(t)e^{i\omega t}$. The time integrals therefore give
\begin{equation}
\begin{aligned}
&\int_0^T\mathrm dt\int_0^T\mathrm ds\,
e^{-i\omega(t-s)}\operatorname{tr}[Y_p(t)Y_q(s)]\\
&\quad=T^2\operatorname{tr}\!\left[
\mathcal A_p^{(\mathrm D)}(\omega)^\dagger
\mathcal A_q^{(\mathrm D)}(\omega)\right].
\end{aligned}
\end{equation}
Consequently, we have the expression for the average gate infidelity:
\begin{equation}
\begin{aligned}
1-F_{\mathrm{avg}}
&\simeq\frac{T^2}{\dimension+1}\Bigg\{
\left\|\sum_p\epsilon_{p,0}
\mathcal A_p^{(\mathrm D)}(0)\right\|^2\\
&\quad+\int_0^\infty\mathrm{d}\nu\sum_{p,q}
S_{pq}(\nu)\operatorname{tr}\!\left[
\mathcal A_p^{(\mathrm D)\dagger}(\nu)
\mathcal A_q^{(\mathrm D)}(\nu)\right]\Bigg\}.
\end{aligned}
\label{eqapp:environment-spectral-infidelity}
\end{equation}
Here, the spectral matrix $S(\omega)$ is Hermitian and positive semidefinite, while the first sum runs over the static perturbation directions $X_p$ with $0\in\Omega_p$. For uncorrelated coupling directions, $S_{pq}(\omega)=\delta_{pq}S_p(\omega)$. Using $\chi_p^{(\mathrm D)}(\omega)=\|\mathcal A_p^{(\mathrm D)}(\omega)\|$, we have the simplified leading-order effect of the environment
\begin{equation}
\frac{T^2}{\dimension+1}\sum_p\int_0^\infty\mathrm d\omega\,
S_p(\omega)\left[\chi_p^{(\mathrm D)}(\omega)\right]^2.
\end{equation}
For two correlated coupling directions with $S_{11}=S_{22}=S_\gamma$, the positivity of the spectral matrix $S(\omega)$ gives $|S_{12}|\leq S_\gamma$. Consequently, we have the inequality
\begin{equation}
\begin{aligned}
\sum_{p,q=1}^2S_{pq}\,
\operatorname{tr}\!\left[
\mathcal A_p^{(\mathrm D)\dagger}\mathcal A_q^{(\mathrm D)}\right]&\leq S_\gamma
\left[\chi_1^{(\mathrm D)}+\chi_2^{(\mathrm D)}\right]^2\\
&\leq2S_\gamma
\left\{\left[\chi_1^{(\mathrm D)}\right]^2+
\left[\chi_2^{(\mathrm D)}\right]^2\right\},
\end{aligned}
\label{eqapp:correlated-spectrum-bound}
\end{equation}
where all quantities are evaluated at the same frequency. This inequality gives the upper bound of the leading-order effect when the bath is correlated, as we show in Eq.~\eqref{eq:bath-leading-infidelity-correlated}.

For discrete spectral lines, we have
\begin{equation}
S_{pq}(\omega)=\sum_jG_{pq}^{(j)}\delta(\omega-\omega_j),
\qquad\omega_j>0,
\end{equation}
where each matrix $G^{(j)}$ is positive semidefinite. For each selected line, we require $\omega_j\in\Omega_p\cup\Omega_q$ whenever $G_{pq}^{(j)}\neq0$, and impose $\chi_p^{(\mathrm D)}(\omega)=0$ for every $\omega\in\Omega_p$. Each term with $G_{pq}^{(j)}\neq0$ then contains at least one response operator equal to zero and therefore vanishes, for both $p=q$ and $p\neq q$. For a continuous spectrum, zeros at finitely many selected frequencies do not, in general, eliminate the integral in Eq.~\eqref{eqapp:environment-spectral-infidelity}. The suppression depends on the spectral weight over the frequency intervals where the susceptibilities are small.

For the thermal bosonic environment used in Sec.~\ref{sec:numerical-B}, we take
\begin{equation}
H_{\mathrm B}=\sum_k\nu_k b_k^\dagger b_k,\qquad
B_p=\sum_k(g_{pk}b_k+g_{pk}^*b_k^\dagger),
\end{equation}
where $\nu_k>0$ and $[b_k,b_l^\dagger]=\delta_{kl}$. The initial state is $\rho_{\mathrm B}=e^{-\beta_{\mathrm B}H_{\mathrm B}}/Z_{\mathrm B}$, with inverse temperature $\beta_{\mathrm B}$ and partition function $Z_{\mathrm B}$. With $n_k=(e^{\beta_{\mathrm B}\nu_k}-1)^{-1}$, the correlation function is
\begin{equation}
\begin{aligned}
\langle B_p(\tau)B_q(0)\rangle_{\mathrm B}
=\sum_k\bigl[&(n_k+1)g_{pk}g_{qk}^*e^{-i\nu_k\tau}\\
&+n_kg_{pk}^*g_{qk}e^{i\nu_k\tau}\bigr].
\end{aligned}
\end{equation}
After substituting this expression into Eq.~\eqref{eqapp:one-sided-spectrum}, we obtain
\begin{equation}
\begin{aligned}
J_{pq}(\omega)
&=\lambda_p\lambda_q\sum_kg_{pk}g_{qk}^*\delta(\omega-\nu_k),\\
S_{pq}(\omega)
&=J_{pq}(\omega)\coth\!\left(\frac{\beta_{\mathrm B}\omega}{2}\right),
\ \omega>0.
\end{aligned}
\end{equation}
By substituting the sum over modes $k$ with an integral, we can describe the continuous spectrum using the same expressions.

\subsection{Lindblad Perturbations}
\label{app:fixed-Lindblad-direction}

In this subsection, we describe the Lindblad evolution in Hilbert--Schmidt space, using the dissipators $\mathcal D_a$, their matrices $D_a$, and the nonnegative rates $\epsilon_a$ defined in the main text. The ideal evolution in this space is $\mathcal U_0(t)=U_0(t)\otimes U_0(t)^*$. In the interaction picture, $|\rho_{\mathrm I}(t)\rangle\rangle=\mathcal U_0^\dagger(t)|\rho(t)\rangle\rangle$ satisfies
\begin{equation}
\frac{\partial|\rho_{\mathrm I}(t)\rangle\rangle}{\partial t}
=\sum_a\epsilon_a\mathcal U_0^\dagger(t)D_a\mathcal U_0(t)
|\rho_{\mathrm I}(t)\rangle\rangle.
\end{equation}
We integrate this equation with the initial state $\rho_0=\rho(0)=\rho_{\mathrm I}(0)$ and keep the linear terms:
\begin{equation}
\begin{aligned}
|\rho_{\mathrm I}(T)\rangle\rangle
&=\left[\identity_{\dimension^2}+T\sum_a\epsilon_a\overline D_a\right]
|\rho_0\rangle\rangle
+O\!\left((T\max_a\epsilon_a)^2\right),\\
\overline D_a
&=\frac1T\int_0^T\mathrm dt\,
\mathcal U_0^\dagger(t)D_a\mathcal U_0(t),
\end{aligned}
\label{eqapp:Lindblad-state-expansion}
\end{equation}
where the noise still stays in the perturbation regime. We use the maximally entangled state $|\phi\rangle$ defined in Eq.~\eqref{eqapp:max-intangled-state}. Under the orthonormal operator basis $E_{jk}=|j\rangle\langle k|$, where $j,k=1,\ldots,\dimension$, we have the expansion
$|\phi\rangle\langle\phi|=\dimension^{-1}\sum_{j,k}E_{jk}\otimes E_{jk}$. After turning to the Hilbert--Schmidt space, the entanglement fidelity after the interaction-picture evolution $F_{\mathrm e}$ takes the form
\begin{equation}
\begin{aligned}
F_{\mathrm e}
&=\frac{1}{\dimension^2}\sum_{j,k}
\langle\langle E_{jk}|
\left[\identity_{\dimension^2}
+T\sum_a\epsilon_a\overline D_a\right]
|E_{jk}\rangle\rangle\\
&\quad\ +O\!\left((T\max_a\epsilon_a)^2\right)\\
&=1+\frac{T}{\dimension^2}
\sum_a\epsilon_a\operatorname{tr}\overline D_a
+O\!\left((T\max_a\epsilon_a)^2\right),
\end{aligned}
\end{equation}
where $\operatorname{tr}\overline D_a=\operatorname{tr}D_a$. By substituting this expression into Eq.~\eqref{eqapp:fidelity-average-entanglement}, we then obtain the average gate infidelity:
\begin{equation}
1-F_{\mathrm{avg}}
=-\frac{T}{\dimension(\dimension+1)}
\sum_a\epsilon_a\operatorname{tr}D_a
+O\!\left((T\max_a\epsilon_a)^2\right).
\label{eqapp:Lindblad-trace-infidelity}
\end{equation}
Here, traces of $D_a$ and its components are ordinary matrix traces in the $\dimension^2$-dimensional representation used in the main text. We can also express the average gate infidelity using the Lindblad operators, since
\begin{equation}
\operatorname{tr}D_a
=\sum_x\left[|\operatorname{tr}L_{ax}|^2
-\dimension\operatorname{tr}(L_{ax}^\dagger L_{ax})\right].
\end{equation}
Thus, at a fixed gate duration, the first-order average infidelity is independent of the coherent control. This conclusion agrees with the weak-dissipation result of Ref.~\cite{Fidlindabad2022}.

We now express this result through the fixed depolarization direction $D_{\mathrm{fix}}$ from the main text. To calculate the projection of $D_a$ onto $D_{\mathrm{fix}}$, we use the definition
\begin{equation}
D_{\mathrm{fix}}
=\frac{1}{\dimension}|\identity_{\dimension}\rangle\rangle
\langle\langle\identity_{\dimension}|-\identity_{\dimension^2},
\end{equation}
which gives
\begin{equation}
\operatorname{tr}D_{\mathrm{fix}}=-(\dimension^2-1),\qquad
\|D_{\mathrm{fix}}\|^2=\dimension^2-1.
\end{equation}
Due to trace preservation, we have $\langle\langle\identity_{\dimension}|D_a=0$. Consequently, the orthogonal projection of $D_a$ onto $D_{\mathrm{fix}}$ gives the coefficient
\begin{equation}
\begin{aligned}
\lambda_{a,\mathrm{fix}}
&=\frac{\operatorname{tr}(D_{\mathrm{fix}}^\dagger D_a)}{\dimension^2-1}
=-\frac{\operatorname{tr}D_a}{\dimension^2-1}\\
&=\frac{\sum_x\left[\dimension\operatorname{tr}(L_{ax}^\dagger L_{ax})
-|\operatorname{tr}L_{ax}|^2\right]}{\dimension^2-1}
\geq0.
\end{aligned}
\label{eqapp:fixed-depolarization-coefficient}
\end{equation}
The inequality follows from $|\operatorname{tr}L|^2\leq\dimension\operatorname{tr}(L^\dagger L)$. The remainder $D_{a,\mathrm{sup}}=D_a-\lambda_{a,\mathrm{fix}}D_{\mathrm{fix}}$ obeys $\operatorname{tr}D_{a,\mathrm{sup}}=0$.

We substitute Eq.~\eqref{eqapp:fixed-depolarization-coefficient} into Eq.~\eqref{eqapp:Lindblad-trace-infidelity} and obtain the average gate infidelity:
\begin{equation}
1-F_{\mathrm{avg}}
=\frac{\dimension-1}{\dimension}T
\sum_a\epsilon_a\lambda_{a,\mathrm{fix}}
+O\!\left((T\max_a\epsilon_a)^2\right).
\label{eqapp:Lindblad-average-infidelity}
\end{equation}
This equation proves the conclusion in Sec.~\ref{sec:theoryA}: we cannot eliminate the leading-order effect of Lindblad noise on the average gate infidelity when the coherent controls are adjustable, but the gate duration $T$ is fixed.

For every $U\in\mathrm{SU}(\dimension)$, we have
\begin{equation}
(U\otimes U^*)^\dagger D_{\mathrm{fix}}(U\otimes U^*)
=D_{\mathrm{fix}}.
\end{equation}
Within the real span of Lindblad dissipators, every matrix fixed by all these conjugations is a real multiple of $D_{\mathrm{fix}}$, as we prove in Appendix~\ref{app:Lindblad-fixed-direction-proof}. In other words, there is no independent fixed direction other than $D_{\mathrm{fix}}$.

To determine the capability of the Lindblad robustness condition $\chi_a^{(\mathrm{ L})}=0$, we consider a pure initial state $P_\psi=|\psi\rangle\langle\psi|$, and the corresponding fidelity is $F_\psi=\langle\langle P_\psi|\rho_{\mathrm I}(T)\rangle\rangle$. We denote the average of $D_{a,\mathrm{sup}}$ along the ideal evolution by
\begin{equation}
\overline D_{a,\mathrm{sup}}
=\frac1T\int_0^T\mathrm dt\,
\mathcal U_0^\dagger(t)D_{a,\mathrm{sup}}\mathcal U_0(t),
\end{equation}
whose Frobenius norm is the susceptibility $\chi_a^{(\mathrm L)}$. The invariance of $D_{\mathrm{fix}}$ implies
\begin{equation}
\overline D_a
=\lambda_{a,\mathrm{fix}}D_{\mathrm{fix}}
+\overline D_{a,\mathrm{sup}}.
\end{equation}
Since
$\langle\langle P_\psi|D_{\mathrm{fix}}|P_\psi\rangle\rangle=-(\dimension-1)/\dimension$, Eq.~\eqref{eqapp:Lindblad-state-expansion} gives the infidelity of the initial state $\ket{\psi}$:
\begin{equation}
\begin{aligned}
1-F_\psi
={}&T\sum_a\epsilon_a\left[
\frac{\dimension-1}{\dimension}\lambda_{a,\mathrm{fix}}
-\langle\langle P_\psi|\overline D_{a,\mathrm{sup}}|P_\psi\rangle\rangle
\right]\\
&+O\!\left((T\max_a\epsilon_a)^2\right).
\end{aligned}
\label{eqapp:Lindblad-state-infidelity}
\end{equation}
The first term is independent of the initial state and the coherent control, while the second term depends on the initial state but has zero average over pure states. The conditions $\chi_a^{(\mathrm L)}=0$ remove the second term for every initial state, giving the equation
\begin{equation}
1-F_{\mathrm{worst}}
=\frac{\dimension-1}{\dimension}T
\sum_a\epsilon_a\lambda_{a,\mathrm{fix}}
+O\!\left((T\max_a\epsilon_a)^2\right),
\end{equation}
where $F_{\mathrm{worst}}=\min_{|\psi\rangle}F_\psi$. The remainder is uniform over pure states in the finite-dimensional system. Since $1-F_{\mathrm{worst}}\geq1-F_{\mathrm{avg}}$, these conditions attain the smallest possible first-order worst-case contribution at fixed $T$. They leave the first-order average infidelity unchanged.

\section{Numerical Optimization of the Robust Gates}
\label{app:numerical-algorithm}

We construct the Fourier controls with Eq.~\eqref{eq:Fourier-control} with fixed gate duration $T$ and harmonic cutoff $\ell_{\max}=64$. The optimization variables are the real Fourier coefficients and, where present, the static coefficient $\staticcoe$. For Hamiltonian robustness conditions, we minimize $\mathcal C_{\mathrm H}$ in Eq.~\eqref{eq:cost-Hamiltonian}, with the specified static DCG, finite-frequency, waveform-selective, and EGG susceptibilities for each example. For Lindblad robustness, we minimize $\mathcal C_{\mathrm L}$ in Eq.~\eqref{eq:cost-Lindblad}, whose susceptibilities correspond to the suppressible components $D_{a,\mathrm{sup}}$. Both cost functions penalize the target-gate error and the selected susceptibilities, with weights $W_{\mathrm G}=W_\chi=1$. To obtain the unprotected controls, we omit the susceptibility terms and use the same gate term and control parameterization.

We first approximate the dynamics on a discrete time grid. On each interval, we replace $H(t)$ by its midpoint value and calculate the propagator. For each iteration, we evaluate the gate error and response integrals from this approximate evolution, which produce the cost function. With this cost and the cost derivatives, we update the optimization variables correspondingly. For robust controls, the initial search also includes the penalty on amplitudes
\begin{equation}
    \frac{\kappa}{2T}\int_0^T\|H(t)\|^2\,\mathrm dt,
\end{equation}
where $\kappa>0$ sets its weight in the cost function. This term discourages large Hamiltonian amplitudes while we search for an initial control.

We then remove the penalty on amplitudes and continue the optimization at higher time resolution. The harmonic cutoff remains fixed, so this step refines the propagation rather than the control parameterization. After the optimization on the discrete grids, we use the resulting coefficients as initial values for a final optimization with the continuous Fourier Hamiltonian. At this stage, we solve the equations for the propagator and the required response integrals together. Their values at $T$ give the gate error and susceptibilities in $\mathcal C_{\mathrm H}$ or $\mathcal C_{\mathrm L}$. The final optimization only uses the cost function defined in Eq.~\eqref{eq:cost-Hamiltonian} or Eq.~\eqref{eq:cost-Lindblad}, without the additional penalty.

For the piecewise-constant single-qubit example, we replace the Fourier parameterization by $128$ equal control segments. We directly optimize the segment Hamiltonian with the same gate and susceptibility terms, without the Hamiltonian-norm penalty. Each segment has a constant Hamiltonian, which gives two optimization coefficients, while each time-dependent control channel contributes $128$ independent coefficients. The optimization process is similar to that of Fourier coefficients.

Once we have obtained a control, we keep it fixed and evaluate its fidelity under the specified perturbations. For Hamiltonian errors, we solve the Schr\"odinger equation with the static or time-dependent perturbations added to $H(t)$. We represent the uniform Hamiltonian-rescaling error by $H(t)\to(1+\delta)H(t)$. We then calculate the average gate infidelity relative to $\target$ through Eq.~\eqref{eqapp:unitary-fidelity}. We apply the same procedure to the robust and unprotected controls. We calculate these errors using exact calculation of evolutions, instead of the leading-order infidelity formulas.

For the thermal-bath examples, we include the static perturbation in the unitary evolution, keeping its full effect. For the system--bath coupling effect, we expand the average gate infidelity in powers of its strength $\lambda$. Using the system Hamiltonian including the static perturbation, we evaluate the average gate infidelity to fourth order in $\lambda$:
\begin{equation}
    1-F_{\mathrm{avg}}
    =c_0+\lambda^2c_2+\lambda^4c_4+O(\lambda^6).
\end{equation}
The coefficient $c_0$ is the infidelity at $\lambda=0$, so it includes the effect of static perturbations. The coefficients $c_2$ and $c_4$ specify the second- and fourth-order bath contributions.

To calculate these coefficients, we take the common coupling-strength dependence $\lambda^2$ out of $J_{pp}(\omega)$ and evaluate the ordered bath correlations $\langle B_p(t)B_p(s)\rangle_{\mathrm B}$, where $t,s\in[0, T]$ are the integration times. In the frequency integral, $e^{-i\omega(t-s)}$ and $e^{i\omega(t-s)}$ have the thermal weights $(1-e^{-\beta_{\mathrm B}\omega})^{-1}$ and $(e^{\beta_{\mathrm B}\omega}-1)^{-1}$, respectively. We integrate the two weighted terms over positive frequencies and keep both the real and imaginary parts of the resulting correlation.

For these Gaussian thermal states, all odd-order bath correlations vanish. We evaluate each four-point correlation from the three possible pairings of the four bath operators. We multiply the two corresponding ordered two-point correlations in each pairing, and then sum the three products.

From the Dyson series of the system--bath propagator and its adjoint, we obtain $c_2$ and $c_4$. We combine terms from the two series according to their total power in $\lambda$. We evaluate the system traces and the thermal averages of bath-operator products separately, then integrate the products of these quantities over the time-ordered domains for the two series. For the numerical evaluation of these integrals, we use midpoint quadrature for the second-order integrals and a scrambled Sobol sequence for the fourth-order integrals.

For different spectral shapes, we recalculate the bath correlations and the coefficients $c_2$ and $c_4$. At the Brownian-bath parameter points that we test, we estimate the sixth-order contribution, whose magnitude is approximately $5\%$ or less of the infidelity calculated through fourth order.

For the Lindblad calculation, we solve the full Lindblad equation in Hilbert--Schmidt space without an expansion in $\gamma$. The equation contains both the suppressible and fixed depolarizing components. The final quantum channel directly gives the average gate infidelity. We estimate $F_{\mathrm{worst}}$ as the smallest $F_\psi$ over a set of pure input states. The same input-state set applies to both the robust and unprotected controls, and we enlarge the set until the minimum fidelity stabilizes.

\begin{figure*}[t]
\centering
\includegraphics[width=1\linewidth]{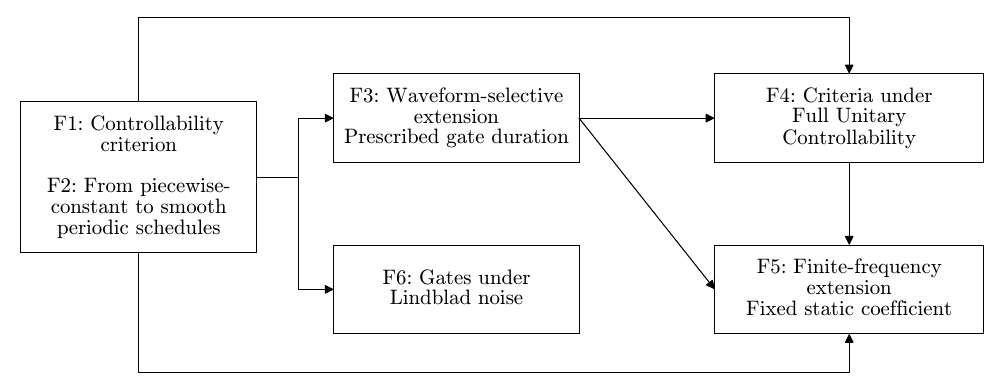}
\caption{\textbf{Proof structure of Appendix~\ref{app:reachability-results}}. Arrows indicate the main dependencies between the proofs. \ref{app:Constrained-Dynamics-and-the-Reachability-Criterion} states the controllability criterion for piecewise-constant controls, and \ref{app:control-regularity} extends the results to smooth periodic controls. \ref{app:waveform-finite-frequency-extension} constructs waveform-selective DCGs at a prescribed duration $T$, with finite-frequency robustness as a special case. \ref{app:full-su-regime} derives the static DCG, EGG, joint static DCG--EGG, finite-frequency, and waveform-selective existence criteria under full unitary controllability. \ref{app:fixed-static-frequency} treats finite-frequency robustness under full unitary controllability with a fixed static coefficient and variable $T$. \ref{app:Lindblad-fixed-direction-proof} proves the Lindblad results.}
\label{fig:appendix-proof-structure}
\end{figure*}

\section{Proofs of the Existence Criteria}
\label{app:reachability-results}

This appendix proves the existence criteria stated in Sec.~\ref{sec:theoretical}. In Fig.~\ref{fig:appendix-proof-structure}, we summarize the proof structure of this appendix. For the zero-frequency constructions, including the static DCG conditions, the EGG condition, and the Lindblad robustness condition, we encode the physical propagator and the accumulated responses as one dynamical state on a semidirect-product group. The group coordinates record the propagator, while the vector coordinates record the responses. A zero vector coordinate at the final time imposes the selected robustness conditions. We use a right-action convention because it reproduces the Schr\"odinger evolution directly.

The first subsection formulates the constrained dynamics and states the controllability criterion for piecewise-constant controls. The second extends the same results to smooth periodic controls. The third constructs waveform-selective DCGs at a prescribed gate duration $T$ and recovers finite-frequency robustness as a special case. We then derive the static DCG, EGG, joint static DCG--EGG, finite-frequency, and waveform-selective existence criteria under full unitary controllability. The fifth subsection derives the finite-frequency result under full unitary controllability for a fixed static coefficient $\staticcoe$. In that setting, $\staticcoe$ is the same for all gate implementations, and we allow the gate duration $T$ to vary. The final subsection proves the Lindblad result. 

For static DCG, EGG, and Lindblad robustness, a global time rescaling connects the fixed-$T$ with the fixed-$\staticcoe$ setting; finite-frequency robustness requires separate proofs for each setting because the rescaling changes the selected physical frequencies. For waveform-selective robustness, we only discuss the fixed-$T$ setting. The same existence results hold for continuous controls, with or without periodicity.

\subsection{Constrained Dynamics and the Reachability Criterion}
\label{app:Constrained-Dynamics-and-the-Reachability-Criterion}

We first prove the general criterion in Sec.~\ref{sec:theoryC} for piecewise-constant controls. Let $K$ be a connected Lie group with Lie algebra $\mathfrak{k}$, and let $V$ be a finite-dimensional real vector space of operators or tuples of operators on which $K$ acts by $v\mapsto k^{-1}vk$. For tuples, conjugation acts componentwise. The semidirect product $G=K\rtimes V$ has the multiplication
\begin{equation}
    (k_1,v_1)(k_2,v_2)
    =
    \bigl(k_1k_2,v_2+k_2^{-1}v_1k_2\bigr).
    \label{eqapp:right-product}
\end{equation}
The group $G$ is a Lie group, and its Lie algebra $\mathfrak{g}=\mathfrak{k}\rtimes V$ has the bracket
\begin{equation}
\begin{aligned}
    \relax[(a_1,v_1),(a_2,v_2)]
    =\bigl(&[a_1,a_2],
    [a_1,v_2]-[a_2,v_1]\bigr).
\end{aligned}
    \label{eqapp:right-bracket}
\end{equation}
For the right-invariant dynamics, we define the trajectory $\varphi(t)=(k(t),v(t))$ on the group $G$, with $\dot\varphi(t)=R(t)\varphi(t)$ and $R(t)=(a(t),w(t))$. For a trajectory starting from $(I,0)$, Eq.~\eqref{eqapp:right-product} gives the evolution of the trajectory components
\begin{equation}
    \dot k(t)=a(t)k(t),
    \qquad
    \dot v(t)=k^{-1}(t)w(t)k(t).
    \label{eqapp:component-dynamics}
\end{equation}
With $a(t)=-iH(t)$ and $k(t)=U_0(t)$, the first component $k(t)$ obeys the Schr\"odinger equation, while the second component $v(t)$ accumulates the response in the toggling frame.

For the static DCG conditions, the trajectory introduced in Sec.~\ref{sec:theoretical} has the components
\begin{equation}
\begin{aligned}
    \varphi_{\mathrm{DCG}}(t)&=(U_0(t),v_{\mathrm{DCG}}(t)),\\
    v_{\mathrm{DCG}}(t)
    &=\left(
    \cdots,
    \int_0^t U_0^\dagger(\tau)X_pU_0(\tau)\,\mathrm d\tau,
    \cdots
    \right).
\end{aligned}
    \label{eqapp:dcg}
\end{equation}
For the EGG condition, the corresponding trajectory is
\begin{equation}
\begin{aligned}
    \varphi_{\mathrm{EGG}}(t)&=(U_0(t),v_{\mathrm{EGG}}(t)),\\
    v_{\mathrm{EGG}}(t)
    &=\int_0^t
    U_0^\dagger(\tau)H(\tau)U_0(\tau)\,\mathrm d\tau.
\end{aligned}
    \label{eqapp:egg-}
\end{equation}
The joint static DCG--EGG construction keeps both response coordinates,
\begin{equation}
    \varphi_{\mathrm{joint}}(t)
    =(U_0(t),v_{\mathrm{DCG}}(t),v_{\mathrm{EGG}}(t)).
    \label{eqapp:joint-}
\end{equation}
Thus, for the Hamiltonian constructions, a trajectory from $(I,0)$ to $(\target,0)$ implements the target gate and makes the selected responses vanish. After removing the fixed depolarizing component, the Lindblad construction has the same form in Hilbert--Schmidt space, but its gate coordinate at the target is $\target\otimes\target^*$. We treat that case separately below.

The dynamics above form a right-invariant control system with driven schedules $u_j(t)\in\mathbb R$ and generator $R(t)\in\Gamma$, where
\begin{equation}
    \Gamma=
    \left\{
    A+\sum_{j=1}^{m}u_jB_j
    \ \middle|\
    u_j\in\mathbb R
    \right\}
    \subset\mathfrak{g}.
    \label{eqapp:control-system}
\end{equation}
Here, $A$ is the time-independent part and the terms $u_j(t)B_j$ are the driven contributions. For piecewise-constant controls, the functions $u_j(t)$ are constant on each control segment, and hence $R(t)$ is a fixed element of $\Gamma$ on that segment. We use Sachkov's Theorem~6.7~\cite{Sachkov2009theorems}, originally proved by Bonnard et al.~\cite[Theorem~1]{Bonnard1982semidirect}:\footnote{Theorem~6.7 writes the vector coordinate first and considers left-invariant systems. The isomorphism $\Xi(k,v)=(kvk^{-1},k)$ maps Eq.~\eqref{eqapp:right-product} to that semidirect-product convention. The inversion map $g\mapsto g^{-1}$ then maps our right-invariant system generated by $\Gamma$ to a left-invariant system generated by $-\Gamma$. Since $\operatorname{Lie}(-\Gamma)=\operatorname{Lie}(\Gamma)$, this change leaves the criterion unchanged.}
\begin{theorem}[Ref.~\cite{Sachkov2009theorems}, Theorem~6.7]
Let $K$ be a compact connected Lie group acting linearly on a vector space $V$, and let $G=K\rtimes V$. Assume that the action of $K$ on $V$ has no nonzero fixed point. The right-invariant system generated by $\Gamma$ is controllable on $G$ if and only if $\operatorname{Lie}(\Gamma)=\mathfrak{g}$.
\end{theorem}
Here, controllability means that, for any $g_0,g_1\in G$, one can find $J\geq1$, elements $R_1,\ldots,R_J\in\Gamma$, and durations $t_1,\ldots,t_J\geq0$ such that
\begin{equation}
    g_1=\exp(R_Jt_J)\cdots\exp(R_1t_1)g_0.
\end{equation}
This product applies $R_j$ for a duration $t_j$ and has total duration $T^*=\sum_{j=1}^Jt_j$.

\textit{Time rescaling.} We first consider the setting with fixed $\staticcoe$ and variable $T$ stated in the main text. Here, $\staticcoe$ has the same fixed value for all target gates, while the gate duration may depend on the target. The controllability theorem above guarantees the existence of a piecewise-constant schedule from $(I,0)$ to $(k_{\ast},0)$ for every $k_{\ast}\in K$, with a variable duration $T^*$ for different gates. The next subsection shows that we can replace this schedule by a smooth periodic schedule without changing $\staticcoe$ or the endpoint.

For the zero-frequency conditions, a global time rescaling relates this setting directly to the fixed-$T$, variable-$\staticcoe$ setting. If $H(t)$ reaches the required endpoint in time $T^*$, then $H_T(t)=\frac{T^*}{T}H\!\left(\frac{T^*}{T}t\right)$ reaches the same endpoint at any prescribed time $T>0$. This transformation multiplies every static DCG or Lindblad response by $T/T^*$ and leaves the EGG response unchanged, so it preserves all corresponding zero-response conditions. It also multiplies $\staticcoe$ and all driven schedules by $T^*/T$. Hence, for static DCG and EGG robustness, the fixed-$\staticcoe$, variable-$T$ and fixed-$T$, variable-$\staticcoe$ settings are directly equivalent. This equivalence assumes no upper bound on the Hamiltonian amplitudes. Time rescaling need not preserve the robustness conditions for selected frequencies or waveforms. We therefore treat waveform-selective and finite-frequency robustness at fixed $T$ in Sec.~\ref{app:waveform-finite-frequency-extension}, and finite-frequency robustness at fixed $\staticcoe$ in Sec.~\ref{app:fixed-static-frequency}.

\subsection{From Piecewise-Constant to Smooth Periodic Schedules}
\label{app:control-regularity}

The controllability theorem constructs schedules that remain constant on each segment, while the main text mostly uses smooth periodic schedules, especially in Sec.~\ref{sec:theoretical} and~\ref{sec:numerical}. Directly smoothing the jumps generally changes the final gate and the accumulated responses. We first choose a piecewise-constant schedule whose endpoint map has full rank at the selected segment durations; here and below, full rank means that the derivative of the endpoint map is surjective. After smoothing the jumps, we can restore the original endpoint exactly by making small adjustments to these durations.

\begin{lemma}
\label{lem:full-rank-endpoint-map}
Fix one of the control systems above and an initial point $x_0$. Each constant choice of the $u_j$ gives an element $R=A+\sum_j u_jB_j\in\Gamma$ and hence a right-invariant vector field. If the Lie algebra generated by these vector fields spans the tangent space at $x_0$, then, with $J=\dim G$, there exist $R_1, \ldots, R_J\in\Gamma$ and strictly positive segment durations $\boldsymbol\tau_*=(\tau_{1,*},\ldots,\tau_{J,*})$ such that the endpoint map
\begin{equation}
\mathcal E(\boldsymbol\tau)
=\exp(\tau_JR_J)\cdots\exp(\tau_1R_1)x_0
\end{equation}
has full rank at $\boldsymbol\tau_*$.
\end{lemma}

\textit{Proof.}
For each constant control choice $R\in\Gamma$, the associated right-invariant vector field generates the flow $x\mapsto\exp(tR)x$. We choose a smooth coordinate chart around $x_0$, which provides local coordinates in an open subset of $\mathbb R^J$. These vector fields, expressed in the chosen coordinates, satisfy the hypothesis of Lemma~4.2.8 of Ref.~\cite{Sontag1998control}, since coordinate changes preserve Lie brackets and map tangent spaces isomorphically. Lemma~4.2.8 then gives $J$ vector fields associated with controls $R_1,\ldots, R_J\in\Gamma$, with the durations chosen sufficiently small that the corresponding flows remain within the domain of this chart. By Definition~4.2.7 of the same reference, the endpoint map defined by their composed flow has rank $J$ at some nonnegative duration vector $\boldsymbol\tau_0$. This endpoint map is the coordinate representation of $\mathcal E$. Since the change of coordinates preserves rank, $\mathcal E$ also has full rank at $\boldsymbol\tau_0$.

The derivative of $\mathcal E$ depends continuously on the durations, so sufficiently small changes in $\boldsymbol\tau_0$ preserve full rank. We can therefore replace any zero entries by small positive values to obtain the required vector $\boldsymbol\tau_*$. \hfill$\square$

\begin{lemma}
\label{lem:smooth-endpoint-correction}
Consider a finite piecewise-constant schedule with strictly positive segment durations $\boldsymbol\tau_*$, as well as an endpoint map that has full rank at $\boldsymbol\tau_*$. For every sufficiently small smoothing width $\eta>0$, we can make the driven schedules smooth near each jump to obtain a $C^\infty$ schedule. We can then slightly adjust the segment durations to exactly recover the original endpoint of the trajectory. The adjusted durations remain positive, the endpoint map remains full rank, and the time-independent element $A$ is unchanged. We can also choose any endpoint near the original one, with the controls and durations depending smoothly on that choice. The same conclusions hold when we add finitely many endpoint coordinates, provided that their equations depend smoothly on the same control schedule and the augmented endpoint map has full rank at $\boldsymbol\tau_*$.

Furthermore, we can choose the driven schedules to take the same prescribed constant values near the initial and final times. If the system on $G$ is controllable, then for every target endpoint $g_f\in G$ we can choose a $C^\infty$ periodic schedule under which the trajectory starts at $(I,0)$ and ends at $g_f$, while $A$ remains unchanged.
\end{lemma}

\textit{Proof.}
We can vary the segment durations $\boldsymbol\tau$ near $\boldsymbol\tau_*$ while keeping the segment values of the original piecewise-constant schedule fixed. Now, we smooth each internal jump over an interval of width $\eta$ centered at the jump, changing the driven schedules from the preceding segment values to the following segment values. We also prescribe the same driven-schedule values at the beginning and the end of the schedule. During the first $\eta$ units of time, the driven schedules change smoothly from the prescribed values to the first segment values. During the final $\eta$ units, they change smoothly from the last segment values back to the prescribed values. For $0<\eta<\tfrac12\min_j\tau_{j,*}$ and $\boldsymbol\tau$ sufficiently close to $\boldsymbol\tau_*$, these intervals are disjoint. Outside these intervals, the driven schedules keep their original segment values. Only the driven schedules change, so $A$ is unchanged.

On each interval where the driven schedules change, we rescale time linearly to $s\in[0,1]$ and choose those schedules as $C^\infty$ functions of $s$, constant near both ends of $[0,1]$ and independent of $(\boldsymbol\tau,\eta)$. In local coordinates, the state $x$ then obeys
\begin{equation}
    \frac{\mathrm d x}{\mathrm d s}=\eta F(s,x),
\end{equation}
where $F$ is the smooth vector field determined by these schedules. The equation is smooth in $\eta$ on an open interval containing zero, and the smoothing construction uses only $\eta>0$. By Theorem~2.11 of Ref.~\cite{Teschl2012ODE}, the local solution maps, which are defined for the smoothing intervals, depend smoothly on the initial state and on $\eta$. Outside these intervals, the same theorem gives smooth dependence on the initial state and on the duration, which also depends smoothly on $(\boldsymbol\tau,\eta)$. Consequently, the endpoint map $\mathcal E_\eta(\boldsymbol\tau)$ depends smoothly on $(\boldsymbol\tau,\eta)$, since it is a finite composition of these local solution maps.

At $\eta=0$, the equation above leaves the state unchanged, and each interval with constant driven schedules has its original duration, so $\mathcal E_0=\mathcal E$. 

To adjust the endpoint while keeping $\eta$ fixed, we define
\begin{equation}
\widehat{\mathcal E}(\eta,\boldsymbol\tau)
=\bigl(\eta,\mathcal E_\eta(\boldsymbol\tau)\bigr).
\end{equation}
Since $D_{\boldsymbol\tau}\mathcal E_0(\boldsymbol\tau_*)$ has full rank and the first component of $\widehat{\mathcal E}(\eta,\boldsymbol\tau)$ is $\eta$, $D\widehat{\mathcal E}(0,\boldsymbol\tau_*)$ is surjective. By continuity, $D\widehat{\mathcal E}$ remains surjective near $(0,\boldsymbol\tau_*)$, so $\widehat{\mathcal E}$ is a submersion there. The local section theorem therefore gives a smooth local section through this point~\cite[Theorem~4.26]{Lee2012Smooth}. Because $\widehat{\mathcal E}$ leaves its first coordinate unchanged, the section has the form $(\eta,\boldsymbol\tau(\eta,g))$, with $\boldsymbol\tau(0,\mathcal E(\boldsymbol\tau_*))=\boldsymbol\tau_*$. It satisfies
\begin{equation}
\mathcal E_\eta\bigl(\boldsymbol\tau(\eta,g)\bigr)=g
\end{equation}
for $(\eta,g)$ near $(0,\mathcal E(\boldsymbol\tau_*))$. By taking $g=\mathcal E(\boldsymbol\tau_*)$, we recover the original endpoint exactly; by varying $g$, we obtain controls and durations that depend smoothly on $g$. The durations stay positive because they stay near $\boldsymbol\tau_*$. The derivative $D_{\boldsymbol\tau}\mathcal E_\eta$ is continuous in $(\boldsymbol\tau,\eta)$, so it remains full rank for $(\boldsymbol\tau,\eta)$ near $(\boldsymbol\tau_*,0)$. The same argument applies to the augmented endpoint map and proves the conclusions for the additional endpoint coordinates.

Assume now that the system on $G$ is controllable and fix $g_f\in G$. We apply Lemma~\ref{lem:full-rank-endpoint-map} to obtain a piecewise-constant schedule whose trajectory starts at $(I,0)$ and ends at $g_*=\mathcal E(\boldsymbol\tau_*)$. Its endpoint map has full rank at $\boldsymbol\tau_*$. By controllability, we choose a second piecewise-constant schedule with positive durations whose trajectory starts at $g_*$ and ends at $g_f$. We keep the controls and durations of the second schedule fixed. Starting from $(I,0)$, the trajectory of the second schedule ends at $h$. Since the dynamics are right-invariant, the trajectory starting at any $g$ ends at $hg$ under the same schedule. By appending the second schedule, we have the endpoint map
\begin{equation}
    \widetilde{\mathcal E}(\boldsymbol\tau)
    =h\,\mathcal E(\boldsymbol\tau),
    \qquad
    hg_*=g_f .
\end{equation}
Left multiplication by $h$ is a diffeomorphism, so $\widetilde{\mathcal E}$ has full rank at $\boldsymbol\tau_*$. We smooth the complete schedule while keeping the second schedule's durations fixed. Since $\widetilde{\mathcal E}$ has full rank with respect to $\boldsymbol\tau$, we can adjust the first schedule's durations slightly to restore $g_f$ exactly for every sufficiently small $\eta>0$. The adjusted durations remain positive, and $A$ remains unchanged. The driven schedules take the same prescribed constant values near the initial and final times, and their values and derivatives of all orders agree at these times. The schedule therefore has a $C^\infty$ periodic extension. \hfill$\square$

\subsection{Finite-Frequency and Waveform-Selective Extensions}
\label{app:waveform-finite-frequency-extension}

As we show above, the controllability condition on $G=K\rtimes V$ is sufficient for the static DCG conditions. In this subsection, we prove the sufficiency for waveform-selective DCGs with the same perturbation operators and the prescribed gate duration $T$. The result holds for both smooth $T$-periodic controls and piecewise-constant controls; the piecewise-constant case follows from the smooth construction, as we show in the footnote later in this subsection. In both cases, we allow a finite collection of selected smooth real functions $f_{p,\ell}(t)$ on $[0, T]$ in each direction, as in Eq.~\eqref{eq:waveform-selective-perturbation}. Finite-frequency robustness follows as a special case. Our proof explains the absence of extra conditions for finite-frequency DCGs in Secs.~\ref{sec:theoryB} and~\ref{sec:theoryC}. Throughout this subsection, $\staticcoe$ may be chosen separately for each protocol but is constant within it.

If we rescale a protocol in time, the robustness conditions for the selected functions may no longer hold. We therefore construct blocks whose trajectories return the gate coordinate to $I$ and have adjustable integrated responses, with the target block coming last. To satisfy the robustness conditions for the original selected functions, we vary the block responses and rescale the complete sequence to duration $T$.

For all included perturbation operators $X_p$, we define the instantaneous response vector
\begin{equation}
 r(t)=\left(\cdots,U_0^\dagger(t)X_pU_0(t),\cdots\right)\in V.
 \label{eqapp:finite-frequency-response}
\end{equation}
The vector space $V$ is the DCG response space for these operators and does not depend on the selected functions. The accumulated response corresponding to the static DCG conditions is
\begin{equation}
 v(T)=\int_0^T r(t)\,\mathrm dt.
 \label{eqapp:finite-frequency-zero-response}
\end{equation}
For sufficiency, we require $\int_0^T f_{p,\ell}(t)r(t)\,\mathrm dt=0$ for every $p,\ell$, together with $v(T)=0$. These conditions require every direction $X_p$ to satisfy the robustness conditions for all functions in $\operatorname{span}_{\mathbb R}\{1,f_{p,\ell}\}$. Here and below, this span includes the selected functions from every direction. The original requirements may specify different functions for different directions and need not include the static DCG condition for all directions. It is therefore enough to prove the existence of controls satisfying these stronger conditions.

Here, we choose $n_f$ linearly independent smooth real functions on $[0, T]$. They satisfy
\begin{equation}
\begin{aligned}
 \phi_0(t)&=1,\\
 \operatorname{span}_{\mathbb R}\{\phi_\alpha\}_{\alpha=0}^{n_f-1}
 &=\operatorname{span}_{\mathbb R}\{1,f_{p,\ell}\}.
\end{aligned}
 \label{eqapp:finite-frequency-weights}
\end{equation}
The vector of weight functions is $\Phi=(\phi_0,\ldots,\phi_{n_f-1})^{\mathsf T}$, and the weighted responses $M_\alpha\in V$ must vanish:
\begin{equation}
 M_\alpha=\int_0^T\phi_\alpha(t)r(t)\,\mathrm dt=0,
 \qquad \alpha=0,\ldots,n_f-1.
 \label{eqapp:finite-frequency-moments}
\end{equation}
By linearity in the weight function, these equations imply the response conditions for every selected $f_{p,\ell}$ and the zero-frequency condition. The Gram matrix of the weight functions is
\begin{equation}
 \mathsf G=\int_0^T\Phi(t)\Phi(t)^{\mathsf T}\,\mathrm dt
 \in\mathbb R^{n_f\times n_f}.
 \label{eqapp:finite-frequency-continuous-Gram}
\end{equation}
The chosen basis functions are continuous and linearly independent on $[0, T]$. Consequently, every nonzero real vector $c\in\mathbb R^{n_f}$ satisfies
\begin{equation*}
 c^{\mathsf T}\mathsf Gc
 =\int_0^T\left[\sum_{\alpha=0}^{n_f-1}c_\alpha\phi_\alpha(t)\right]^2\,\mathrm dt>0.
\end{equation*}
Therefore, $\mathsf G$ is positive definite.

\textit{Control blocks with adjustable responses.} With initial point at $(I,0)$, Lemma~\ref{lem:full-rank-endpoint-map} gives a piecewise-constant schedule whose endpoint map has full rank at strictly positive segment durations. We use these durations and denote the endpoint of the trajectory by $g_*\in G$. By controllability, we can append a fixed piecewise-constant return schedule, whose trajectory starts at $g_*$ and ends at $(I,0)$. Due to right invariance, the same schedule generates a trajectory from any $g\in G$ to $g_*^{-1}g$. Since left multiplication by $g_*^{-1}$ is a diffeomorphism, the endpoint map of the complete piecewise-constant schedule has full rank with respect to the first schedule's durations, while the return durations remain fixed.

We omit any zero-duration segments in the return schedule, so every segment of the complete piecewise-constant schedule has positive duration. We then apply Lemma~\ref{lem:smooth-endpoint-correction} to this schedule, taking only the durations of the first schedule as adjustable parameters. Since the endpoint map has full rank with respect to these durations, the lemma gives smooth schedules reaching every endpoint in a neighborhood of $(I,0)$ while the return durations remain fixed. For $z\in V$ near zero, we denote by $N(z)$ the schedule whose trajectory starts at $(I,0)$ and ends at $(I,z)$, and by $t_N(z)>0$ its duration. The gate coordinate returns to $I$, while the integrated DCG response equals $z$, and the trajectory returns to $(I,0)$ only when $z=0$.

For the target gate $\target\in K$, the controllable case of Lemma~\ref{lem:smooth-endpoint-correction} gives a smooth schedule whose trajectory starts at $(I,0)$ and ends at $(\target,0)$. We denote this schedule by $B_f$ and its duration by $t_f>0$. The control blocks have the endpoints and durations
\begin{equation}
\begin{aligned}
 N(z)&:(I,0)\longmapsto(I,z),& t_0&=t_N(0)>0,\\
 B_f&:(I,0)\longmapsto(\target,0),& t_f&>0.
\end{aligned}
 \label{eqapp:finite-frequency-neutral-block}
\end{equation}
All blocks use the same static coefficient $\staticcoe$. We can choose the driven schedules to take the same prescribed constant values near the initial and final times of each block. The duration $t_N(z)$ depends smoothly on $z$, as does the instantaneous response within $N(z)$ when time is divided by $t_N(z)$. If $(k,v)$ is the state generated within a block from $(I,0)$ and $v'\in V$ is the response accumulated before the block, the group multiplication gives
\begin{equation*}
    (k,v)(I,v')=(k,v+v').
\end{equation*}
Therefore, the response accumulated before a block does not change the gate evolution or the response generated within that block.

\textit{Block positions and the common time scale.} We first place $L$ copies of $N(0)$ before $B_f$, where $L$ is a positive integer that will be chosen sufficiently large. The total duration is initially $Lt_0+t_f$, and we rescale this sequence to duration $T$. For $k=1,\ldots,L$, the common factor $\lambda_L^0$, the width $h_L$ of each $N(0)$, and the midpoint $\theta_k^*$ of block $k$ are respectively
\begin{equation}
\begin{aligned}
 \lambda_L^0&=\frac{T}{Lt_0+t_f},& h_L&=\lambda_L^0t_0,\\
 \theta_k^*&=\left(k-\frac12\right)h_L,& k&=1,\ldots,L.
\end{aligned}
\label{eqapp:finite-frequency-reference-grid}
\end{equation}
We keep the times $\theta_k^*$ fixed and introduce the adjustable parameters $b=(b_0,\ldots,b_{n_f-1})\in V^{n_f}$. For $k=1,\ldots,L$, the assigned response $z_k(b)\in V$ is defined by
\begin{equation}
\begin{aligned}
 z_k(b)&=t_0\Phi(\theta_k^*)^{\mathsf T}b\\
 &=t_0\sum_{\beta=0}^{n_f-1}\phi_\beta(\theta_k^*)b_\beta.
\end{aligned}
 \label{eqapp:finite-frequency-assigned-response}
\end{equation}
The values $\phi_\beta(\theta_k^*)$ determine how each parameter $b_\beta$ changes the responses across the blocks. For $b$ near zero, we replace the $k$th copy of $N(0)$ by $N(z_k(b))$, for $k=1,\ldots,L$, and keep $B_f$ as the final block. Since every $N(z_k(b))$ has gate endpoint $I$, each following block starts with gate coordinate $I$, and the complete sequence still implements $\target$.

Since the durations $t_N(z_k(b))$ depend on $b$, we define the total unscaled duration $S_L(b)$ and the common rescaling factor $\lambda_L(b)$ for each $b$. After this rescaling, $a_k(b)$, $w_k(b)$, and $\theta_k(b)$ denote the start time, width, and midpoint of block $k$, respectively:
\begin{equation}
\begin{aligned}
 S_L(b)&=t_f+\sum_{k=1}^Lt_N(z_k(b)),& \lambda_L(b)&=\frac{T}{S_L(b)},\\
 a_k(b)&=\lambda_L(b)\sum_{j<k}t_N(z_j(b)),& w_k(b)&=\lambda_L(b)t_N(z_k(b)),\\
 \theta_k(b)&=a_k(b)+\frac12w_k(b).&&
\end{aligned}
 \label{eqapp:finite-frequency-exact-times}
\end{equation}
At $b=0$, $\lambda_L(0)=\lambda_L^0$, $w_k(0)=h_L$, and $\theta_k(0)=\theta_k^*$. For nonzero $b$, the block midpoints $\theta_k(b)$ may differ from the fixed times $\theta_k^*$ in Eq.~\eqref{eqapp:finite-frequency-assigned-response}. The exact response equations include the $b$-dependence of $a_k(b)$, $w_k(b)$, and $\theta_k(b)$.

We denote the Hamiltonian of the concatenated block sequence before the common rescaling by $H_{L,b}^{\mathrm{con}}$, defined on $[0, S_L(b)]$. The Hamiltonian on $[0,T]$ is
\begin{equation}
 H_{L,b}(t)=\lambda_L(b)^{-1}H_{L,b}^{\mathrm{con}}
 \left(\frac{t}{\lambda_L(b)}\right).
 \label{eqapp:finite-frequency-rescaled-Hamiltonian}
\end{equation}
This Hamiltonian implements the gate $\target$ in time $T$, with the same static coefficient $\staticcoe/\lambda_L(b)$ in every block. The coefficient may vary between protocols, as allowed in the fixed-$T$ setting. We assume no upper bound on Hamiltonian amplitudes; a larger $L$ can require larger amplitudes.

\textit{The exact weighted responses and their error bounds.} For $H_{L,b}(t)$, we denote the instantaneous DCG response by $r_{L,b}(t)$. For the selected weight functions, its weighted responses $\Psi_{L,\alpha}(b)$ and the map $\Psi_L$ are
\begin{equation}
\begin{aligned}
 \Psi_{L,\alpha}(b)&=\int_0^T\phi_\alpha(t)r_{L,b}(t)\,\mathrm dt,\\
 \Psi_L(b)&=(\Psi_{L,\alpha}(b))_{\alpha=0}^{n_f-1}\in V^{n_f}.
\end{aligned}
 \label{eqapp:finite-frequency-exact-moment-map}
\end{equation}
We seek $b$ with $\Psi_L(b)=0$.

We fix Euclidean norms on $V$ and $V^{n_f}$, and define $B_\rho=\{b\in V^{n_f}:\|b\|\leq\rho\}$ for $\rho>0$. We choose $\rho$ small enough that all $z_k(b)$ lie in a fixed closed ball contained in the domain of $N(z)$ whenever $b\in B_\rho$. The same $\rho$ works for every $L$ and $k$ because $t_0$ is fixed and $\Phi$ is bounded on $[0,T]$. On the closed ball containing all $z_k(b)$, $t_N(z)$ is bounded above and below by positive constants, and its first two derivatives with respect to $z$ are bounded. Thus, uniformly on $B_\rho$, $S_L(b)$ is bounded above and below by positive constants times $L$, and its first two $b$-derivatives are $O(L)$. Here $O_{C^2}$ means that the stated order bound holds for a function and its first two derivatives with respect to $b$, uniformly on $B_\rho$. Equation~\eqref{eqapp:finite-frequency-exact-times} then gives
\begin{equation}
\begin{aligned}
 \lambda_L(b)&=O_{C^2}(L^{-1}),& w_k(b)&=O_{C^2}(L^{-1}),\\
 a_k(b),\theta_k(b)&=O_{C^2}(1).&&
\end{aligned}
 \label{eqapp:finite-frequency-block-scaling}
\end{equation}
The constants in these estimates may depend on $T$, the selected weight functions, and the chosen schedules $N(z)$ and $B_f$, but not on $L$ or $k$.

Within $N(z)$, $x\in[0,1]$ denotes the elapsed time divided by $t_N(z)$, and $\widetilde r_N(x,z)$ denotes the instantaneous response in this variable. Since the integrated DCG response is $z$, we have
\begin{equation*}
 t_N(z)\int_0^1\widetilde r_N(x,z)\,\mathrm dx=z.
\end{equation*}
For block $N(z_k(b))$, $\mathcal M_{L,\alpha}^{(k)}(b)$ denotes its contribution to $\Psi_{L,\alpha}(b)$, and $\mathcal R_{L,\alpha}^{(k)}(b)$ denotes the error when we replace the weight by its midpoint value:
\begin{equation}
\begin{aligned}
 \mathcal M_{L,\alpha}^{(k)}(b)
 &=w_k(b)\int_0^1\phi_\alpha(a_k(b)+w_k(b)x)\\
 &\qquad {}\times \widetilde r_N(x,z_k(b))\,\mathrm dx,\\
 \mathcal M_{L,\alpha}^{(k)}(b)
 &=\lambda_L(b)\phi_\alpha(\theta_k(b))z_k(b)
 +\mathcal R_{L,\alpha}^{(k)}(b),\\
 \mathcal R_{L,\alpha}^{(k)}(b)
 &=w_k(b)\int_0^1\bigl[\phi_\alpha(a_k(b)+w_k(b)x)\\
 &\qquad-\phi_\alpha(a_k(b)+\tfrac12w_k(b))\bigr]\\
 &\qquad {}\times \widetilde r_N(x,z_k(b))\,\mathrm dx.
\end{aligned}
 \label{eqapp:finite-frequency-single-block-remainder}
\end{equation}
The second equality uses the integrated response of $N(z_k(b))$ and $w_k(b)=\lambda_L(b)t_N(z_k(b))$. We write the exact weight difference as
\begin{equation*}
\begin{aligned}
 &\phi_\alpha(a_k(b)+w_k(b)x)-\phi_\alpha(\theta_k(b))\\
 &\quad=w_k(b)\left(x-\frac12\right)\\
 &\qquad\times\int_0^1\phi_\alpha'\!\left(\theta_k(b)+s w_k(b)\left(x-\frac12\right)\right)\,\mathrm ds,
\end{aligned}
\end{equation*}
where $\phi_\alpha'$ is the time derivative of $\phi_\alpha$. The factor $w_k(b)$ and its first two $b$-derivatives are $O(L^{-1})$, while $\theta_k(b)$ and its first two derivatives are bounded. The fixed weight functions have bounded first three derivatives on $[0, T]$, so the integral above and its first two $b$-derivatives are bounded. Consequently, the weight difference is $O_{C^2}(L^{-1})$. The response $\widetilde r_N(x,z_k(b))$ and its first two $b$-derivatives are uniformly bounded on $[0,1]\times B_\rho$. The prefactor $w_k(b)=O_{C^2}(L^{-1})$ in Eq.~\eqref{eqapp:finite-frequency-single-block-remainder} then gives
$\mathcal R_{L,\alpha}^{(k)}(b)=O_{C^2}(L^{-2})$, with a bound independent of the block index $k$.

The target block $B_f$ has width $w_f(b)=\lambda_L(b)t_f=O_{C^2}(L^{-1})$ and midpoint $T-w_f(b)/2$. Its integrated DCG response is zero, so the constant weight $\phi_\alpha(T-w_f(b)/2)$ contributes zero. We subtract this constant inside the response integral for $B_f$, which gives an $O_{C^2}(L^{-2})$ bound by the same calculation. The total error $\mathcal R_L(b)\in V^{n_f}$ is the sum of the errors from the $L$ blocks $N(z_k(b))$ and the weighted response of $B_f$. With $\Phi(t)\otimes z=(\phi_\alpha(t)z)_{\alpha=0}^{n_f-1}$, the complete weighted response is
\begin{equation}
 \Psi_L(b)=\lambda_L(b)\sum_{k=1}^L
 \Phi(\theta_k(b))\otimes z_k(b)+\mathcal R_L(b).
 \label{eqapp:finite-frequency-moment-decomposition}
\end{equation}
The $L$ block errors are $O_{C^2}(L^{-2})$, as is the error from $B_f$, so
\begin{equation}
 \|\mathcal R_L\|_{C^2(B_\rho)}=O(L^{-1}).
 \label{eqapp:finite-frequency-remainder-bound}
\end{equation}
For $\phi_0(t)=1$, every weight difference vanishes, so the zero-frequency component of $\mathcal R_L(b)$ is exactly zero.

\textit{An exact solution of the response equations.} Although $z_k(0)=0$ holds for every $k$, the nonconstant weights can leave $\Psi_L(0)\neq0$, with norm $O(L^{-1})$ by Eq.~\eqref{eqapp:finite-frequency-remainder-bound}. We now adjust $b$ to obtain $\Psi_L(b)=0$.

We differentiate Eq.~\eqref{eqapp:finite-frequency-moment-decomposition} at $b=0$. In the explicit sum, every term containing $D\lambda_L(0)$ or $D\theta_k(0)$ also contains $z_k(0)$ and is zero. Only the terms from $D z_k(0)$ remain, in addition to $D\mathcal R_L(0)=O(L^{-1})$. We define the matrix $\mathsf G_L$ by the sum below and obtain
\begin{equation}
\begin{aligned}
 D\Psi_L(0)&=\mathsf G_L\otimes I_V+O(L^{-1}),\\
 \mathsf G_L&=h_L\sum_{k=1}^L
 \Phi(\theta_k^*)\Phi(\theta_k^*)^{\mathsf T}
 =\mathsf G+O(L^{-1}),
\end{aligned}
 \label{eqapp:finite-frequency-linearization}
\end{equation}
where $I_V$ is the identity operator on $V$. The midpoint sum for $\mathsf G_L$ approximates the integral on $[0,Lh_L]$ with error $O(L^{-2})$. At $b=0$, $B_f$ occupies $[Lh_L,T]$, whose length is $\lambda_L^0t_f=O(L^{-1})$; the integral of $\Phi(t)\Phi(t)^{\mathsf T}$ over this interval is therefore $O(L^{-1})$. These bounds give $\mathsf G_L=\mathsf G+O(L^{-1})$.

Since $\mathsf G$ is positive definite, $D\Psi_L(0)$ is invertible for sufficiently large $L$, with a uniformly bounded inverse. In Eq.~\eqref{eqapp:finite-frequency-moment-decomposition}, $\lambda_L(b)$ multiplies a sum of $L$ terms with uniformly bounded $C^2$ norms. The bound $\lambda_L(b)=O_{C^2}(L^{-1})$ and the estimate for $\mathcal R_L(b)$ therefore give a uniform bound on $D^2\Psi_L(b)$. Thus there is a constant $C>0$, independent of $L$, such that
\begin{equation}
\begin{aligned}
 \|\Psi_L(0)\|&\leq\frac{C}{L},&
 \|D\Psi_L(0)^{-1}\|&\leq C,\\
 \sup_{b\in B_\rho}\|D^2\Psi_L(b)\|&\leq C.
\end{aligned}
 \label{eqapp:finite-frequency-uniform-bound}
\end{equation}
We define $A_L=D\Psi_L(0)$ and $\mathcal T_L(b)=b-A_L^{-1}\Psi_L(b)$. A fixed point of $\mathcal T_L$ solves the exact equation $\Psi_L(b)=0$. Since $D\mathcal T_L(b)=A_L^{-1}[A_L-D\Psi_L(b)]$, Eq.~\eqref{eqapp:finite-frequency-uniform-bound} gives
\begin{equation*}
 \|\mathcal T_L(0)\|\leq\frac{C^2}{L},
 \qquad
 \|D\mathcal T_L(b)\|\leq C^2\|b\|.
\end{equation*}
We choose a constant $R>2C^2$, then take $L$ sufficiently large that $R/L<\rho$ and $C^2R/L\leq1/2$. The closed ball $\|b\|\leq R/L$ lies in $B_\rho$, and the derivative bound makes $\mathcal T_L$ a contraction there. Moreover,
\begin{equation*}
 \|\mathcal T_L(b)\|
 \leq \frac{C^2}{L}+\frac12\|b\|
 <\frac{R}{L},
\end{equation*}
so $\mathcal T_L$ maps this closed ball into itself. The contraction theorem~\cite[Theorem~2.1]{Teschl2012ODE} gives a fixed point $b^{(L)}$ in this ball, with
\begin{equation}
\begin{aligned}
 \Psi_L(b^{(L)})&=0,& \|b^{(L)}\|&=O(L^{-1}),\\
 \max_k\|z_k(b^{(L)})\|&=O(L^{-1}).&&
\end{aligned}
 \label{eqapp:finite-frequency-exact-solution}
\end{equation}
Thus every chosen response $z_k(b^{(L)})$ lies in the domain of $N(z)$. It suffices to fix one sufficiently large $L$.

For $\phi_0(t)=1$, the solution $b^{(L)}$ satisfies
\begin{equation*}
 \Psi_{L,0}(b^{(L)})
 =\lambda_L(b^{(L)})\sum_{k=1}^L z_k(b^{(L)})=0.
\end{equation*}
The trajectory of each $N(z_k(b^{(L)}))$ ends with gate coordinate $I$, and $B_f$ has zero integrated DCG response. The zero-frequency equation therefore gives the full endpoint $(\target,0)$. The other components of $\Psi_L(b^{(L)})=0$ impose the remaining weighted-response conditions, so $H_{L,b^{(L)}}(t)$ satisfies
\begin{equation}
 U_0(T)=\target,\qquad M_\alpha=0,
 \quad \alpha=0,\ldots,n_f-1.
 \label{eqapp:finite-frequency-final-conditions}
\end{equation}
In particular, linearity in the weight function gives $\mathcal A_p^{(\mathrm D)}[f_{p,\ell}]=0$ for every $p$ and $\ell$.

The driven schedules have the same prescribed constant values near the initial and final times of every block. After the common rescaling, these constants all acquire the same factor $\lambda_L(b^{(L)})^{-1}$. The values and all derivatives therefore agree between the end of each block and the start of the next, and also between $t=T$ and $t=0$. Since every block is smooth, the driven schedules extend to $C^\infty$ $T$-periodic functions.

If the joint zero-frequency static DCG--EGG system is controllable, its endpoint records the gate, the DCG response, and the EGG response, in that order; $z\in V$ still denotes only the DCG response. We apply the two lemmas to this system and obtain smooth schedules whose trajectories end at $(I,z,0)$ for the adjustable blocks and at $(\target,0,0)$ for the target block. The adjustable blocks have smooth $z$-dependent durations and instantaneous DCG responses, with integrated DCG response $z$; the target block has zero integrated DCG response. The trajectories of the adjustable blocks end with gate coordinate $I$, and every block has zero integrated EGG response, so the total EGG response is zero. Under the common time rescaling, the factor in $\mathrm dt$ cancels its reciprocal in $H$ and leaves the EGG response unchanged. We apply the weighted-response estimates and contraction argument to the DCG coordinates of these schedules, which gives the robustness conditions for all selected functions without changing their zero EGG response.\footnote{For piecewise-constant controls at fixed $T$, with or without the EGG condition, the invertible response derivative, together with variations of the endpoint of $B_f$, gives a full-rank endpoint map. Approximating this smooth family by piecewise-constant controls on fixed time partitions gives convergence of the endpoint map and its first parameter derivatives. Local endpoint correction then restores the gate and all required zero responses exactly.}

\textit{Static and finite-frequency perturbations.} By selecting the constant function $f(t)=1$, we obtain the static DCG conditions. For finite-frequency robustness, the distinct positive frequencies in the finite sets $\Omega_p$ form the set $\Omega^+=\{\omega_1,\ldots,\omega_Q\}$. We select the cosine and sine functions for these frequencies, together with the constant function. With $n_f=1+2Q$, we can choose the basis
\begin{equation*}
\begin{aligned}
 \phi_0(t)&=1,\\
 \phi_{2q-1}(t)&=\cos(\omega_qt),\\
 \phi_{2q}(t)&=\sin(\omega_qt),\qquad q=1,\ldots,Q.
\end{aligned}
\end{equation*}
The functions $1$, $\cos(\omega_qt)$, and $\sin(\omega_qt)$ are linearly independent on every interval of nonzero length, no matter whether the frequencies are commensurate. For each positive frequency, the cosine and sine equations are equivalent to the complex exponential condition because the response operators are Hermitian. The constructed controls therefore satisfy the finite-frequency DCG conditions for every original set $\Omega_p$, whether or not it contains zero. These sets may differ between perturbation operators.

\subsection{Criteria under Full Unitary Controllability}
\label{app:full-su-regime}

We prove the Hamiltonian criteria in Sec.~\ref{sec:theoryB}. The proofs require a compact simple Hamiltonian Lie algebra, and $\mathfrak{su}(\dimension)$ is a typical example. Specifically, we use the Hamiltonian algebra $\mathfrak{k}$ defined in the main text,
\begin{equation}
\mathfrak{k}=\operatorname{Lie}\bigl(\{i\staticterm\}\cup
\{i\scheduleterm\mid j=1,\ldots,m\}\bigr),
\end{equation}
and assume that $\mathfrak{k}$ is simple and $K$ is compact. Here, we only consider target gates in $K$. For this Hamiltonian algebra, we define the common symmetry algebra of the driven Hamiltonian operators by \begin{equation*} 
\mathcal C_{\mathrm{dyn}}= \{Y\in\mathfrak{k}\mid [Y,\scheduleterm]=0,\ j=1,\ldots,m\}. 
\end{equation*} 
For $\mathfrak{k}=\mathfrak{su}(\dimension)$, this definition agrees with Eq.~\eqref{eq:C_dyn}. We use the response spaces and control systems defined in Sec.~\ref{sec:theoryC}. Their coordinates are written as $(a,v_{\mathrm{DCG}})$, $(a,v_{\mathrm{EGG}})$, and $(a,v_{\mathrm{DCG}},v_{\mathrm{EGG}})$ for the static DCG, EGG, and joint robustness conditions, respectively. In these coordinates, $a\in\mathfrak{k}$ is anti-Hermitian, while the response coordinates consist of Hermitian operators or tuples of Hermitian operators.

For each zero-frequency construction, we use its response space $V$, control system $\Gamma$, and enlarged algebra $\mathfrak{g}=\mathfrak{k}\rtimes V$ from Sec.~\ref{sec:theoryC}. Since $K$ is compact, the remaining sufficient conditions in Eq.~\eqref{eq:general-criteria} are
\begin{equation*}
V^K=\{0\},\qquad \operatorname{Lie}(\Gamma)=\mathfrak{g}.
\end{equation*}
The controllability theorem then gives a trajectory from $(I,0)$ to $(\target,0)$ for every $\target\in K$: the gate component reaches the target, and the zero response enforces the robustness conditions. Necessity, when claimed, requires separate arguments and is proved below.

For $Y,Z,C\in\mathfrak{k}$, we use the invariant inner product
\begin{equation}
\begin{aligned}
\langle Y,Z\rangle&=-\operatorname{tr}(YZ),\\
\|Y\|^2&=\langle Y,Y\rangle,\\
\langle[C,Y],Y\rangle&=0.
\end{aligned}
\label{eqapp:full-su-inner-product}
\end{equation}
In this equation, the norm is the Frobenius norm. Since $\mathfrak{k}$ is simple, its center is zero and its only ideals are $\{0\}$ and $\mathfrak{k}$. Since $K$ is connected, a real subspace of $\mathfrak{k}$ is invariant under conjugation if and only if it is an ideal. We also use the standard fact that every derivation of the simple Lie algebra $\mathfrak{k}$ is inner~\cite[Proposition~6.8]{Kirillov2008Lie}: if a real-linear map $f:\mathfrak{k}\to\mathfrak{k}$ satisfies
\begin{equation}
f([Y,Z])=[f(Y),Z]+[Y,f(Z)]
\end{equation}
for all $Y,Z\in\mathfrak{k}$, then there exists $C\in\mathfrak{k}$ such that
\begin{equation}
f(Y)=[C,Y]
\end{equation}
for every $Y\in\mathfrak{k}$.

For each zero-frequency sufficiency proof below, we fix an arbitrary finite $\staticcoe\neq0$ and verify Eq.~\eqref{eq:general-criteria} for this same coefficient. The controllability theorem then gives a piecewise-constant schedule for every target in $K$ at some finite duration, and Sec.~\ref{app:control-regularity} replaces it with a smooth periodic schedule without changing the endpoint or $\staticcoe$. The zero-frequency time rescaling established above gives the corresponding result at any prescribed $T>0$, while allowing the finite static coefficient to vary between different protocols. No upper bound on the Hamiltonian amplitudes is assumed. The necessity arguments in this subsection also apply when $\staticcoe=0$.

\textit{The single static DCG condition: $X=\staticterm$.} We first prove that this condition can be satisfied without any additional condition on the Hamiltonian control channels. In Eq.~\eqref{eq:smeidirect-DCG}, we take $r=1$ and $X_1=\staticterm$. If $\staticterm=0$, the response vanishes identically, giving $V_{\mathrm{DCG}}=\{0\}$, and the driven generators generate $\mathfrak{k}$. The compact-group controllability criterion then guarantees the existence of every robust target. Here, we only consider $\staticterm\neq0$. Since $\staticterm\neq0$ and $\mathfrak{k}$ is simple, we have 
\begin{equation}
    \operatorname{span}_{\mathbb R}\{k^{-1}(-i\staticterm)k\mid k\in K\}=\mathfrak{k},
\end{equation}
hence $-iV_{\mathrm{DCG}}=\mathfrak{k}$.

For this single static DCG condition, we must verify $V_{\mathrm{DCG}}^K=\{0\}$ and $\operatorname{Lie}(\Gamma_{\mathrm{DCG}})=\mathfrak{g}_{\mathrm{DCG}}$. If $v_0\in V_{\mathrm{DCG}}$ is fixed by every conjugation in $K$, then $-iv_0\in\mathfrak{k}$ commutes with every element of $\mathfrak{k}$. The zero center therefore gives $v_0=0$, which proves the condition $V_{\mathrm{DCG}}^K=\{0\}$.

To prove the Lie-algebra equality, we use the static and driven control elements specified by Eq.~\eqref{eq:smeidirect-DCG} with $r=1$ and $X_1=\staticterm$:
\begin{equation*}
\begin{aligned}
A_{\mathrm{DCG}}&=(-i\staticcoe\staticterm,\staticterm),\\
B_{\mathrm{DCG},j}&=(-i\scheduleterm,0),\qquad j=1,\ldots,m.
\end{aligned}
\end{equation*}
The element $A_{\mathrm{DCG}}$ belongs to $\Gamma_{\mathrm{DCG}}$, while each $B_{\mathrm{DCG},j}$ is a difference of two elements of $\Gamma_{\mathrm{DCG}}$; together they generate $\operatorname{Lie}(\Gamma_{\mathrm{DCG}})$. Since $\staticcoe\neq0$, their Hamiltonian components generate $\mathfrak{k}$; every $a\in\mathfrak{k}$ therefore occurs in some $(a,v)\in\operatorname{Lie}(\Gamma_{\mathrm{DCG}})$. We first obtain one nonzero element $(0,v)$, then use simplicity to generate the whole response space with zero Hamiltonian component.

Suppose no such nonzero $(0,v)$ exists. If $(a,v)$ and $(a,w)$ belong to $\operatorname{Lie}(\Gamma_{\mathrm{DCG}})$, their difference $(0,v-w)$ forces $v=w$. Each $a\in\mathfrak{k}$ therefore determines a unique response $v$, and we use its anti-Hermitian form to define $f_0:\mathfrak{k}\to\mathfrak{k}$ by $f_0(a)=-iv$. Linear combinations of the pairs $(a,v)$ show that $f_0$ is real-linear, and their Lie bracket gives
\begin{equation*}
f_0([a,b])=[f_0(a),b]+[a,f_0(b)],\qquad a,b\in\mathfrak{k}.
\end{equation*}
Thus $f_0$ is a derivation, so $f_0(a)=[C_0,a]$ for some $C_0\in\mathfrak{k}$. At $A_{\mathrm{DCG}}$, this identity implies
\begin{equation}
-i\staticterm=[C_0,-i\staticcoe\staticterm],
\end{equation}
and Eq.~\eqref{eqapp:full-su-inner-product} gives
\begin{equation*}
\|-i\staticterm\|^2
=\staticcoe\langle[C_0,-i\staticterm],-i\staticterm\rangle=0.
\end{equation*}
This contradicts $\staticterm\neq0$, so a nonzero $(0,v)$ must belong to $\operatorname{Lie}(\Gamma_{\mathrm{DCG}})$.

For any $a\in\mathfrak{k}$, an element $(a,w)\in\operatorname{Lie}(\Gamma_{\mathrm{DCG}})$ satisfies
\begin{equation*}
[(a,w),(0,v)]=(0,[a,v]).
\end{equation*}
Consequently, the real subspace
\begin{equation*}
\{-iv\mid(0,v)\in\operatorname{Lie}(\Gamma_{\mathrm{DCG}})\}
\end{equation*}
is a nonzero ideal of $\mathfrak{k}$ and therefore equals $\mathfrak{k}$. For any $(a,v)\in\operatorname{Lie}(\Gamma_{\mathrm{DCG}})$, we can therefore subtract $(0,v)\in\operatorname{Lie}(\Gamma_{\mathrm{DCG}})$ to obtain $(a,0)$. Hence
\begin{equation}
\operatorname{Lie}(\Gamma_{\mathrm{DCG}})
=\mathfrak{k}\rtimes V_{\mathrm{DCG}}
=\mathfrak{g}_{\mathrm{DCG}}.
\end{equation}
Thus the single condition $X=\staticterm$ admits every target with the fixed-$\staticcoe$ and fixed-$T$ control settings above. \hfill$\square$

\textit{All-static DCG conditions.} We prove that dynamically corrected gates against all static Hamiltonian perturbations $X$ with $iX\in\mathfrak{k}$ exist for every target gate $\target\in K$ with a given finite period $T>0$, if and only if there is no common symmetry of the driven Hamiltonian operators $\scheduleterm$, i.e., $\mathcal C_{\mathrm{dyn}}=\{0\}$.

To prove sufficiency, we assume $\mathcal C_{\mathrm{dyn}}=\{0\}$. We take $r=\dim\mathfrak{k}$ and choose $\{iX_p\}_{p=1}^r$ as a complete basis of $\mathfrak{k}$, with all elements linearly independent. For any fixed $\staticcoe\neq0$, Eq.~\eqref{eq:smeidirect-DCG} gives the control elements
\begin{equation}
\begin{aligned}
A_{\mathrm{DCG}}&=(-i\staticcoe\staticterm,(X_1,\ldots,X_r)),\\
B_{\mathrm{DCG},j}&=(-i\scheduleterm,0),\quad j=1,\ldots,m,
\end{aligned}
\label{eqapp:full-su-DCG-generators}
\end{equation}
where $0$ is the zero response tuple. These elements generate $\operatorname{Lie}(\Gamma_{\mathrm{DCG}})$, whose Hamiltonian component ranges over $\mathfrak{k}$ by the definition of $\mathfrak{k}$. The proof below also covers $\staticterm=0$, since the definition of $\mathfrak{k}$ then gives $\operatorname{Lie}(\{i\scheduleterm\mid j=1,\ldots,m\})=\mathfrak{k}$.

Suppose these elements do not generate every tuple $(a,v_1,\ldots,v_r)$ with $a,iv_1,\ldots,iv_r\in\mathfrak{k}$. Then some response components must depend linearly on $a$ and the other response components. We choose the largest possible number $s$, with $0\leq s<r$, of coordinates $iv_p$ that vary independently over $\mathfrak{k}$ together with $a$. We put the corresponding $X_p$ first and write these coordinates as $\varphi_\mu=iv_\mu$, $\mu=1,\ldots,s$. When $a$ and all $\varphi_\mu$ vanish, brackets show that the possible values of any remaining coordinate form an ideal of $\mathfrak{k}$. Simplicity and the choice of $s$ force this ideal to be zero. Thus, for $q=s+1,\ldots,r$, we write
\begin{equation*}
\begin{aligned}
iv_q&=\psi_q(\varphi_1,\ldots,\varphi_s,a)\\
&=\sum_{\mu=1}^s\psi_q^{(\mu)}(\varphi_\mu)+f_q(a),
\end{aligned}
\end{equation*}
where $\psi_q^{(\mu)}$ and $f_q$ are real-linear maps from $\mathfrak{k}$ to $\mathfrak{k}$.

Taking brackets and using the independent choices of $a$ and the $\varphi_\mu$, we obtain
\begin{equation*}
\begin{aligned}
\psi_q^{(\mu)}([a,y])&=[a,\psi_q^{(\mu)}(y)],\\
f_q([a,b])&=[f_q(a),b]+[a,f_q(b)],
\end{aligned}
\end{equation*}
for $a,b,y\in\mathfrak{k}$. Schur's lemma on the simple complexification of $\mathfrak{k}$ gives $\psi_q^{(\mu)}(y)=\lambda_q^{(\mu)}y$, with $\lambda_q^{(\mu)}\in\mathbb R$~\cite[Lemma~4.23]{Kirillov2008Lie}. The second identity makes $f_q$ a derivation, so $f_q(a)=[C_q,a]$ for some $C_q\in\mathfrak{k}$. Hence every generated element satisfies
\begin{equation}
iv_q-\sum_{\mu=1}^s\lambda_q^{(\mu)}\varphi_\mu=[C_q,a].
\label{eqapp:full-su-dependent-components}
\end{equation}

At the driven and static control elements in Eq.~\eqref{eqapp:full-su-DCG-generators}, this relation gives
\begin{equation}
\begin{aligned}
\relax[C_q,-i\scheduleterm]&=0,\qquad j=1,\ldots,m,\\
iX_q-\sum_{\mu=1}^s\lambda_q^{(\mu)}iX_\mu
&=[C_q,-i\staticcoe\staticterm].
\end{aligned}
\label{eqapp:full-su-DCG-relations}
\end{equation}
The driven relations imply $C_q\in\mathcal C_{\mathrm{dyn}}$, so $C_q=0$. The static relation then contradicts the linear independence of $\{iX_p\}_{p=1}^r$. Therefore, these control elements generate every tuple, which gives
\begin{equation}
\begin{aligned}
iV_{\mathrm{DCG}}&=\mathfrak{k}^r,\\
\operatorname{Lie}(\Gamma_{\mathrm{DCG}})&=\mathfrak{g}_{\mathrm{DCG}}.
\end{aligned}
\end{equation}
A tuple invariant under every conjugation in $K$ has each $iv_p$ in the center of $\mathfrak{k}$. The zero center therefore gives $V_{\mathrm{DCG}}^K=\{0\}$. We therefore prove the sufficiency of $\mathcal C_{\mathrm{dyn}}=\{0\}$ for the all-static DCG condition.

For necessity, we first show how a nonzero common symmetry constrains the generated algebra, then obtain a constraint on the target gate. Suppose $\mathcal C_{\mathrm{dyn}}$ has dimension $s_{\mathrm C}>0$, with basis $Z_1,\ldots,Z_{s_{\mathrm C}}$. For $\staticcoe\neq0$, the map $Z\mapsto[Z,-i\staticcoe\staticterm]$ is injective on this space: a vector in its null space commutes with every available Hamiltonian and hence lies in the zero center of $\mathfrak{k}$. We define real coefficients $\beta_{\ell p}$ by
\begin{equation*}
[Z_\ell,-i\staticcoe\staticterm]=\sum_{p=1}^r\beta_{\ell p}iX_p,
\qquad \ell=1,\ldots,s_{\mathrm C}.
\end{equation*}
Injectivity and independence of the $Z_\ell$ imply that the coefficient rows $(\beta_{\ell p})_{p=1}^r$ are linearly independent.

The control elements $A_{\mathrm{DCG}}$ and $B_{\mathrm{DCG},j}$ satisfy
\begin{equation*}
\sum_{p=1}^r\beta_{\ell p}iv_p=[Z_\ell,a],\qquad \ell=1,\ldots,s_{\mathrm C}.
\end{equation*}
Linearity and the Jacobi identity preserve these relations under all real linear combinations and Lie brackets. We complete the coefficient rows to an invertible real change of response coordinates; its first $s_{\mathrm C}$ responses are fixed by $a$ through the displayed relations. Thus $\operatorname{Lie}(\Gamma_{\mathrm{DCG}})$ is a proper subalgebra of $\mathfrak{g}_{\mathrm{DCG}}$.

To show that some target gates are excluded, we choose $0\neq C\in\mathcal C_{\mathrm{dyn}}$ and set $Y_C(t)=U_0^\dagger(t)CU_0(t)$. For a target-reaching protocol, $Y_C(0)=C$ and $Y_C(T)=\target^\dagger C\target$. The Schr\"odinger equation and $[C,\scheduleterm]=0$ give
\begin{equation}
\dot Y_C(t)=\staticcoe U_0^\dagger(t)[C,-i\staticterm]U_0(t).
\end{equation}
Since $[C,-i\staticterm]\in\operatorname{span}_{\mathbb R}\{iX_1,\ldots,iX_r\}$, zero DCG responses imply
\begin{equation*}
\int_0^T U_0^\dagger(t)[C,-i\staticterm]U_0(t)\,\mathrm dt=0.
\end{equation*}
The target must therefore satisfy
\begin{equation}
\target^\dagger C\target-C=\int_0^T\dot Y_C(t)\,\mathrm dt=0.
\end{equation}
This identity also holds when $\staticcoe=0$, without division by that coefficient. Some targets in $K$ do not commute with $C$, since an element fixed by all of $K$ would be central in $\mathfrak{k}$. These targets cannot satisfy the static DCG conditions at any duration; individual targets commuting with $C$ are not ruled out by this argument.

Thus $\mathcal C_{\mathrm{dyn}}=\{0\}$ is necessary and sufficient for an all-static DCG family covering every target at every prescribed $T>0$. By linearity of the response integrals, controls suppressing the full basis also suppress every smaller perturbation space within its real span. If $\operatorname{Lie}(\{i\scheduleterm\mid j=1,\ldots,m\})=\mathfrak{k}$, every common symmetry in $\mathfrak{k}$ is central and therefore zero; the all-static DCG criterion is then automatically satisfied. \hfill$\square$

\textit{EGG and joint static DCG--EGG conditions.} Under the assumptions of this subsection that $\mathfrak{k}$ is simple and $K$ is compact, we prove that any static DCG construction satisfying the general criteria in Eq.~\eqref{eq:general-criteria} also allows us to impose the EGG condition simultaneously with the static DCG conditions. For the same fixed $\staticcoe$, we assume that
\begin{equation*}
V_{\mathrm{DCG}}^K=\{0\},\quad
\operatorname{Lie}(\Gamma_{\mathrm{DCG}})=\mathfrak{g}_{\mathrm{DCG}},
\end{equation*}
and seek to prove
\begin{equation*}
V_{\mathrm{joint}}^K=\{0\},\quad
\operatorname{Lie}(\Gamma_{\mathrm{joint}})=\mathfrak{g}_{\mathrm{joint}}.
\end{equation*}
Here $\{iX_p\}$ need not be a basis of $\mathfrak{k}$, and the individual $iX_p$ need not belong to $\mathfrak{k}$; under the standing assumptions on $\mathfrak{k}$ and $K$, we require only that the given static DCG construction satisfy the general criteria in Eq.~\eqref{eq:general-criteria}. We use $-iV_{\mathrm{EGG}}=\mathfrak{k}$ from the definition in Sec.~\ref{sec:theoryC}. In the all-static DCG setting corresponding to Sec.~\ref{sec:theoryB}, full unitary controllability gives $\mathfrak{k}=\mathfrak{su}(\dimension)$, and $\mathcal C_{\mathrm{dyn}}=\{0\}$ guarantees that the static DCG construction satisfies these general criteria. 

To generate the joint algebra, we use the control elements from Eq.~\eqref{eq:semidirect-DCG+EGG},
\begin{equation*}
\begin{aligned}
A_{\mathrm{joint}}&=(-i\staticcoe\staticterm,(X_1,\ldots,X_r),\staticcoe\staticterm),\\
B_{\mathrm{joint},j}&=(-i\scheduleterm,0,\scheduleterm),\qquad j=1,\ldots,m.
\end{aligned}
\end{equation*}
To cancel the DCG component of $A_{\mathrm{joint}}$, we first construct $(0,(X_1,\ldots,X_r))$ from $A_{\mathrm{DCG}}$ and $B_{\mathrm{DCG},j}$ by finite applications of Lie brackets and real linear combinations; this is possible because $\operatorname{Lie}(\Gamma_{\mathrm{DCG}})=\mathfrak{g}_{\mathrm{DCG}}$. We perform the same operations on $A_{\mathrm{joint}}$ and $B_{\mathrm{joint},j}$. The Hamiltonian and DCG components obey the same bracket relations and have the same input values, so the operations produce
\begin{equation}
(0,(X_1,\ldots,X_r),\zeta)\in\operatorname{Lie}(\Gamma_{\mathrm{joint}}),
\end{equation}
where $\zeta\in V_{\mathrm{EGG}}$ is the EGG response produced by these operations. We subtract the constructed element from $A_{\mathrm{joint}}$, without imposing any condition on $\zeta$:
\begin{equation}
\begin{aligned}
&A_{\mathrm{joint}}-(0,(X_1,\ldots,X_r),\zeta)\\
&\qquad=(-i\staticcoe\staticterm,0,\staticcoe\staticterm-\zeta)
\in\operatorname{Lie}(\Gamma_{\mathrm{joint}}).
\end{aligned}
\end{equation}
This element and the $B_{\mathrm{joint},j}$ have zero DCG components, and their Hamiltonian components generate $\mathfrak{k}$. Thus for every $a\in\mathfrak{k}$, at least one $(a,0,v_{\mathrm{EGG}})$ belongs to $\operatorname{Lie}(\Gamma_{\mathrm{joint}})$.

Suppose $\operatorname{Lie}(\Gamma_{\mathrm{joint}})$ contained no nonzero $(0,0,v_{\mathrm{EGG}})$. Then the difference of $(a,0,v_{\mathrm{EGG}})$ and $(a,0,w_{\mathrm{EGG}})$ would force $v_{\mathrm{EGG}}=w_{\mathrm{EGG}}$, so every $a$ determines a unique EGG response. This defines a mapping $f_{\mathrm E}:\mathfrak{k}\to\mathfrak{k}$ by $f_{\mathrm E}(a)=-iv_{\mathrm{EGG}}$; linear combinations of the zero-DCG elements make this map real-linear. The bracket of two zero-DCG elements gives
\begin{equation*}
f_{\mathrm E}([a,b])=[f_{\mathrm E}(a),b]+[a,f_{\mathrm E}(b)].
\end{equation*}
Since $\mathfrak{k}$ is simple, this derivation is always inner, so we have $f_{\mathrm E}(a)=[C_{\mathrm E}, a]$ for some $C_{\mathrm E}\in\mathfrak{k}$. At the driven elements $B_{\mathrm{joint},j}$, this forces
\begin{equation}
-i\scheduleterm=[C_{\mathrm E},-i\scheduleterm],\qquad j=1,\ldots,m,
\end{equation}
and consequently,
\begin{equation*}
\|-i\scheduleterm\|^2=\langle[C_{\mathrm E},-i\scheduleterm],-i\scheduleterm\rangle=0.
\end{equation*}
This means all driven Hamiltonians would vanish, which is impossible for a simple algebra $\mathfrak{k}$. Therefore a nonzero $(0,0,v_{\mathrm{EGG}})$ exists.

For any element $(a,v_{\mathrm{DCG}},w_{\mathrm{EGG}})$ of the joint algebra,
\begin{equation*}
[(a,v_{\mathrm{DCG}},w_{\mathrm{EGG}}),(0,0,v_{\mathrm{EGG}})]
=(0,0,[a,v_{\mathrm{EGG}}]).
\end{equation*}
The responses $-iv_{\mathrm{EGG}}$ obtainable with both other components zero form a nonzero ideal, so simplicity gives every $(0,0,v_{\mathrm{EGG}})$ with $v_{\mathrm{EGG}}\in V_{\mathrm{EGG}}$. These pure EGG elements allow us to remove the EGG component, producing all $(a,v_{\mathrm{DCG}},0)$. Thus,
\begin{equation}
\operatorname{Lie}(\Gamma_{\mathrm{joint}})
=\mathfrak{k}\rtimes(V_{\mathrm{DCG}}\oplus V_{\mathrm{EGG}})
=\mathfrak{g}_{\mathrm{joint}}.
\end{equation}
For the fixed-vector condition, $-iV_{\mathrm{EGG}}=\mathfrak{k}$ and the zero center give $V_{\mathrm{EGG}}^K=\{0\}$. Therefore
\begin{equation}
V_{\mathrm{joint}}^K=V_{\mathrm{DCG}}^K\oplus V_{\mathrm{EGG}}^K=\{0\}.
\end{equation}
The general criterion therefore guarantees simultaneous static DCG and EGG robustness for every target $\target\in K$ when $\mathfrak{k}$ is simple. By taking $V_{\mathrm{DCG}}=\{0\}$, we have the EGG-only case, proving the existence of an EGG for every target $\target\in K$ when $\mathfrak{k}$ is simple. \hfill$\square$

\textit{Finite-frequency and waveform-selective criteria.} We first present the finite-frequency case, for which we also establish a necessity result below. We then state the more general waveform-selective sufficiency result. For prescribed finite sets $\Omega_p\subset[0,\infty)$, we seek a smooth $T$-periodic or piecewise-constant protocol satisfying
\begin{equation*}
\begin{aligned}
U_0(T)&=\target,\\
\mathcal A_p^{(\mathrm D)}(\omega)
&=\frac1T\int_0^T U_0^\dagger(t)X_pU_0(t)e^{i\omega t}\,\mathrm dt=0,
\end{aligned}
\end{equation*}
for every $p=1,\ldots,r$ and $\omega\in\Omega_p$. We use the same $X_p$ to define $V_{\mathrm{DCG}}$ and $\Gamma_{\mathrm{DCG}}$ as in Eq.~\eqref{eq:smeidirect-DCG}. If
\begin{equation*}
V_{\mathrm{DCG}}^K=\{0\},\qquad
\operatorname{Lie}(\Gamma_{\mathrm{DCG}})=\mathfrak{g}_{\mathrm{DCG}},
\end{equation*}
then Eq.~\eqref{eq:general-criteria} makes this zero-frequency system controllable on $K\rtimes V_{\mathrm{DCG}}$. Under these conditions, Sec.~\ref{app:waveform-finite-frequency-extension} proves the existence of the required finite-frequency DCG for every $\target\in K$, every prescribed finite collection $\{\Omega_p\}_{p=1}^r$, and every prescribed $T>0$. The frequency sets may differ between operators, may contain zero, and need not be commensurate.

Moreover, the static DCG--EGG proof makes the joint zero-frequency system controllable under the same conditions. The joint construction in Sec.~\ref{app:waveform-finite-frequency-extension} then imposes all selected frequencies while preserving the EGG condition. The same sufficient conditions also guarantee waveform-selective robustness for any finite collection of selected smooth real functions $f_{p,\ell}(t)$ on $[0, T]$. We can also impose the EGG condition simultaneously.

For a full basis $\{iX_p\}_{p=1}^r$ of $\mathfrak{k}$, with $r=\dim\mathfrak{k}$, the all-static DCG result gives the sufficient condition $\mathcal C_{\mathrm{dyn}}=\{0\}$. For finite-frequency robustness, we now prove its necessity when $\bigcap_{p=1}^r\Omega_p\neq\varnothing$.

We choose a common frequency $\omega$ and suppose that $0\neq C\in\mathcal C_{\mathrm{dyn}}$. Both $C$ and $[C,-i\staticterm]$ are in the space $\operatorname{span}_{\mathbb R}\{iX_1,\ldots,iX_r\}$. We use $Y_C(t)=U_0^\dagger(t)CU_0(t)$, with $\dot Y_C(t)=\staticcoe U_0^\dagger(t)[C,-i\staticterm]U_0(t)$ as in the proof of the all-static DCG conditions. At the common frequency $\omega$, every basis response vanishes, so the linearity of the response integral gives
\begin{equation*}
\int_0^T e^{i\omega t}Y_C(t)\,\mathrm dt=0,\qquad
\int_0^T e^{i\omega t}\dot Y_C(t)\,\mathrm dt=0.
\end{equation*}
With $Y_C(0)=C$ and $Y_C(T)=\target^\dagger C\target$, we integrate by parts:
\begin{equation}
\begin{aligned}
0&=\int_0^T e^{i\omega t}\dot Y_C(t)\,\mathrm dt\\
&=e^{i\omega T}\target^\dagger C\target-C
-i\omega\int_0^T e^{i\omega t}Y_C(t)\,\mathrm dt.
\end{aligned}
\end{equation}
Every target implemented by such a robust protocol must therefore obey
\begin{equation}
e^{i\omega T}\target^\dagger C\target=C.
\label{eqapp:finite-frequency-common-symmetry}
\end{equation}
The identity remains valid for $\staticcoe=0$.

The Hermitian adjoint of Eq.~\eqref{eqapp:finite-frequency-common-symmetry}, with $C^\dagger=-C$, gives the same equality with $e^{-i\omega T}$. Since $C\neq0$, the phase must therefore be $1$ or $-1$; any nonreal phase excludes every target. For phase $1$, targets not commuting with $C$ are excluded, and such targets exist because $C$ is not central. For phase $-1$, continuity of $U\mapsto U^\dagger CU$ at $I$ excludes a neighborhood of $I$, including nonidentity targets, because $C\neq-C$. Thus, for every prescribed $T>0$, a nonzero common symmetry excludes a family covering all targets. The same conclusion holds when $T$ is chosen separately for each target: a target $\target$ near $I$ with $[C,\target]\neq0$ satisfies $\target^\dagger C\target\neq\pm C$, so Eq.~\eqref{eqapp:finite-frequency-common-symmetry} fails for every $T>0$. This completes the necessity proof.

Together, these results show that the full-basis condition $\mathcal C_{\mathrm{dyn}}=\{0\}$ is sufficient for arbitrary finite frequency assignments and necessary when the sets have a common element. We make no necessity claim when their intersection is empty. The EGG condition can be added under the sufficient condition; it cannot remove the necessary endpoint constraint, since any joint protocol must still satisfy the finite-frequency conditions. If the driven Hamiltonians alone generate $\mathfrak{k}$, the common symmetry space is zero and all these constraints can be imposed simultaneously. \hfill$\square$

For $\mathfrak{k}=\mathfrak{su}(\dimension)$, we have $r=\dimension^2-1$. Together with the zero-frequency results above, this proves the Hamiltonian criteria in Sec.~\ref{sec:theoryB}, including Eq.~\eqref{eq:reachability-all-DCG} and the driven-only case in Eq.~\eqref{EQ:full-driving-term}.

The zero-frequency algebra equalities hold for each prescribed finite $\staticcoe\neq0$. Section~\ref{app:fixed-static-frequency} uses them for the fixed-$\staticcoe$ finite-frequency implementation, with variable $T$ and unchanged physical frequencies.

\subsection{Finite-Frequency Extension for a Fixed Static Coefficient}
\label{app:fixed-static-frequency}

Under full unitary controllability, we prove the statement in Sec.~\ref{sec:theoryB} that the Hamiltonian in Eq.~\eqref{eq:control_resources} can satisfy the selected finite-frequency conditions at any prescribed finite $\staticcoe=\bar u_0\neq0$, with the gate duration free to vary. We assume that $K$ is compact and connected, $\mathfrak{k}$ is simple, and $\{iX_p\}_{p=1}^r$ is a basis of $\mathfrak{k}$, where $r=\dim\mathfrak{k}$. For the same $\bar u_0$, the zero-frequency DCG generators are assumed to generate the full algebra $\mathfrak{k}\rtimes\mathfrak{k}^r$ after the response coordinates are multiplied by $i$. The zero center of $\mathfrak{k}$ then gives zero-frequency controllability by the compact-group criterion. For every $\target\in K$ and every family of finite sets $\Omega_p\subset[0,\infty)$, we construct a finite $T>0$ and smooth $T$-periodic controls satisfying
\begin{equation*}
\begin{aligned}
U_0(T)&=\target,\\
\mathcal A_p^{(\mathrm D)}(\omega)&=0,\quad
\omega\in\Omega_p\cup\{0\},\quad p=1,\ldots,r.
\end{aligned}
\end{equation*}
We can also impose the EGG condition on the same controls. We first prove the finite-frequency statement for commensurate frequencies, then remove the commensurability assumption, and finally include EGG.

\textit{Commensurate frequencies.} We collect the distinct positive frequencies in $\Omega^+=\{\omega_1,\ldots,\omega_Q\}$, where $Q$ is their number. If $Q=0$, zero-frequency controllability and Lemma~\ref{lem:smooth-endpoint-correction} give the DCG statement directly. For $Q\geq1$, we first assume $\omega_q=n_q\omega_*$, with $\omega_*>0$ and distinct positive integers $n_q$. We encode the common phase and all responses in a semidirect product, and verify the hypotheses of the compact-group controllability theorem.

We keep the zero-frequency DCG responses and add a cosine and a sine response at each selected positive frequency. In anti-Hermitian response coordinates, the zero-frequency group is $G_0=K\rtimes V_0$, with $V_0=\mathfrak{k}^r$ and Lie algebra $\mathfrak g_0=\mathfrak{k}\rtimes V_0$. We denote the image of $\Gamma_{\mathrm{DCG}}$ under $(a,v)\mapsto(a,iv)$ by $\Gamma_0$. For each frequency $\omega_q$, the set $S_q=\{p:\omega_q\in\Omega_p\}$ specifies the included perturbation operators. Their response tuple and its conjugation span are
\begin{equation*}
\boldsymbol X_{S_q}=(iX_p)_{p\in S_q},\qquad
V_q=\operatorname{span}_{\mathbb R}
\{U^\dagger\boldsymbol X_{S_q}U\mid U\in K\}.
\end{equation*}
Conjugation acts on every operator in a tuple. The space of positive-frequency responses is
\begin{equation*}
V_\Omega=\bigoplus_{q=1}^Q(V_{q,c}\oplus V_{q,s}),
\qquad V_{q,c}\simeq V_{q,s}\simeq V_q.
\end{equation*}

The phase must be recorded because a control segment's weighted response depends on its starting time. We use elapsed time $\tau$, reduced modulo $2\pi/\omega_*$, as the coordinate on $S^1$. With $r_q(t)=U_0^\dagger(t)\boldsymbol X_{S_q}U_0(t)$, the added responses $w_{q,c}\in V_{q,c}$ and $w_{q,s}\in V_{q,s}$ obey
\begin{equation}
\begin{aligned}
\dot w_{q,c}&=r_q(t)\cos(\omega_q\tau),\\
\dot w_{q,s}&=r_q(t)\sin(\omega_q\tau),\qquad \dot\tau=1,
\end{aligned}
\label{eqapp:fixed-u0-augmented-dynamics}
\end{equation}
with all response and phase coordinates initially zero. The cosine and sine responses vanish together exactly when the corresponding complex-frequency response vanishes. A time shift $s$, at fixed initial gate coordinate, rotates the response pairs through
\begin{equation*}
\rho_\Omega(s)=\bigoplus_{q=1}^Q
\begin{pmatrix}
\cos(\omega_qs)&-\sin(\omega_qs)\\
\sin(\omega_qs)&\cos(\omega_qs)
\end{pmatrix}\otimes I_{V_q},
\end{equation*}
where $I_{V_q}$ is the identity on $V_q$. The period $2\pi/\omega_*$ makes $\rho_\Omega$ an action of the common circle. The augmented group is $(K\times S^1)\rtimes\widehat V_\Omega$, where $\widehat V_\Omega=V_0\oplus V_\Omega$; the circle fixes $V_0$, while $K$ conjugates every response operator.

We now prove full Lie generation by separating the Hamiltonian, phase, and response components. The derivative of $\rho_\Omega$ at zero defines $D_\Omega$ on $V_\Omega$, and we define $D_\Omega=0$ on $V_0$. Thus
\begin{equation}
D_\Omega(c_q,s_q)=\omega_q(-s_q,c_q),
\qquad(c_q,s_q)\in V_{q,c}\oplus V_{q,s}.
\label{eqapp:fixed-u0-clock-generator}
\end{equation}
The augmented algebra is $\widehat{\mathfrak g}_\Omega=\mathfrak{k}\oplus\widehat V_\Omega\oplus\mathbb R$. For $Y,Z\in\mathfrak{k}$, $y,z\in\widehat V_\Omega$, and $\sigma_1,\sigma_2\in\mathbb R$, the right-action convention gives
\begin{equation}
\begin{aligned}
&[(Y,y,\sigma_1),(Z,z,\sigma_2)]\\
&\quad=\bigl([Y,Z],[Y,z]-[Z,y]
-\sigma_1D_\Omega z+\sigma_2D_\Omega y,0\bigr).
\end{aligned}
\label{eqapp:fixed-u0-augmented-bracket}
\end{equation}
We write $a_{\mathrm d}=-i\bar u_0\staticterm$, $a_j=-i\scheduleterm$, and $\boldsymbol X=(iX_1,\ldots,iX_r)$. The generators of $\Gamma_0$ are $A_0=(a_{\mathrm d},\boldsymbol X)$ and $B_{0,j}=(a_j,0)$. The augmented control system $\widehat\Gamma_\Omega$ has generators
\begin{equation*}
\begin{aligned}
\widehat A&=(a_{\mathrm d},w_*,1),\qquad
\widehat B_j=(a_j,0,0),\\
w_*&=\bigl(\boldsymbol X,(\boldsymbol X_{S_q},0)_{q=1}^Q\bigr).
\end{aligned}
\end{equation*}
In an augmented element, the middle coordinate contains all zero- and positive-frequency responses, and the last coordinate is the phase component. We denote the generated algebra by $\widehat{\mathfrak h}=\operatorname{Lie}(\widehat\Gamma_\Omega)$.

To generate $(Y,0,0)$ for every $Y\in\mathfrak{k}$, we first remove the phase component from an element with Hamiltonian component $a_{\mathrm d}$. At least one $a_j$ is nonzero, because one static element cannot generate a nonabelian simple algebra. The ideal generated by $a_1,\ldots,a_m$ is therefore $\mathfrak{k}$. Consequently, $a_{\mathrm d}$ is a real linear combination of nested brackets in $a_{\mathrm d},a_1,\ldots,a_m$, with a driven element in every term. We evaluate the expression for $a_{\mathrm d}$ on $\widehat A,\widehat B_1,\ldots,\widehat B_m$. Each driven element and every bracket have zero phase component, so the result has the form $\widetilde A=(a_{\mathrm d},\widetilde w,0)\in\widehat{\mathfrak h}$, where $\widetilde w\in\widehat V_\Omega$ need not equal $w_*$.

For each $Y\in\mathfrak{k}$, full zero-frequency generation provides a finite expression $P_Y$, formed from Lie brackets and real linear combinations, such that
\begin{equation*}
P_Y(A_0,B_{0,1},\ldots,B_{0,m})=(Y,0).
\end{equation*}
For a single response input $\eta\in\mathfrak{k}$, we define $T_Y:\mathfrak{k}\to\mathfrak{k}$ by
\begin{equation*}
P_Y\bigl((a_{\mathrm d},\eta),(a_1,0),\ldots,(a_m,0)\bigr)
=(Y,T_Y(\eta)).
\end{equation*}
The map $T_Y$ is real-linear: Eq.~\eqref{eqapp:fixed-u0-augmented-bracket} contains no products of responses, and the driven responses are zero, so every nested response expression remains linear in $\eta$. The zero-frequency calculation gives $T_Y(iX_p)=0$ for every basis element, so $T_Y=0$.

We now evaluate $P_Y$ on $\widetilde A,\widehat B_1,\ldots,\widehat B_m$. All phase components are zero, eliminating the $D_\Omega$ terms from every bracket, and every response component of $\widetilde w$ lies in $\mathfrak{k}$. The same map $T_Y$ therefore annihilates every component of $\widetilde w$, giving
\begin{equation*}
P_Y(\widetilde A,\widehat B_1,\ldots,\widehat B_m)
=(Y,0,0)\in\widehat{\mathfrak h}.
\end{equation*}

We next separate the phase from the responses. Subtraction of $(a_{\mathrm d},0,0)$ from $\widehat A$ gives $(0,w_*,1)$. Since $\widehat{\mathfrak h}$ contains every $(Y,0,0)$ and $K$ is connected, conjugation by $K$ preserves $\widehat{\mathfrak h}$. The normalized group average of $(0,w_*,1)$ belongs to $\widehat{\mathfrak h}$, a finite-dimensional closed subspace. Every averaged response operator commutes with $K$ and belongs to $\mathfrak{k}$, whose center is zero. The average is therefore $(0,0,1)$, and subtraction from $(0,w_*,1)$ also gives $(0,w_*,0)$.

The pure phase element now lets us separate the frequency responses. Repeated brackets with $(0,0,1)$ and real linear combinations give $(0,D_\Omega^\ell w_*,0)$ for every integer $\ell\geq0$. Since $\widehat{\mathfrak h}$ is a finite-dimensional closed subspace, the phase conjugates $(0,e^{sD_\Omega}w_*,0)$ also belong to $\widehat{\mathfrak h}$. The coefficients of the zero-frequency, cosine, and sine tuples in $e^{sD_\Omega}w_*$ are $1$, $\cos(\omega_qs)$, and $\sin(\omega_qs)$, respectively. These functions are linearly independent, so some $1+2Q$ values of $s$ give an invertible matrix of coefficient values. Real linear combinations of the corresponding conjugates isolate each cosine and sine tuple; conjugation by $K$ then spans each copy of $V_q$.

The remaining task is to separate the zero-frequency coordinates. The projection of $\widehat{\mathfrak h}$ onto $\mathfrak g_0$ is surjective because the projected generators are $A_0,B_{0,j}$. For any prescribed element of $\mathfrak g_0$, we choose a generated augmented element with that projection and subtract its phase and positive-frequency components. Thus
\begin{equation}
\operatorname{Lie}(\widehat\Gamma_\Omega)
=\widehat{\mathfrak g}_\Omega.
\label{eqapp:fixed-u0-lifted-rank}
\end{equation}

For commensurate frequencies, full Lie generation completes the controllability proof. The factor $K\times S^1$ is compact and connected, and no nonzero response is fixed under the action of $K$, because every response operator lies in the algebra $\mathfrak{k}$ with zero center. The controllability theorem stated earlier in this appendix therefore provides any target gate with all zero- and positive-frequency responses zero at a finite time. The final phase is unrestricted by the robustness conditions. Lemma~\ref{lem:smooth-endpoint-correction} gives smooth periodic schedules with the same endpoint and the same $\bar u_0$.

\textit{Extension to arbitrary finite frequencies.} We now remove the common-period assumption and use the elapsed-time coordinate $\tau\in\mathbb R$. The Lie-generation proof remains valid: the construction of $\widetilde A$ and the evaluation of $P_Y$ require no relation between the frequencies, and the frequency-separation step uses only the linear independence of $1$, $\cos(\omega_qs)$, and $\sin(\omega_qs)$. Section~\ref{app:waveform-finite-frequency-extension} establishes that independence for arbitrary distinct positive frequencies. Equation~\eqref{eqapp:fixed-u0-lifted-rank} therefore holds for the augmented group $(K\times\mathbb R)\rtimes\widehat V_\Omega$. The factor $K\times\mathbb R$ is not compact, so we replace the global controllability argument with local endpoint control and a finite concatenation.

We use blocks that return the gate to $I$ and every zero-frequency response to zero, while allowing a small change of the positive-frequency response. We will show that repetition at a suitable common duration leaves a bounded total response. We can then distribute the required cancellation among enough blocks to keep every correction within the local adjustment range.

\textit{A locally adjustable block.} We require a smooth block with fixed duration and independently adjustable positive-frequency response. Lemma~\ref{lem:full-rank-endpoint-map} gives a positive-duration piecewise-constant schedule whose endpoint map has full rank in the $G_0$, positive-frequency, and elapsed-time coordinates. The full-rank schedule ends at $(g_*,c_*,t_*)$, where $g_*\in G_0$, $c_*\in V_\Omega$, and $t_*>0$. Zero-frequency controllability supplies a fixed return schedule from $g_*$ to the identity $e_0=(I,0)$ of $G_0$. We append the return schedule to the full-rank schedule. The return schedule acts on the augmented endpoint by a diffeomorphism, so the endpoint map of the concatenation remains full rank. Lemma~\ref{lem:smooth-endpoint-correction} then gives a smooth family realizing every endpoint near the concatenation's endpoint $(e_0,c,t_{\mathrm c})$, where $c\in V_\Omega$ and $t_{\mathrm c}>0$.

Because the local family covers a neighborhood in all endpoint coordinates, we may keep the $G_0$ endpoint at $e_0$ and the positive-frequency response at $c$ while varying the duration near $t_{\mathrm c}$. We choose a positive duration $t_0$ satisfying
\begin{equation}
e^{i\omega_qt_0}\neq1,\qquad q=1,\ldots,Q.
\label{eqapp:fixed-u0-nonresonant-duration}
\end{equation}
For finitely many nonzero frequencies, the excluded times form a locally finite union of discrete sets and cannot fill a time interval. We now fix both $t_0$ and the $G_0$ endpoint. We use the product Frobenius norm on response tuples. The local family still allows every positive-frequency response $c+z$ with $\|z\|<\varepsilon$, for some $\varepsilon>0$. We denote the corresponding control block by $N(z)$:
\begin{equation}
N(z):(e_0,0,0)\longmapsto(e_0,c+z,t_0),
\qquad \|z\|<\varepsilon.
\label{eqapp:fixed-u0-local-block}
\end{equation}
The positive-frequency response $c$ need not vanish; only the freedom to adjust $c$ locally is required.

The blocks $N(z)$ return the gate to $I$, so we append a separate block to implement $\target$. Zero-frequency controllability and Lemma~\ref{lem:smooth-endpoint-correction} give a smooth target block $B_f$ with $G_0$ endpoint $(\target,0)$ and duration $t_f>0$. We denote the positive-frequency response of $B_f$, measured from starting time zero, by $b\in V_\Omega$. Every block uses $\bar u_0$, and the smooth-endpoint lemma allows the same prescribed constant driven-schedule values near both ends of every block.

\textit{Exact cancellation by concatenation.} For an integer $L\geq1$, we apply $N(z_0),\ldots,N(z_{L-1})$ and then $B_f$. The $k$th adjustable block starts at time $kt_0$, with gate coordinate $I$; a starting-time shift therefore rotates its positive-frequency response by $\rho_\Omega(kt_0)$ without an additional gate conjugation. The total positive-frequency response is
\begin{equation}
M_L=\sum_{k=0}^{L-1}\rho_\Omega(kt_0)(c+z_k)
+\rho_\Omega(Lt_0)b.
\label{eqapp:fixed-u0-total-response}
\end{equation}
We denote the uncorrected response by $w_L$. With $I_{V_\Omega}$ the identity on $V_\Omega$, Eq.~\eqref{eqapp:fixed-u0-nonresonant-duration} makes $I_{V_\Omega}-\rho_\Omega(t_0)$ invertible, and the finite geometric sum gives
\begin{equation*}
\begin{aligned}
w_L&=\sum_{k=0}^{L-1}\rho_\Omega(kt_0)c+\rho_\Omega(Lt_0)b\\
&=[I_{V_\Omega}-\rho_\Omega(Lt_0)]
[I_{V_\Omega}-\rho_\Omega(t_0)]^{-1}c
+\rho_\Omega(Lt_0)b.
\end{aligned}
\end{equation*}
Each $\rho_\Omega(s)$ is orthogonal, so
\begin{equation*}
\|w_L\|\leq
2\|[I_{V_\Omega}-\rho_\Omega(t_0)]^{-1}c\|+\|b\|.
\end{equation*}
The bound is independent of $L$. We choose the local corrections as
\begin{equation}
z_k=-\frac1L\rho_\Omega(-kt_0)w_L,
\qquad k=0,\ldots,L-1.
\label{eqapp:fixed-u0-corrections}
\end{equation}
Every rotated correction equals $-w_L/L$, hence
\begin{equation*}
\sum_{k=0}^{L-1}\rho_\Omega(kt_0)z_k=-w_L,
\qquad M_L=0.
\end{equation*}
Moreover, $\max_k\|z_k\|=\|w_L\|/L$ tends to zero. One sufficiently large finite $L$ therefore places every $z_k$ inside the ball of radius $\varepsilon$, where all blocks $N(z_k)$ exist.

The concatenation reaches $(\target,0)$ in $G_0$ and cancels every selected positive-frequency response exactly. Its duration is $T=Lt_0+t_f<\infty$, with $\staticcoe=\bar u_0$ unchanged. The constant driven-schedule values near each block endpoint agree, and every derivative of positive order vanishes there. The coefficients consequently join smoothly between blocks and across $t=0$ and $t=T$, giving the required $C^\infty$ $T$-periodic controls.

\textit{The EGG condition.} We prove the simultaneous EGG statement by establishing full generation after adding the EGG response. We use the coordinate order $(Y,v_0,-iv_{\mathrm{EGG}},w,\sigma)$, where $Y\in\mathfrak{k}$, $v_0\in V_0$, $w\in V_\Omega$, and $\sigma\in\mathbb R$. The augmented static generator has EGG component $a_{\mathrm d}$, and the $j$th augmented driven generator has EGG component $a_j$; all other components agree with $\widehat A$ and $\widehat B_j$. We repeat on the joint generators the expression that produced $(a_{\mathrm d},0,0)$. The expression produces an element $(a_{\mathrm d},0,\eta,0,0)$ for some $\eta\in\mathfrak{k}$. Together with the driven elements $(a_j,0,a_j,0,0)$, this element generates every Hamiltonian component while keeping all static DCG, frequency, and phase components zero.

Suppose the algebra generated by these elements contains no nonzero pure EGG response. Each Hamiltonian component $Y$ then has a unique EGG component, which we denote by $f(Y)$; linear combinations make $f:\mathfrak{k}\to\mathfrak{k}$ real-linear. The bracket gives $f([Y,Z])=[f(Y),Z]+[Y,f(Z)]$, so $f$ is a derivation. Since $\mathfrak{k}$ is simple, there exists $C\in\mathfrak{k}$ such that $f(Y)=[C, Y]$ for every $Y\in\mathfrak{k}$~\cite[Proposition~6.8]{Kirillov2008Lie}. Each driven joint generator requires $f(a_j)=a_j$, hence
\begin{equation*}
\|a_j\|^2=-\operatorname{tr}\bigl(a_j[C,a_j]\bigr)=0.
\end{equation*}
All driven Hamiltonians would vanish, contradicting the generation of the nonabelian algebra $\mathfrak{k}$. A nonzero pure EGG response therefore exists.

Brackets with arbitrary Hamiltonian components make the pure EGG responses a nonzero ideal, so simplicity gives the whole EGG copy. We lift any augmented DCG element by repeating its generating expression on the joint generators, then subtract the EGG component. Thus the joint augmented algebra is full. If $Q=0$, we begin with $(a_{\mathrm d},0)\in\mathfrak g_0$ instead of the augmented element; the same EGG argument applies without frequency or phase coordinates.

Projection onto the gate, static DCG, and EGG coordinates gives full joint zero-frequency generation. The joint zero-frequency system is controllable by the compact-group criterion, since both response spaces have zero fixed subspace. For $Q\geq1$, we apply the local-endpoint construction with EGG included among the zero-frequency responses in $G_0$. Each $N(z)$ and $B_f$ then has zero EGG response, while the parameters $z_k$ change only positive-frequency DCG responses. The same finite concatenation therefore imposes EGG and all selected frequencies simultaneously.

For other zero-frequency response spaces and any fixed static coefficient, the local-block construction requires only controllability of $\Gamma_0$ on $G_0$ and full augmented Lie rank. For the full-unitary case with $\mathcal C_{\mathrm{dyn}}=\{0\}$, the zero-frequency DCG criterion supplies the opening hypothesis for every prescribed finite $\bar u_0\neq0$. The result therefore applies to every $\mathrm{SU}(\dimension)$ target and every finite direction-dependent frequency assignment, with or without EGG.

The piecewise-constant endpoint maps also provide the local freedom in $N(z)$, and zero-frequency controllability provides a piecewise-constant $B_f$. The same concatenation therefore gives piecewise-constant $T$-periodic controls with $\staticcoe=\bar u_0$, with or without EGG.

For all existence criteria, we impose no upper bound on the Hamiltonian amplitudes. For a fixed target and frequency assignment, the corrections remain in a fixed compact subset of the smooth family $N(z)$ for all sufficiently large $L$, so the block amplitudes stay bounded while $T=Lt_0+t_f$ grows linearly. By contrast, the fixed-$T$ construction in Sec.~\ref{app:waveform-finite-frequency-extension} uses a common amplitude factor of order $L$. \hfill$\square$

\subsection{The Fixed Depolarization Direction}
\label{app:Lindblad-fixed-direction-proof}

We first identify the fixed subspace under all coherent unitary conjugations and explain what this implies for the general group $K_{\mathrm{Lind}}$ in Sec.~\ref{sec:theoryC}. Under Eq.~\eqref{EQ:full-driving-term}, we then verify the conditions of the general criterion and obtain the Lindblad result in Sec.~\ref{sec:theoryB}.

\textit{Fixed direction under all coherent unitary conjugations.} We prove that $D_{\mathrm{fix}}$ spans the fixed subspace in the real span of Lindblad dissipators, without a controllability assumption. For $U\in\mathrm{SU}(\dimension)$, the matrix $\mathcal U_U=U\otimes U^*$ represents the map $X\mapsto UXU^\dagger$ in Hilbert--Schmidt space.

The fixed depolarization direction from Sec.~\ref{sec:theoryA} acts on a system operator $X$ as
\begin{equation}
D_{\mathrm{fix}}(X)
=\frac{\operatorname{tr}(X)}{\dimension}\identity_{\dimension}-X.
\end{equation}
It is itself a Lindblad dissipator. For an orthonormal basis $\{|j\rangle\}_{j=1}^{\dimension}$, we define $L_{jk}=|j\rangle\langle k|/\sqrt{\dimension}$ for $j,k=1,\ldots,\dimension$ and obtain
\begin{equation}
\begin{aligned}
&\sum_{j,k=1}^{\dimension}\left(L_{jk}XL_{jk}^\dagger
-\frac12\{L_{jk}^\dagger L_{jk},X\}\right)\\
&\qquad=\frac{\operatorname{tr}(X)}{\dimension}\identity_{\dimension}-X.
\end{aligned}
\end{equation}
Coherent conjugation leaves this map fixed.

For $\dimension\geq2$, we denote the space of linear operators on $\mathbb C^{\dimension}$ by $\mathcal B(\mathbb C^{\dimension})$ and its complex subspace of traceless operators by $\mathcal B_0$. Coherent conjugation preserves the decomposition
\begin{equation}
\mathcal B(\mathbb C^{\dimension})
=\operatorname{span}_{\mathbb C}\{\identity_{\dimension}\}
\oplus\mathcal B_0.
\end{equation}
It acts trivially on the first subspace. For any invariant complex subspace of $\mathcal B_0$, we differentiate the conjugation action. The subspace is then closed under commutators with all traceless anti-Hermitian matrices and, by complex linearity, with all traceless complex matrices. It is therefore an ideal of the simple Lie algebra of traceless complex matrices, so it is either zero or $\mathcal B_0$. Thus the action on $\mathcal B_0$ is irreducible.

We take $M$ in the real linear span of Lindblad dissipators and assume
\begin{equation}
\mathcal U_U^\dagger M\mathcal U_U=M
\qquad
\text{for every }U\in\mathrm{SU}(\dimension).
\end{equation}
Such an $M$ is complex-linear, preserves Hermiticity, and satisfies $\operatorname{tr}[M(X)]=0$ for every system operator $X$. Its invariance implies that $M(\identity_{\dimension})$ commutes with every $U\in\mathrm{SU}(\dimension)$, so it is proportional to $\identity_{\dimension}$. The trace condition then gives $M(\identity_{\dimension})=0$ and also ensures that $M$ maps $\mathcal B_0$ into itself. Schur's lemma gives a scalar $\beta\in\mathbb C$ such that $M(X)=\beta X$ for every $X\in\mathcal B_0$~\cite[Lemma~4.23]{Kirillov2008Lie}. Hermiticity preservation implies $\beta\in\mathbb R$. Therefore, for every system operator $X$,
\begin{equation}
M(X)=\beta\left(X-\frac{\operatorname{tr}(X)}{\dimension}
\identity_{\dimension}\right)
=-\beta D_{\mathrm{fix}}(X).
\end{equation}
Since $D_{\mathrm{fix}}$ is itself a fixed Lindblad dissipator, the fixed subspace in the real span of Lindblad dissipators is exactly
\begin{equation}
\operatorname{span}_{\mathbb R}\{D_{\mathrm{fix}}\}.
\label{eqapp:unique-fixed-Lindblad-direction}
\end{equation}
\textit{The general group $K_{\mathrm{Lind}}$.} We now distinguish invariance under $K_{\mathrm{Lind}}$ from invariance under all coherent unitary conjugations. Every element of $K_{\mathrm{Lind}}$ represents such a conjugation, so it fixes $D_{\mathrm{fix}}$. The averaged response therefore satisfies
\begin{equation*}
\frac1T\int_0^T\mathrm dt\,
\mathcal U_0^\dagger(t)D_{\mathrm{fix}}\mathcal U_0(t)
=D_{\mathrm{fix}}\neq0,
\qquad T>0.
\end{equation*}

By definition, each $D_{a,\mathrm{sup}}$ lies in the real span of Lindblad dissipators and is orthogonal to $D_{\mathrm{fix}}$ under the Hilbert--Schmidt inner product on superoperators. Coherent conjugation preserves this real span and this inner product, and fixes $D_{\mathrm{fix}}$. Every component of every vector in $V_{\mathrm{Lind}}$ therefore lies in that real span and remains orthogonal to $D_{\mathrm{fix}}$. 

For a restricted control group, a superoperator may be fixed by every element of $K_{\mathrm{Lind}}$ without being fixed by all coherent unitary conjugations. Orthogonality to $D_{\mathrm{fix}}$ therefore does not by itself imply $V_{\mathrm{Lind}}^{K_{\mathrm{Lind}}}=\{0\}$. The general criterion in Eq.~\eqref{eq:general-criteria} still requires this condition, together with compactness of $K_{\mathrm{Lind}}$ and $\operatorname{Lie}(\Gamma_{\mathrm{Lind}})=\mathfrak{g}_{\mathrm{Lind}}$.

\textit{When the driven Hamiltonians generate $\mathfrak{su}(\dimension)$.} We now assume Eq.~\eqref{EQ:full-driving-term} and verify the three conditions in Eq.~\eqref{eq:general-criteria}. Under this assumption, $K_{\mathrm{Lind}}$ is the full group of coherent unitary conjugations:
\begin{equation}
K_{\mathrm{Lind}}
=\{\mathcal U_U\mid U\in\mathrm{SU}(\dimension)\}.
\label{eqapp:Lindblad-group}
\end{equation}
The group $K_{\mathrm{Lind}}$ is compact and connected because it is the continuous image of $\mathrm{SU}(\dimension)$.

If a response vector is fixed by every element of $K_{\mathrm{Lind}}$, each component is a real multiple of $D_{\mathrm{fix}}$ by Eq.~\eqref{eqapp:unique-fixed-Lindblad-direction}. The only such multiple orthogonal to $D_{\mathrm{fix}}$ is zero. Thus
\begin{equation}
V_{\mathrm{Lind}}^{K_{\mathrm{Lind}}}=\{0\}.
\label{eqapp:Lindblad-no-fixed-point}
\end{equation}
Equation~\eqref{EQ:full-driving-term} also implies that $C(\scheduleterm)$, $j=1,\ldots,m$, generate $\mathfrak{k}_{\mathrm{Lind}}$. We take differences of elements of $\Gamma_{\mathrm{Lind}}$ and obtain $(C(\scheduleterm),0)$ in $\operatorname{Lie}(\Gamma_{\mathrm{Lind}})$. Their Lie brackets and real linear combinations give $(a,0)$ for every $a\in\mathfrak{k}_{\mathrm{Lind}}$, including $(\staticcoe C(\staticterm),0)$. We subtract this element from $A_{\mathrm{Lind}}$, which was defined in the main text as
\begin{equation}
A_{\mathrm{Lind}}
=(\staticcoe C(\staticterm),d_{\mathrm{sup}}),
\end{equation}
and obtain $(0,d_{\mathrm{sup}})\in\operatorname{Lie}(\Gamma_{\mathrm{Lind}})$.

The vector coordinates of $(0,d_{\mathrm{sup}})$ and its repeated commutators with $(a,0)$ span the smallest real $\mathfrak{k}_{\mathrm{Lind}}$-invariant subspace containing $d_{\mathrm{sup}}$. Because $K_{\mathrm{Lind}}$ is connected, a real subspace is invariant under $K_{\mathrm{Lind}}$ if and only if it is invariant under $\mathfrak{k}_{\mathrm{Lind}}$. This smallest subspace is therefore the real span of all coherent conjugates of $d_{\mathrm{sup}}$, which is $V_{\mathrm{Lind}}$ by definition. Thus $\operatorname{Lie}(\Gamma_{\mathrm{Lind}})$ contains $(0,v)$ for every $v\in V_{\mathrm{Lind}}$, as well as $(a,0)$ for every $a\in\mathfrak{k}_{\mathrm{Lind}}$. Hence
\begin{equation}
\operatorname{Lie}(\Gamma_{\mathrm{Lind}})
=\mathfrak{k}_{\mathrm{Lind}}\rtimes V_{\mathrm{Lind}}
=\mathfrak{g}_{\mathrm{Lind}}.
\end{equation}
Together with the compactness of $K_{\mathrm{Lind}}$ and Eq.~\eqref{eqapp:Lindblad-no-fixed-point}, this equality satisfies the general criterion in Eq.~\eqref{eq:general-criteria} and proves the Lindblad statement in Sec.~\ref{sec:theoryB}.

\addtocontents{toc}{\protect\let\protect\contentsline\protect\skipcontentsline}
\bibliography{ref}

\end{document}